\documentclass[twoside,twocolumn,9pt]{article}
\usepackage{extsizes}
\usepackage[super,sort&compress,comma]{natbib}
\usepackage[left=1.5cm, right=1.5cm, top=1.785cm, bottom=2.0cm]{geometry}
\usepackage{balance}
\usepackage{mathptmx}
\usepackage{sectsty}
\usepackage{graphicx}
\usepackage{lastpage}
\usepackage{mathtools}
\usepackage{cuted}
\usepackage[format=plain,justification=justified,singlelinecheck=false,font={stretch=1.125,small,sf},labelfont=bf,labelsep=space]{caption}
\usepackage{float}
\usepackage{fancyhdr}
\usepackage{fnpos}
\usepackage[english]{babel}
\addto{\captionsenglish}{%
  \renewcommand{\refname}{Notes and references}
}
\usepackage{array}
\usepackage{droidsans}
\usepackage{charter}
\usepackage[T1]{fontenc}
\usepackage[usenames,dvipsnames]{xcolor}
\usepackage{setspace}
\usepackage[compact]{titlesec}
\usepackage{hyperref}

\usepackage{epstopdf}
\usepackage{extarrows}
\definecolor{cream}{RGB}{222,217,201}
\newcommand{\ib}[1]{{\color{black}#1}}

\newcommand{\figname}{Fig.~}
\newcommand{\Figname}{Fig.~}

\newcommand{\figsname}{Figs.~}
\newcommand{\secname}{Section~}
\newcommand{\secsname}{Sections~}
\newcommand{\gl}{eqn}
\newcommand{\gls}{eqn}
\newcommand{\Gl}{Eqn}

\newcommand{\alt}{\raisebox{-0.3ex}{$\stackrel{<}{\sim}$}}
\newcommand{\agt}{\raisebox{-0.3ex}{$\stackrel{>}{\sim}$}}
\usepackage[version=4]{mhchem} 
\usepackage{threeparttable}
\DeclareGraphicsExtensions{.jpg, .pdf, .png}

\begin{document}

\pagestyle{fancy}
\thispagestyle{plain}
\fancypagestyle{plain}{
\renewcommand{\headrulewidth}{0pt}
}
\makeFNbottom
\makeatletter
\renewcommand\LARGE{\@setfontsize\LARGE{15pt}{17}}
\renewcommand\Large{\@setfontsize\Large{12pt}{14}}
\renewcommand\large{\@setfontsize\large{10pt}{12}}
\renewcommand\footnotesize{\@setfontsize\footnotesize{7pt}{10}}
\makeatother

\renewcommand{\thefootnote}{\fnsymbol{footnote}}
\renewcommand\footnoterule{\vspace*{1pt}%
\color{cream}\hrule width 3.5in height 0.4pt \color{black}\vspace*{5pt}}
\setcounter{secnumdepth}{5}

\makeatletter
\renewcommand\@biblabel[1]{#1}
\renewcommand\@makefntext[1]%
{\noindent\makebox[0pt][r]{\@thefnmark\,}#1}
\makeatother
\renewcommand{\figurename}{\small{Fig.}~}
\sectionfont{\sffamily\Large}
\subsectionfont{\normalsize}
\subsubsectionfont{\bf}
\setstretch{1.125} 
\setlength{\skip\footins}{0.8cm}
\setlength{\footnotesep}{0.25cm}
\setlength{\jot}{10pt}
\titlespacing*{\section}{0pt}{4pt}{4pt}
\titlespacing*{\subsection}{0pt}{15pt}{1pt}
\fancyfoot{}
\fancyfoot[RO]{\footnotesize{\sffamily{1--\pageref{LastPage} ~\textbar  \hspace{2pt}\thepage}}}
\fancyfoot[LE]{\footnotesize{\sffamily{\thepage~\textbar\hspace{4.65cm} 1--\pageref{LastPage}}}}
\fancyhead{}
\renewcommand{\headrulewidth}{0pt}
\renewcommand{\footrulewidth}{0pt}
\setlength{\arrayrulewidth}{1pt}
\setlength{\columnsep}{6.5mm}
\setlength\bibsep{1pt}
\makeatletter
\newlength{\figrulesep}
\setlength{\figrulesep}{0.5\textfloatsep}

\newcommand{\topfigrule}{\vspace*{-1pt}%
\noindent{\color{cream}\rule[-\figrulesep]{\columnwidth}{1.5pt}} }

\newcommand{\botfigrule}{\vspace*{-2pt}%
\noindent{\color{cream}\rule[\figrulesep]{\columnwidth}{1.5pt}} }

\newcommand{\dblfigrule}{\vspace*{-1pt}%
\noindent{\color{cream}\rule[-\figrulesep]{\textwidth}{1.5pt}} }

\makeatother
\twocolumn[
  \begin{@twocolumnfalse}
\vspace{1em}
\sffamily
\begin{tabular}{m{4.5cm} p{13.5cm} }

  & \noindent\LARGE{\textbf{Can tunneling current in molecular junctions be so strongly temperature dependent to challenge a hopping mechanism?
  Analytical formulas answer this question and provide important insight into large area junctions}} \\
\vspace{0.3cm} & \vspace{0.3cm} \\

 & \noindent\large{Ioan B\^aldea\textit{$^{a}$}} \\

& \noindent\normalsize{Analytical equations like Richardson-Dushman's or Shockley's provided a general, if simplified conceptual background,
which was widely accepted in conventional electronics and made a fundamental contribution to advances in the field.
In the attempt to develop a (highly desirable, but so far missing) counterpart for molecular electronics,
in this work, we deduce a general analytical formula for the tunneling current through molecular junctions
mediated by a single level that is valid for any bias voltage and temperature.
Starting from this expression, which is exact and obviates cumbersome numerical integration,
in the low and high temperature limits    
we also provide analytical formulas expressing the current in terms of elementary functions.
They are accurate for broad model parameter ranges relevant for real molecular junctions.
Within this theoretical framework we show that: (i) by varying the temperature, the tunneling current can vary
by several orders of magnitude, thus debunking the myth that a strong temperature dependence of the current
is evidence for a hopping mechanism, (ii) real molecular junctions can undergo
a gradual (Sommerfeld-Arrhenius) transition 
from a weakly temperature dependent to a strongly (``exponential'') temperature dependent current
that can be tuned by the applied bias, 
\ib{and (iii) important insight into large area molecular junctions with eutectic gallium indium alloy (EGaIn)
  top electrodes can be gained. \emph{E.g.},
  merely based on transport data, we estimate that the current carrying molecules represent only a fraction of
  $ f \approx 4 \times 10^{-4}$ out of the total number of molecules
  in a large area \ce{Au-S-(CH2)13-CH_3 / EGaIn} junction.}} \\

\end{tabular}

 \end{@twocolumnfalse} \vspace{0.6cm}
  ]
\renewcommand*\rmdefault{bch}\normalfont\upshape
\rmfamily
\section*{}
\vspace{-1cm}

\footnotetext{\textit{$^{a}$~Theoretical Chemistry, Heidelberg University, Im Neuenheimer Feld 229, D-69120 Heidelberg, Germany. Fax: +49 6221 545221; Tel: +49 6221 545219; E-mail: ioan.baldea@pci.uni-heidelberg.de}}

\section{Introduction}
\label{sec:intro}
Although representing only a small fraction compared to the overwhelming majority of studies conducted at room temperature,
a number of investigations on charge transport in molecular junctions at variable temperature have been carried out in the past.
\cite{Poot:06,Tao:07c,Choi:08,Reed:09,Choi:10b,Tao:10,Tao:11,Lambert:11,Zandvliet:12,McCreery:13a,McCreery:13b,Amdursky:13b,Asadi:13,Lewis:13,Ferreira:14,Chiechi:15b,Tao:15,Pace:15,Gilbert:15,Frisbie:16a,Frisbie:16d,Tao:16a,Tao:16b,McCreery:16a,Nijhuis:16a,Nijhuis:16b,Nijhuis:16d,Guo:16a,Guo:16b,Guo:17,McCreery:17a,Baldea:2018a,Valianti:19,Guo:21,Park:22}
Charge flow through molecular junctions can occur
via a single-step tunneling mechanism or via a two-step hopping mechanism. 
Whether the measured current depends on temperature or not is taken by conventional wisdom as an expedient criterion
to assign a specific case to hopping or tunneling, respectively.
Presumably the reason why experimentalists resort(ed) to this pragmatic criterion can easily be understood. 
Most model theoretical approaches 
\cite{Caroli:71a,Caroli:71b,Combescot:71,Schmickler:86,Metzger:01b,Stafford:96,Buttiker:03,Baldea:2012a}
were developed for zero temperature ($T = 0$) and, if at all and even if deduced for $T = 0$, compact analytical formulas  
enabling straightforward $I$-$V$ data fitting are rather scarce.\cite{Simmons:63,Schmickler:86,Baldea:2012a,Baldea:2015c}

A temperature dependence of the tunneling current ($I$) can arise due to
the thermal broadening of the electronic Fermi distribution in electrodes.
Recognizing this fact, several investigations drew attention that temperature dependent transport properties
are compatible with a tunneling mechanism.
\cite{Poot:06,Lambert:11,Tao:16b,Nijhuis:16b,Baldea:2017d,Baldea:2017g,Baldea:2018a,Baldea:2018c,Baldea:2021b,Baldea:2021d,Baldea:2022b,Baldea:2022c,Baldea:2022j}

Still, a fundamental question remains to be addressed: 
is the smearing of electrodes Fermi distribution sufficient to cause a temperature dependence
strong enough to compete with that generated by hopping, which is known to yield currents 
exhibiting an exponential Arrhenius-type variation over several orders of magnitude? 
From a pragmatic standpoint, equally important is another question: is it possible
to meet the experimentalists' legitimate need of having compact theoretical formulas expressing the
tunneling current at arbitrary temperature obviating the cumbersome numerical integration (see \gl~(\ref{eq-I}) below)?

In \secname\ref{sec:model}, an analytical exact formula will be presented
which enables to express the tunneling current ($I$) mediated by a single level at any temperature and bias
in terms of digamma functions of complex argument.
Based on that formula, we will show that $I$ can indeed vary over several orders of magnitudes (\secname\ref{sec:tunneling-vs-hopping}),
making discrimination between tunneling and hopping a challenging matter (see Conclusion).
We will also deduce analytical formulas for the current expressed in terms of elementary functions valid in the low and high temperature limits
(\secname\ref{sec:j-0K-T}); they turn out to be very accurate in very broad model parameter ranges
(\secname\ref{sec:analytic-approx}), complementary covering virtually all situations of interest for real molecular junctions.

As an application, starting from experimental current-voltage data measured for a real molecular junction,\cite{Zotti:10}
we will illustrate (\secname\ref{sec:mcbj}) how a (Sommerfeld-Arrhenius \cite{Baldea:2022c}) transition
from a current weakly varying with temperature at low $T$
to a strongly temperature current at high $T$ can be be controlled by tuning the applied bias voltage
within a ``tunneling-throughout'' scenario.

\ib{As a by-product, we show how the presently considered single-level model of transport can provide important insight into
large area molecular junctions with eutectic gallium indium alloy (EGaIn) top electrodes.}
\section{Model and exact formula for the current at arbitrary temperature}
\label{sec:model}

The Keldysh nonequilibrium formalism allows to express the tunneling current through a single molecule mediated by a single level (molecular orbital, MO)
of energy $\varepsilon_{V}$ by \cite{Caroli:71a,HaugJauho}
\begin{equation}
\label{eq-I-gen}
  I = N \frac{2 e}{h} \int_{-\infty}^{\infty}
  \underbrace{\Gamma_{s} \Gamma_t \left\vert \mathcal{G}\left(\varepsilon\right)\right\vert^2}_{\mathcal{T}(\varepsilon)}
  \left[f\left(\varepsilon - \mu_s\right) - f\left(\varepsilon - \mu_t\right)\right] d \varepsilon
\end{equation}
where $f(\varepsilon) = \left[1 + \exp\left(\frac{\varepsilon}{k_B T}\right)\right]^{-1}$
is the Fermi-Dirac distribution, and \(\mu_{s,t} = \pm e V/2\) are the chemical potentials of the
electrodes (``substrate $s$ and ``tip/top'' $t$).

Key quantities in the above equation are the MO-electrode electronic couplings $\Gamma_{s,t}$,
which also enter the retarded Green's function $\mathcal{G}$ of the embedded molecule via Dyson's equation
\begin{equation}
  \label{eq-dyson}
  \mathcal{G}^{-1}(\varepsilon) =
  \varepsilon - \varepsilon_{V} + \underbrace{\frac{i}{2} \left( \Gamma_{s}  + \Gamma_{t} + \Gamma_{env}\right)}_{-\Sigma}
\end{equation}

We emphasize that although for the sake of convenience
we will simply write $\varepsilon_{V}$ almost everywhere,
various forward-backward asymmetries can yield a bias dependent MO energy offset, e.g.
\begin{equation}
  \label{eq-gamma}
  \varepsilon_{V} \to \varepsilon_0 (V) = \epsilon_{0} + \gamma e V
\end{equation}
and all the formula presented below hold for a general bias dependent quantity $\varepsilon_0 (V)$
not even restricted to the (linear in $V$) form mentioned above.

\ib{Notice that, along with electrodes' contribution $\Gamma_{s,t}$, we have also included
  in the Dyson equation (\ref{eq-dyson}) 
  an additional (``extrinsic'') contribution to the self-energy expressed \emph{via} $\Gamma_{env}$.
  This is requested by the physical reality. Except for genuine single-molecule setups---
  \emph{e.g.}, mechanically controlled break junctions (MC-BJ)---,
  the current carrying molecules are placed in an environment.
  In interaction with the surrounding---
  whether molecules of a solvent, neighboring molecules from a self-assembled monolayer,
  possibly also interaction in a cavity---
  the sharp delta-shape MO of an isolated molecule (gas phase)
  acquires a finite width quantified by $\Gamma_{env}$.
  The Green function \(\mathcal{G}\) of \gl~(\ref{eq-dyson}) 
  may refer to a molecule linked to one electrode (\(\Gamma_t \equiv 0\))---
  the case in a ultraviolet photoelectron spectroscopy (UPS) ``half-junction'' setup---
  or linked to two electrodes--- the case of molecular junctions,
  whatever the (coherent one-step tunneling or incoherent two-step sequential) transport mechanism;
  it does not matter, a nonvanishing $\Gamma_{env}$ can exist in all these cases.

  Noteworthily, this reality needs not be accounted for phenomenologically. 
  Whether referring to an UPS setup or to a molecular junction under bias
  the retarded Green's function
  (within an equilibrium formalism \cite{mahan} or in a nonequilibrium Keldysh formalism,\cite{Caroli:71a,HaugJauho}
  respectively) can naturally account for this reality. In the Dyson \gl~(\ref{eq-dyson})
  underlying the present single level model, $\Gamma_{s,t}$ and $\Gamma_{env}$ are accounted for on the same footing.
}

The assumption of a self-energy $\Sigma$ independent
of energy ($\varepsilon$) --- which underlines the ubiquitous Lorentzian transmission $\mathcal{T}(\varepsilon)$ ---
is legitimate in the case of metal electrodes possessing flat energy bands
and environments without special features around the Fermi energy.
\begin{eqnarray}
  \label{eq-I}
  I & = & \frac{2 e}{h} \Gamma^2 \int_{-\infty}^{\infty}
  \frac{f\left(\varepsilon - \mu_s\right) - f\left(\varepsilon - \mu_t\right)}
       {\left(\varepsilon - \varepsilon_{V}\right)^2 + \Lambda^2}
       d \varepsilon \\
       \label{eq-Gamma_g}
\Gamma^2 & \equiv & \Gamma_s \Gamma_t = \Gamma_g^2 \\
       \label{eq-Gamma_a}
       \Lambda & \equiv & \frac{1}{2} \left(\Gamma_s + \Gamma_t + \Gamma_{env}\right) = \Gamma_a + \frac{\Gamma_{env}}{2} 
\end{eqnarray}
Notice that, after replacing $\Gamma_a$ by $\Lambda$, all formulas presented in our earlier works
(e.g., Refs.~\citenum{Baldea:2012a,Baldea:2022c,Baldea:2022j} and \citenum{Baldea:2023a})
in terms of the geometric (``effective'') and arithmetic average ($\Gamma_g$ and $\Gamma_a$) remain valid
although a possible contribution of the surrounding environment to the MO width ($\Gamma_{env}$)
has not been explicitly mentioned.
To avoid possible confusions (see refs.~\citenum{baldea:comment} 
and \citenum{Baldea:2023a}),
we reiterate that the presently defined $\Gamma$'s may differ by a factor of two
from quantities denoted by the same symbol by other authors. 

Expressed as product of a Lorentzian and a difference of Fermi function, the integrand entering \gl~(\ref{eq-I})
is similar to that encountered in other phenomena (e.g., superconductivity \cite{Maki:69} or charge density waves \cite{Baldea:93})
for which contour integration is known to yield exact formulas expressed in terms of Euler's digamma function $\psi$
of complex argument $z$ ($\psi(z) \equiv d\,\Gamma(z)/d\,z$, $\Gamma$ being here Euler's gamma function).
Integration of \gl~(\ref{eq-I}) can be carried out exactly and leads to the following form
\begin{eqnarray}
  \label{eq-Iex}
  I \to I_{exact} & = & \frac{2 e}{h} \frac{\Gamma^2}{\Lambda} \left[
    \mbox{Im}\, \psi\left( \frac{1}{2} + \frac{\Lambda}{2 \pi k_B T} + i \frac{\varepsilon_{V} + e V/2}{2 \pi k_B T}\right) \right . \nonumber \\
    & - & \left .
    \mbox{Im}\, \psi\left( \frac{1}{2} + \frac{\Lambda}{2 \pi k_B T} + i \frac{\varepsilon_{V} - e V/2}{2 \pi k_B T}\right)
    \right]
\end{eqnarray}
Noteworthily, \gl~(\ref{eq-Iex}) is exact at arbitrary temperature $T$ and bias $V$
within the single level model with Lorentzian transmission. Limitations of this model
have been discussed recently \cite{Baldea:2022a,Baldea:2023a} and will not be repeated here.
The above formula generalizes various particular
results deduced for $\Gamma_{env} \equiv 0$
\cite{Stafford:96,Hush:00,Metzger:01b,HaugJauho,Baldea:2010e,Lambert:11,Nijhuis:16a,Baldea:2012a,Ratner:15a},
to which we will also refer below.

\begin{figure*}[htb]
  \centerline{
    \includegraphics[width=0.3\textwidth,height=0.22\textwidth,angle=0]{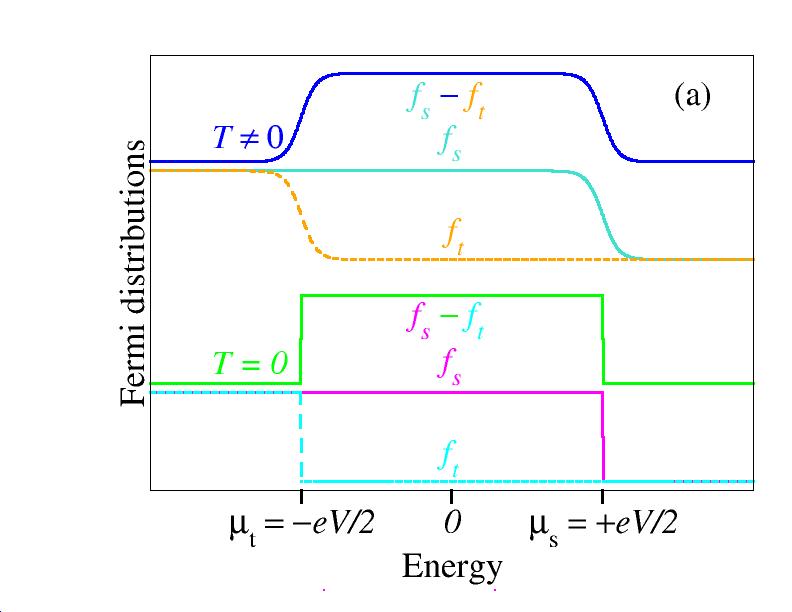} 
    \includegraphics[width=0.3\textwidth,height=0.22\textwidth,angle=0]{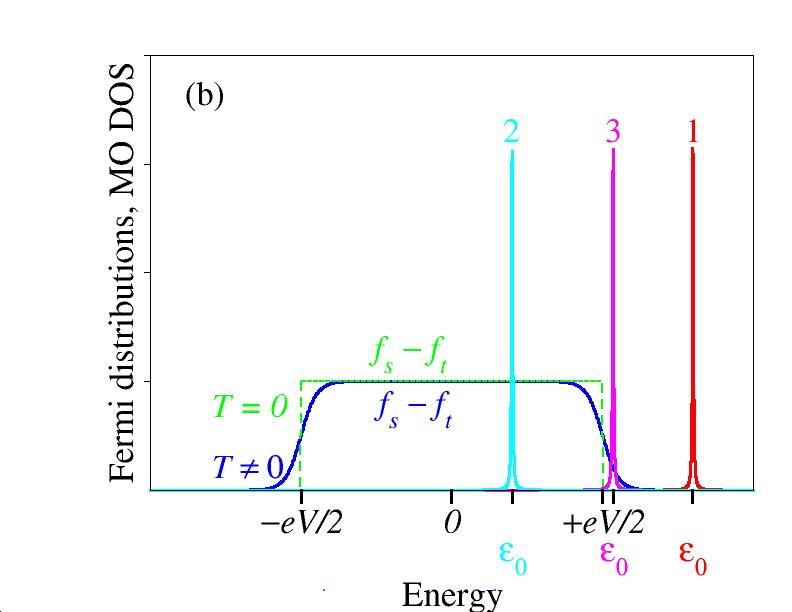} 
    \includegraphics[width=0.3\textwidth,height=0.22\textwidth,angle=0]{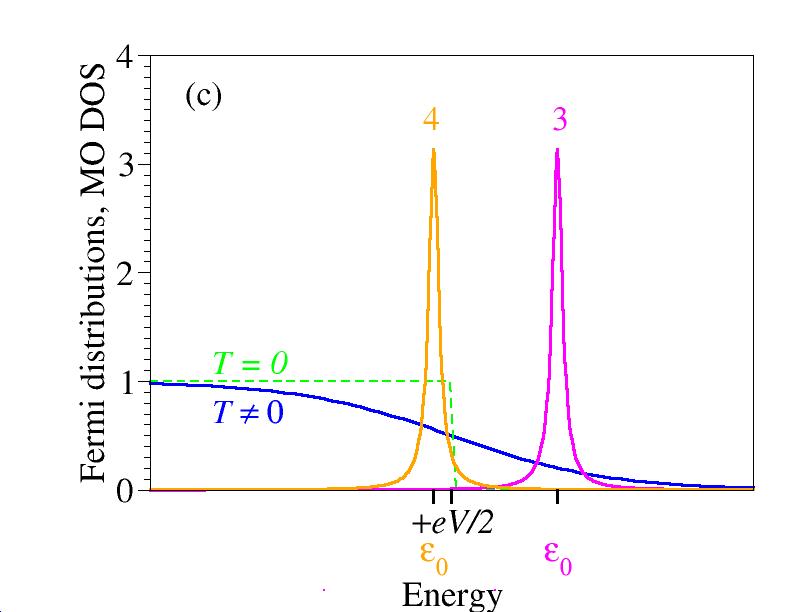} 
  }
  \caption{(a) A nonvanishing temperature smears out the electrode Fermi distributions.
    (b) Negligible far away from resonance (curves 1 and 2), the effect of thermal broadening
    becomes important when the transmission peak comes into resonance with the electrochemical potential
    of one electrode (curve 3). (c) Thermal effects are more pronounced below resonance (curve 4) than above resonance (curve 3)
    because the overlap is larger in the former case.}
  \label{fig:background}
\end{figure*}

Putting in more physical terms, thermal fluctuations give rise to smeared Fermi distributions
$f_{s,t}(\varepsilon) = f\left(\varepsilon - \mu_{s,t}\right)$
in electrodes that appreciably differ from the
zero temperature step (Heaviside) shape 
within ranges of width on the order of $ k_B T$ around the electrochemical potentials $\mu_{s,t} = \pm e V/2$
(\figname\ref{fig:background}a). The temperature has a negligible impact on situations wherein the
energy $\varepsilon_{0}$ of dominant transport channel lies far away from resonance
(curves 1 and 2 in \figname\ref{fig:background}b). However, the tunneling current    
can acquire a significant dependence on temperature in cases where a narrow enough
transmission curve (i.e., sufficiently small $\Lambda$) appreciably overlaps
with one of the two energy ranges affected by thermal smearing
(curves 3 and 4 in \figsname\ref{fig:background}b and c). 
More precisely, the temperature effect is stronger in situations slightly below resonance
($\varepsilon_{V} \alt \mu_s$, curve 4 in \figsname\ref{fig:background}c) than slightly above resonance
($\varepsilon_{V} \agt \mu_s$, curve 3 in \figsname\ref{fig:background}b and c).
As illustrated in \figname\ref{fig:background}c,
in the former case the current $I$($\to I_{exact}$) at temperature $T$ can be substantially larger than the current at $T = 0$ denoted by
$I_{0K}$ below (cf.~\gl~(\ref{eq-I-0K})),
while in the latter case $I$ is at least half the value of $I_{0K}$.
\section{Analytical expressions in the low and high temperature limits}
\label{sec:j-0K-T}
\subsection{Current of low temperatures}
\label{sec:j-0K}
The zero temperature limit ($I \approx I_{0K}$)
of \gl~(\ref{eq-Iex})
\begin{subequations}
  \label{eq-0K}
  \begin{equation}
    I  \xlongequal{T \to 0} I_{0K} = \frac{G_0}{e}
    \ \frac{\Gamma^2}{\Lambda}
    \left(\arctan \frac{\varepsilon_{V} + e V/2}{\Lambda} -
    \arctan \frac{\varepsilon_{V} - e V/2}{\Lambda} \right)
    \label{eq-I-0K}
  \end{equation}
is recovered using the asymptotic expansion (\gl~(6.3.18) in Ref.~\citenum{AbramowitzStegun:64})
%
%
\begin{equation*}
\psi(z) \xlongequal[\vert arg\, z\vert < \pi]{z \to \infty} \ln z + \mathcal{O}\left(1/z\right) \\
\end{equation*}
along with the identity
\begin{equation*}
\mbox{Im}\,\ln (x + i y) = \arctan\frac{y}{x}\\
\end{equation*}
An important case of the low temperature limit is the off-resonant transport, for which \gl~(\ref{eq-I-0K})
acquires a particularly appealing form \cite{Baldea:2012a,Baldea:2023a}
\begin{equation}
  \label{eq-I-0K-off}
  I_{0K} \xlongequal[\left\vert \varepsilon_{V} \pm e V/2\right\vert \gg \Lambda]{\Lambda \ll \left\vert\varepsilon_{0}\right\vert}
  I_{0K,off} = \frac{2 e^2}{h} \frac{\Gamma^2}{\varepsilon_{V}^2 - (e V/2)^2} V
\end{equation}
\end{subequations}

Low temperature (thermal) corrections to current can be deduced by
expanding the digamma function in powers $T$ \cite{JahnkeEmde:45,AbramowitzStegun:64}
or via Sommerfeld expansions \cite{Sommerfeld:33,AshcroftMermin}
\begin{subequations}
  \label{eq-sommerfeld-exp}
  \begin{eqnarray}
  I  & = & I_{l,1} + \mathcal{O}\left(k_B T\right)^4 ; 
  I  = I_{l,2} + \mathcal{O}\left(k_B T\right)^6; \nonumber \\
  I  & = & I_{l,3} + \mathcal{O}\left(k_B T\right)^8 
  \\
\label{eq-I-l1}
   I \approx I_{l,1}  & = & I_{0K} +  \frac{2 e}{h} \Gamma^2 
   \left[ c_1\left(\varepsilon_{V} + e V/2, \Lambda, T\right) \right . \nonumber \\
     & - & \left . c_1\left(\varepsilon_{V} - e V/2, \Lambda, T\right) \right] \left(\pi k_B T\right)^2\\
\label{eq-I-l2}
   I \approx I_{l,2} & = & I_{l,1} + \frac{2 e}{h} \Gamma^2 
   \left[ c_2\left(\varepsilon_{V} + e V/2, \Lambda, T\right) \right . \nonumber \\
     & - & \left . c_2\left(\varepsilon_{V} - e V/2, \Lambda, T\right) \right] \left(\pi k_B T\right)^4\\
 \label{eq-I-l3}
   I \approx I_{l,3} & = & I_{l,2} + \frac{2 e}{h} \Gamma^2
   \left[ c_3\left(\varepsilon_{V} + e V/2, \Lambda, T\right) \right . \nonumber \\
     & - & \left . c_3\left(\varepsilon_{V} - e V/2, \Lambda, T\right) \right] \left(\pi k_B T\right)^6
\end{eqnarray}
  where 
\begin{eqnarray}
  \label{eq-c123}
  c_1(\varepsilon, \Gamma, T) & = & -\frac{1}{3} \frac{\varepsilon}{\left(\varepsilon^2 + \Gamma^2\right)^2} \\
  c_2(\varepsilon, \Gamma, T) & = & - \frac{7}{15}
  \frac{\varepsilon \left(\varepsilon^2 - \Gamma^2\right)}{\left(\varepsilon^2 + \Gamma^2\right)^4} \\
  c_3(\varepsilon, \Gamma, T) & = & - \frac{31}{63} 
                   \frac{\varepsilon \left(3 \varepsilon^4 - 10 \varepsilon^2 \Gamma^2 + 3 \Gamma^4\right)}{\left(\varepsilon^2 + \Gamma^2\right)^6}
\end{eqnarray}

\end{subequations}
\subsection{Current at high temperatures}
\label{sec:j-T}
\begin{figure}[htb]
  \centerline{\includegraphics[width=0.3\textwidth,height=0.22\textwidth,angle=0]{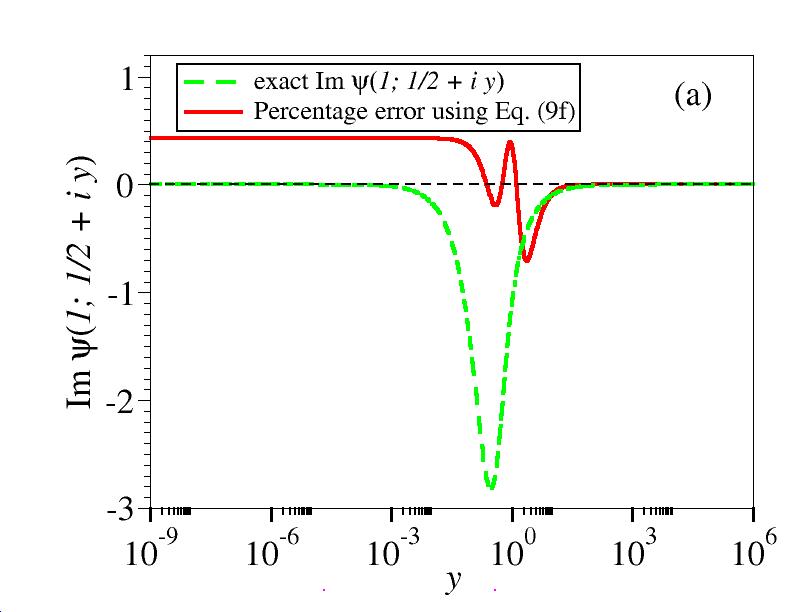}} 
  \centerline{\includegraphics[width=0.3\textwidth,height=0.22\textwidth,angle=0]{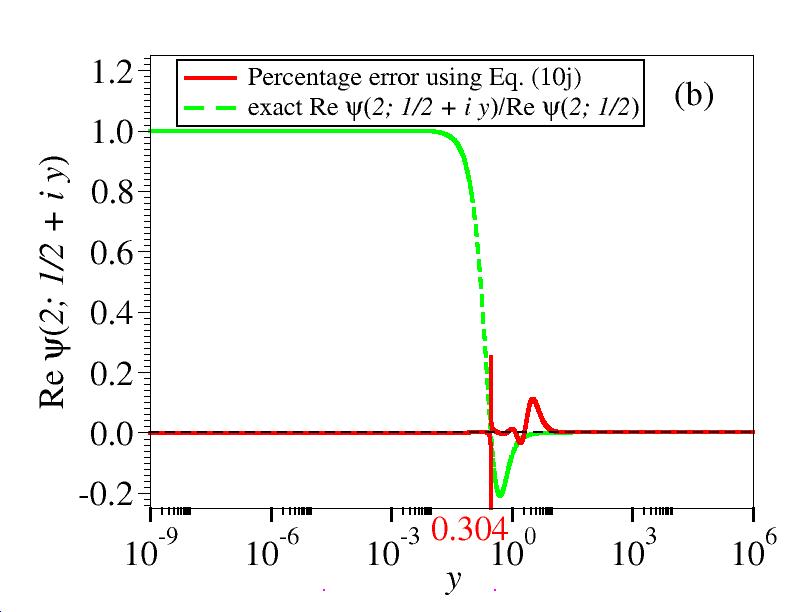}} 
  \caption{The deviations of the numerical interpolation functions (a) $d_2(y)$ of \gl~(\ref{eq-d1})
    and (b) $\varphi(x)$ of \gl~(\ref{eq-interpolation-psi4})
    from Im\,$\psi^\prime(1/2 + i y)$ and Re\,$\psi(2; 1/2 + i y)$, respectively are smaller than half of percent.}
  \label{fig:Im-psi3-Re-psi4}
\end{figure}

The identity (cf.~\gl~(6.3.12) of Ref.~\citenum{AbramowitzStegun:64})
\begin{subequations}
  \label{eq-T}
\begin{equation*}
  \label{eq-psi-1/2}
\mbox{Im}\, \psi\left(\frac{1}{2} + i y\right) = \frac{\pi}{2}\tanh \pi y 
\end{equation*}
yields the formula for current
valid in the high temperature limit ($\Lambda \ll \pi k_B T$, $ I = I_{T} + \mathcal{O}\left(\frac{\Lambda}{k_B T}\right)$) 
\begin{eqnarray}
  \label{eq-I_T}
  & I &  \xlongequal[]{\Lambda \ll \pi k_B T} I_{T} \nonumber \\
  & I_{T} & \equiv \frac{2 e}{h} \frac{\Gamma^2}{\Lambda}\frac{\pi}{2} \left[\tanh\frac{\varepsilon_{V} + eV/2}{2 k_B T} - \tanh\frac{\varepsilon_{V} - eV/2}{2 k_B T} \right] 
  \\
  & I_{T} & \xlongequal[]{e \vert V\vert \ll 2 \left\vert\varepsilon_{0}\right\vert} 
  \frac{2 e}{h} \frac{\Gamma^2}{\Lambda} \pi \frac{\sinh \frac{e V}{2 k_B T}}{\cosh\frac{\varepsilon_{V}}{k_B T}}
  \xlongrightarrow[]{ k_B T \ll e \vert V\vert \ll 2 \left\vert \varepsilon_{0}\right\vert} I_{p.A}
  \nonumber \\
   & I_{p.A} & = \frac{2 e}{h} \frac{\Gamma^2}{\Lambda} \pi \exp \left[- \frac{\left\vert \varepsilon_{V}\right\vert - \vert e V\vert/2}{k_B T}\right] \mbox{sign}\, V
\end{eqnarray}
Above, the subscript $p.A$ stands for ``pseudo-Arrhenius'' to emphasize that, in the limit specified, the tunneling current
exhibits an Arrhenius-type dependence on temperature, which is routinely (or, better, ``abusively'', see Introduction)
considered characteristic for the hopping mechanism.

Series expansion in powers of $x \equiv \Lambda /\left(2 \pi k_B T \right)$ in terms of
polygamma functions $\psi(n; z) \equiv d^n \psi(z)/d z^n$
\begin{eqnarray}
  \label{eq-taylor}
  \mbox{Im}\,  \psi\left( \frac{1}{2} + x + i y\right) & = &
 \frac{\pi}{2}\tanh \pi y
 + x \, \mbox{Im}\,\psi\left(1; \frac{1}{2} + i y\right) \nonumber \\
 & + & \frac{x^2}{2} \mbox{Im}\,\psi\left(2; \frac{1}{2} + i y\right) 
 + \mathcal{O}\left(x^3\right)
\end{eqnarray}
allows to deduce corrections to \gl~(\ref{eq-I_T}).  
Unlike the imaginary part of the tetragamma function ($\psi^{\prime\prime}(z) \equiv \psi(2; z)$),
which can be expressed analytically in terms of elementary functions for $z = 1/2 + i y$ 
\begin{equation}
  \label{eq-d2}
  \mbox{Im}\, \psi\left(2; \frac{1}{2} + i y\right) = d_2(y) \equiv -\pi^3 \frac{\sinh \pi y}{\cosh^3 \pi y}
\end{equation}
the imaginary part of the trigamma function ($\psi^\prime (z) \equiv \psi(1; z)$) entering \gl~(\ref{eq-taylor})
cannot. However, we found that it can be accurately interpolated numerically
(see \figname\ref{fig:Im-psi3-Re-psi4}a)
\begin{equation}
  \label{eq-d1}
  \mbox{Im}\,\psi\left(1; \frac{1}{2} + i y\right) 
  \simeq
  d_{1}(y) \equiv  - y \left[\frac{1}{ 0.958287 + y^2} + \frac{0.944894}{\left(0.244104 + y^2\right)^2}\right]
\end{equation}
We thus arrive at the following expressions applicable for 
MO broadening smaller that the thermal smearing of electrode Fermi functions
(quantified by $\Lambda$ smaller than $k_B T$) characterizing  
transport data that exhibit a pronounced temperature dependence
\begin{eqnarray}
  I & = & I_{h,1} + \mathcal{O}\left(\frac{\Lambda}{k_B T}\right)^2; \ 
  I = I_{h,2} + \mathcal{O}\left(\frac{\Lambda}{k_B T}\right)^3 \\
  \label{eq-I-h1}
  I & \approx & I_{h,1} = I_T + \frac{2 e}{h}\frac{\Gamma^2}{\Lambda} \frac{\Lambda}{2\pi k_B T}
  \left[d_{1}\left(\frac{\varepsilon_{V} + e V/2}{2 \pi k_B T}\right) \right . \nonumber \\
    & & - \left . d_{1}\left(\frac{\varepsilon_{V} - e V/2}{2 \pi k_B T}\right) \right] \\
  \label{eq-I-h2}
  I & \approx & I_{h,2} = I_{h,1} + \frac{2 e}{h}\frac{\Gamma^2}{\Lambda} \left(\frac{\Lambda}{2\pi k_B T}\right)^2
  \left[d_{2}\left(\frac{\varepsilon_{V} + e V/2}{2 \pi k_B T}\right) \right . \nonumber \\
    & - & \left . d_{2}\left(\frac{\varepsilon_{V} - e V/2}{2 \pi k_B T}\right) \right]  
\end{eqnarray}
\end{subequations}

Notice that, similar to $c_{1,2,3}$ (cf.~\gl~(\ref{eq-c123})),
the quantities $d_{1,2}$ are expressed in terms of elementary functions
(cf.~\gls~(\ref{eq-d1}) and (\ref{eq-d2})). This is important for potential applications to
processing transport data exhibiting substantial dependence temperature;
special functions (digamma $\psi$ in this specific case)
of complex argument (cf.~\gl~(\ref{eq-Iex}))) are not implemented in software packages
(like ORIGIN or MATLAB) routinely utilized by experimentalists for data fitting.
\subsection{Analytical expressions for the zero bias conductance}
\label{sec:g-0K-T}
In the zero bias limit ($V \to 0$), \gl~(\ref{eq-Iex}) recovers
the recently deduced formula for the exact ohmic conductance $G$.\cite{Baldea:2022c}
It is expressed in terms of the trigamma function ($\psi^\prime (z) = d \psi(z)/d z$) 
\begin{subequations}
  \begin{equation}
    \label{eq-Gex}
  G \to G_{exact} \equiv \left . \frac{\partial I_{exact}}{\partial V}\right\vert_{V = 0} = G_0 \frac{\Gamma^2}{2 \pi \Lambda k_B T}
  \mbox{Re}\, \psi^\prime \left(\frac{1}{2} + \frac{\Lambda + i \varepsilon_{0}}{2 \pi k_B T}\right)
\end{equation}
Thermal corrections to the zero temperature conductance $G_{0K}$
\begin{equation}
  \label{eq-G-0K}
G \xlongrightarrow{T = 0} G_{0K} \equiv G_0 \frac{\Gamma^2}{\varepsilon_{0}^2 + \Lambda^2} \\ 
\end{equation}
obtained from the asymptotic expansion of the digamma function read
\begin{eqnarray}
  G & = & G_{0K} + \mathcal{O}\left(k_B T\right)^2 \\
  \label{eq-G-l1}
  G & = & G_{l,1} + \mathcal{O}\left(k_B T\right)^4; \nonumber \\
  & & G \approx G_{l,1} = G_{0K} +  G_0 \Gamma^2 \frac{1}{3}
  \frac{3\varepsilon_{0}^2 - \Lambda^2}{\left(\varepsilon_{0}^2 + \Lambda^2\right)^3} \left(\pi k_B T\right)^2 \\
  G & = & G_{l,2} + \mathcal{O}\left(k_B T\right)^6; \\
  & & G \approx G_{l,2} = G_{l,1} + G_0 \Gamma^2 \frac{7}{15}
  \frac{5 \varepsilon_{0}^4 - 10\varepsilon_{0}^2 \Lambda^2 + \Lambda^4}{\left(\varepsilon_{0}^2 + \Lambda^2\right)^5}  \left(\pi k_B T\right)^4 \nonumber \\
  \label{eq-G-l3}
  G & = & G_{l,3} + \mathcal{O}\left(k_B T\right)^8; \\
  & G & \approx G_{l,3} = G_{l,2} \nonumber \\
  & & + G_0 \Gamma^2 \frac{31}{21}
  \frac{7 \varepsilon_{0}^6 - 35\varepsilon_{0}^4 \Lambda^2 + 21 \varepsilon_{0}^2 \Lambda^4 - \Lambda^6}
       {\left(\varepsilon_{0}^2 + \Lambda^2\right)^7} \left(\pi k_B T\right)^6     
\end{eqnarray}
Analytical expressions in the high temperature limit ($\Lambda$ smaller than $\pi k_B T$) 
deduced previously \cite{Baldea:2022c,Baldea:2022j} are included for the reader's convenience ($y \equiv \varepsilon_{0} /\left(2 \pi k_B T\right)$)
\begin{eqnarray}
    \label{eq-G_T}
& G & \xlongequal[]{\Lambda \ll \pi k_B T} G_{T} \equiv G_0 \frac{\pi \Gamma^2}{4 \Lambda k_B T} \mbox{sech}^2 (\pi y) \nonumber \\
      & & G_T \xlongrightarrow[]{\pi k_B T \ll \left\vert\varepsilon_{0}\right\vert} 
             G_0 \frac{\pi \Gamma^2}{\Lambda k_B T} \exp\left(-\frac{\left\vert\varepsilon_{0}\right\vert}{k_B T}\right)  \\
    \label{eq-G-h1}
    &  G & = G_{h,1} + \mathcal{O}\left(\frac{\Lambda}{\pi k_B T}\right); \nonumber \\
    & & G \approx G_{h,1} \equiv G_{T} + G_0 \Gamma^2 \frac{\varphi(y)}{\left(2 \pi k_B T\right)^2} \\
    \label{eq-G-h2}
    G & = & G_{h,2} + \mathcal{O}\left(\frac{\Lambda}{\pi k_B T}\right)^2;\nonumber \\
    & & G \approx G_{h,2} \equiv G_{h,1} +  G_0 \Gamma^2
    \frac{\pi}{16} \frac{\Lambda}{\left(k_B T\right)^3} \nonumber \\
    & & \times \left[2 - \mbox{cosh}\,(2 \pi y)\right]\mbox{sech}^4 (\pi y) \\
    &  & \mbox{Re}\,\psi\left(2; \frac{1}{2} + i y\right) \simeq \varphi(y) \equiv \frac{y^2 - 34.7298}{\left(y^2 + 2.64796\right)^2}
    \label{eq-interpolation-psi4} 
    \\
  & & + 37.262 \frac{y^2 + 1.12874}{\left(y^2 + 2.17786\right)^3} +
  3.01373 \frac{y^2 - 0.082815}{\left(y^2 + 0.25014\right)^3}
  \nonumber 
\end{eqnarray}
\end{subequations}
The high accuracy of the numerical interpolation expressed by \gl~(\ref{eq-interpolation-psi4})
is depicted in \figname\ref{fig:Im-psi3-Re-psi4}b.
\section{Can the temperature impact on the transport by tunneling be as strong as on the transport by hopping?}  
\label{sec:tunneling-vs-hopping}
To reiterate what we have mentioned in \secname\ref{sec:model},
all the above analytical formulas deduced for the tunneling current hold for MO energy offsets $\varepsilon_0(V)$
exhibiting an arbitrary dependence on the applied bias $V$, including in particular the case of nonvanishing Stark strengths $\gamma \neq 0$.
In this section, for simplicity, we confine ourselves to present numerical results for
the bias independent case ($\varepsilon_0(V) = \varepsilon_{0}, \gamma = 0$).
Because the charge conjugation invariance of this model yields equal currents computed for $+\varepsilon_{0}$ and $-\varepsilon_{0}$
and symmetric $I-V$ curves ($I(-V) = - I(V)$), we can restrict ourselves to positive $\varepsilon_{0}$ and $V$ without loss of generality.

As said, traditionally a strongly temperature dependent transport was considered as a fingerprint for a hopping mechanism,
but we showed above that the thermal broadening of electron Fermi distributions in
electrodes makes the tunneling current dependent on temperature.
Can the tunneling current expressed by the above formulas vary over orders of magnitude
if the temperature varies within experimentally relevant ranges, challenging thereby the hopping mechanism or the occurrence of a tunneling-hopping transition?

\begin{figure*}[htb]
    \centerline{
    \includegraphics[width=0.22\textwidth,height=0.3\textwidth,angle=-90]{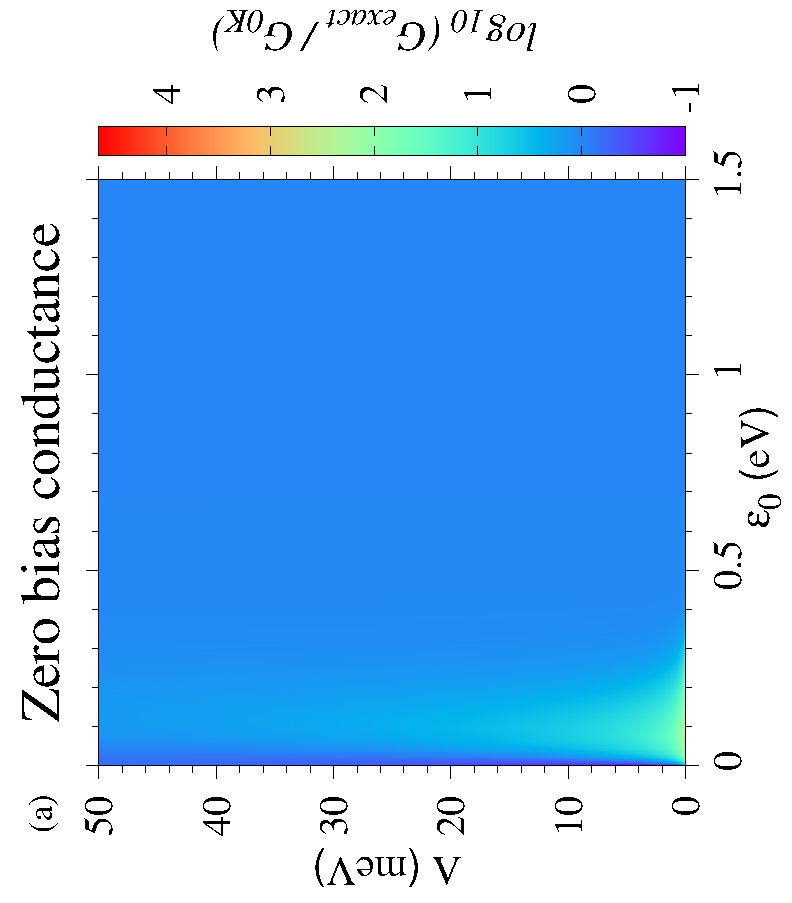} 
    \includegraphics[width=0.22\textwidth,height=0.3\textwidth,angle=-90]{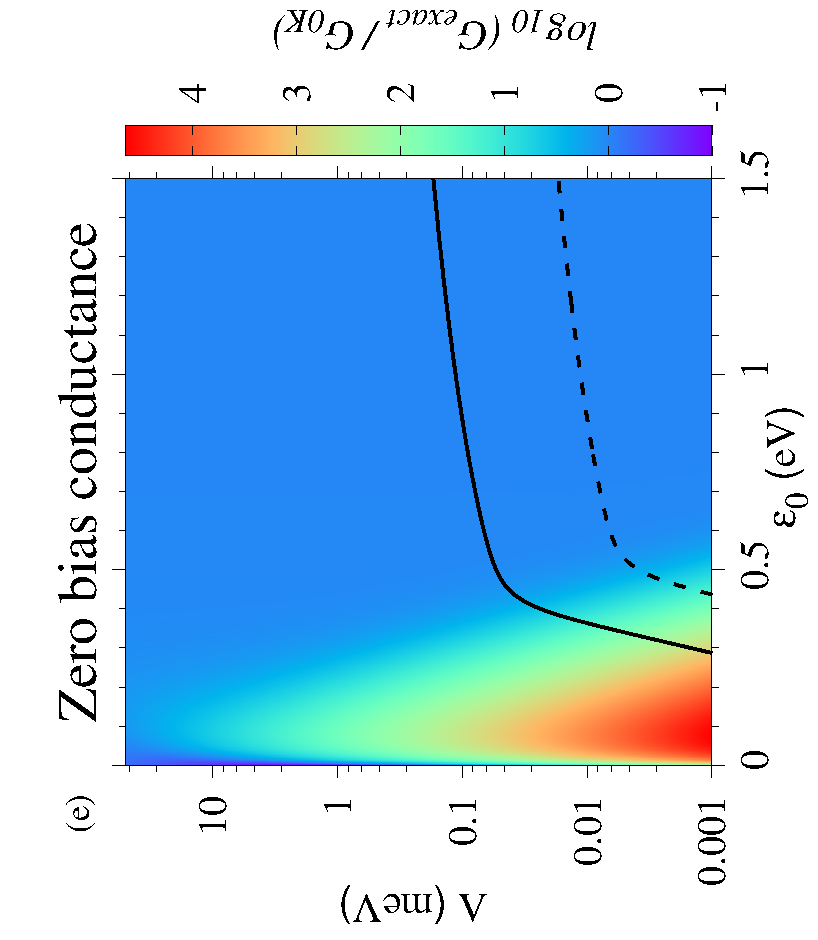} 
    \includegraphics[width=0.205\textwidth,height=0.3\textwidth,angle=-90]{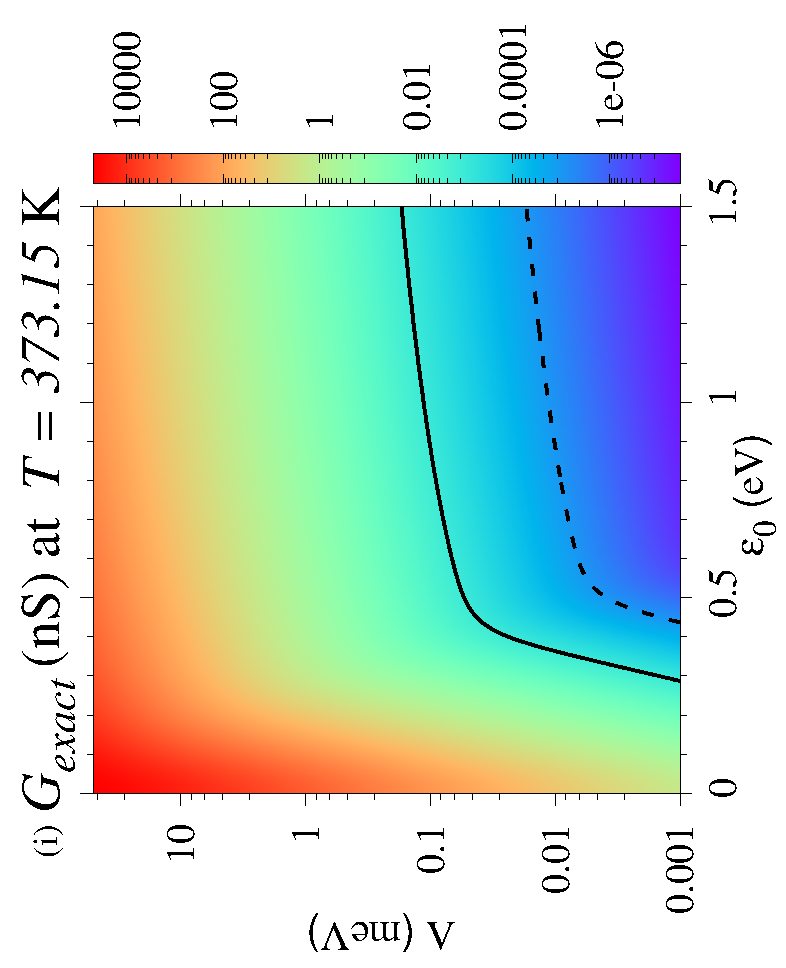} 
    }
  \centerline{
    \includegraphics[width=0.22\textwidth,height=0.3\textwidth,angle=-90]{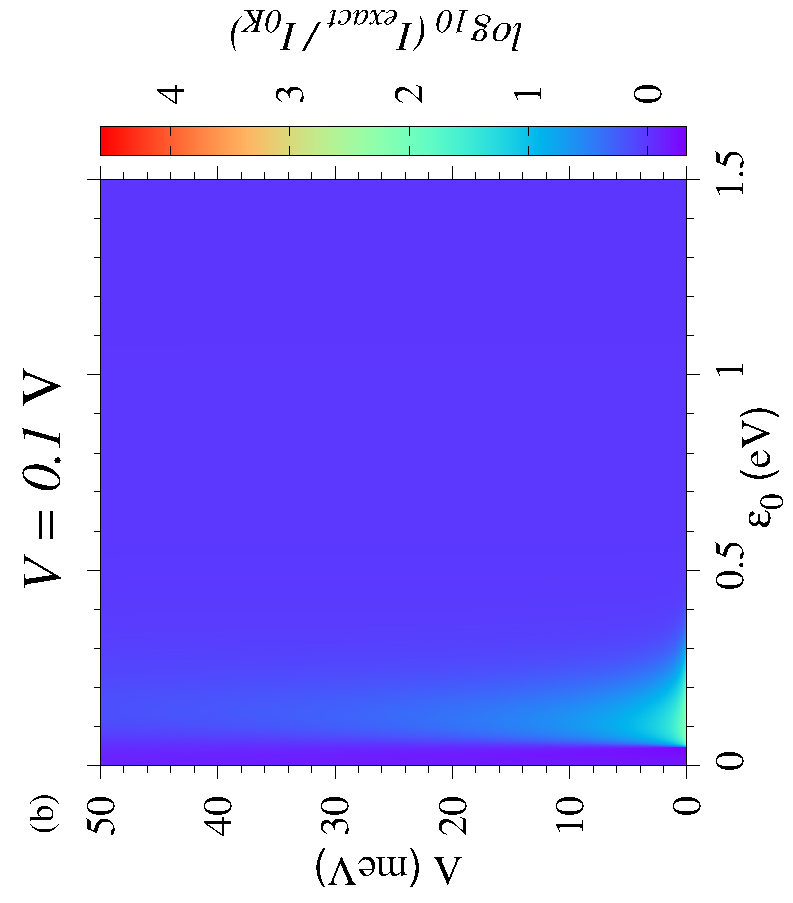} 
    \includegraphics[width=0.22\textwidth,height=0.3\textwidth,angle=-90]{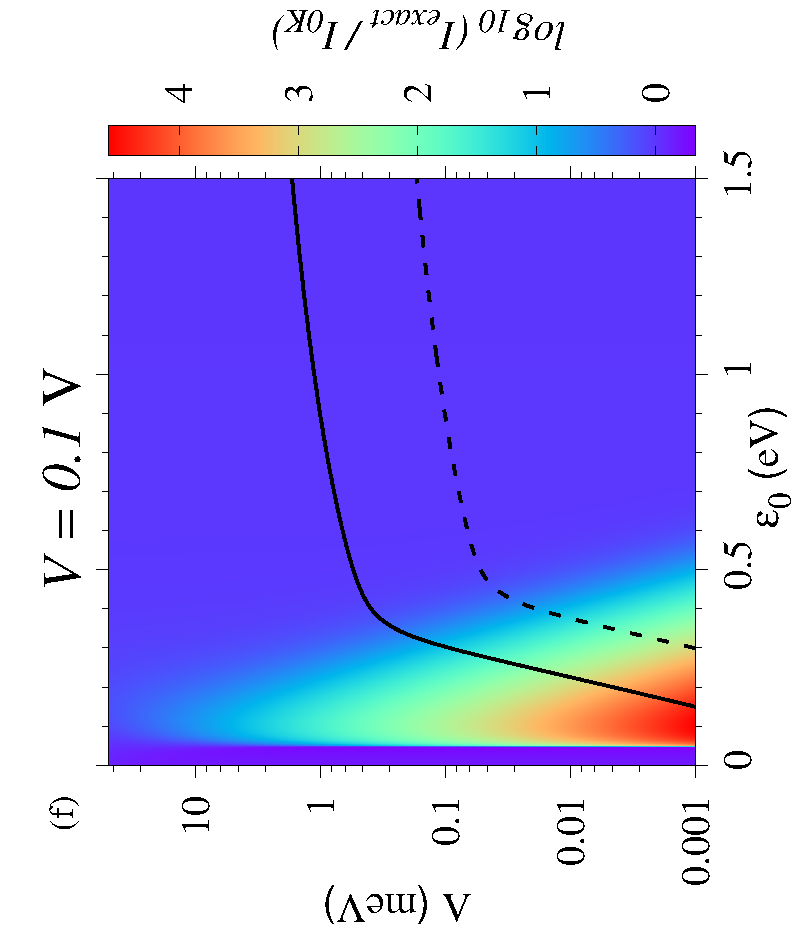} 
    \includegraphics[width=0.22\textwidth,height=0.3\textwidth,angle=-90]{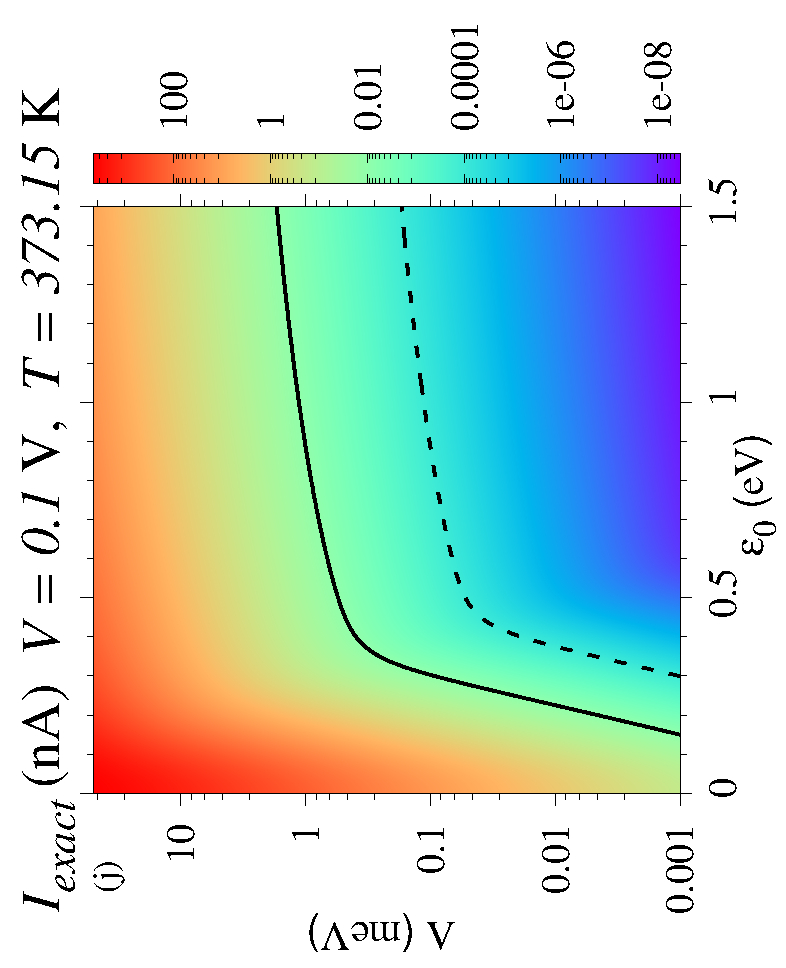} 
  }
  \centerline{
    \includegraphics[width=0.22\textwidth,height=0.3\textwidth,angle=-90]{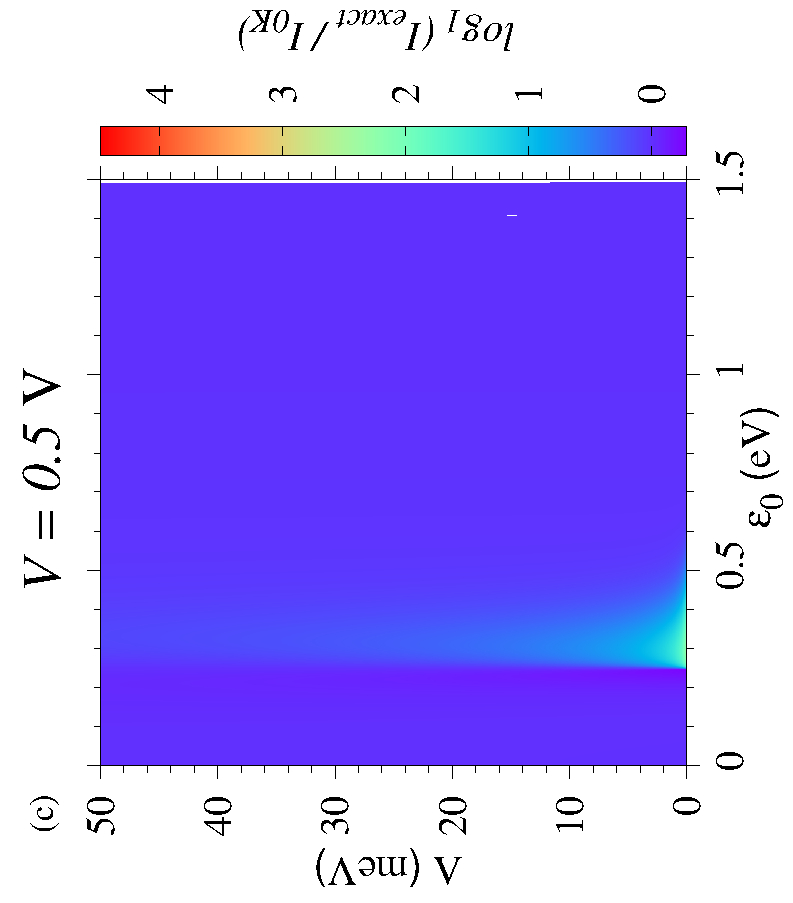} 
    \includegraphics[width=0.22\textwidth,height=0.3\textwidth,angle=-90]{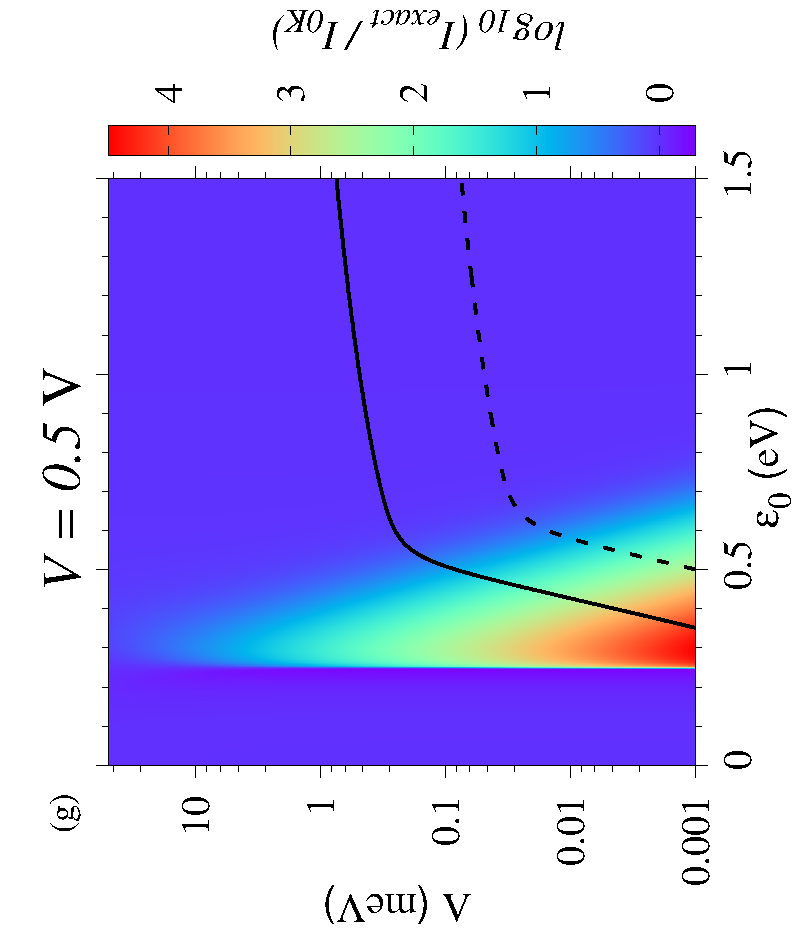} 
    \includegraphics[width=0.22\textwidth,height=0.3\textwidth,angle=-90]{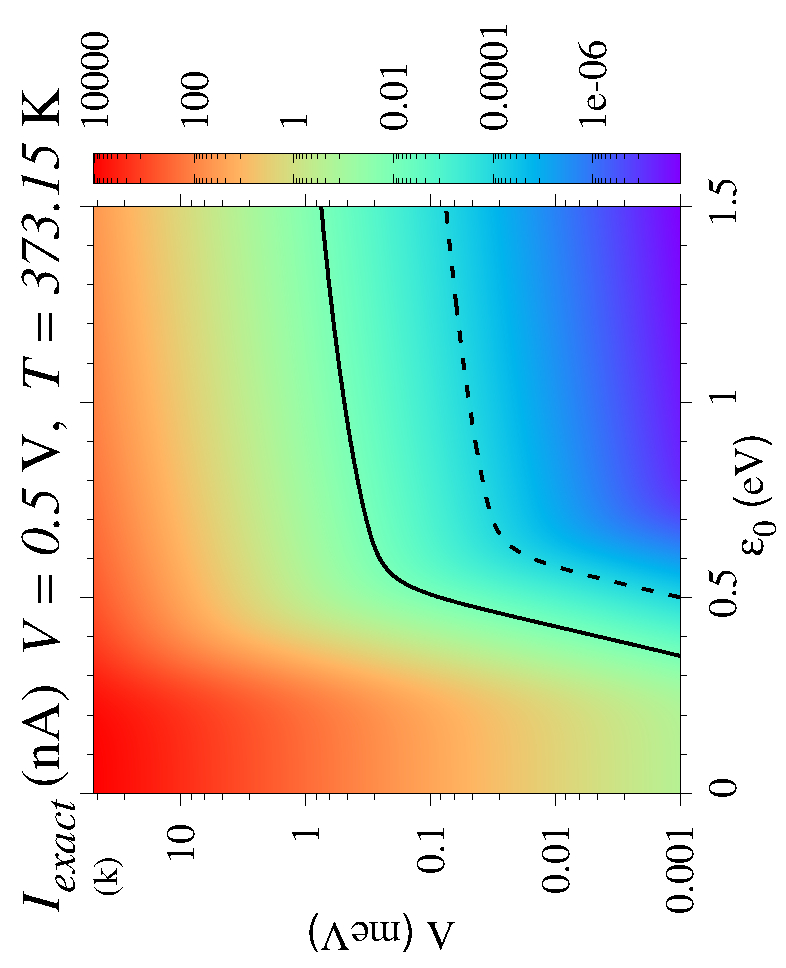} 
  }
    \centerline{
      \includegraphics[width=0.22\textwidth,height=0.3\textwidth,angle=-90]{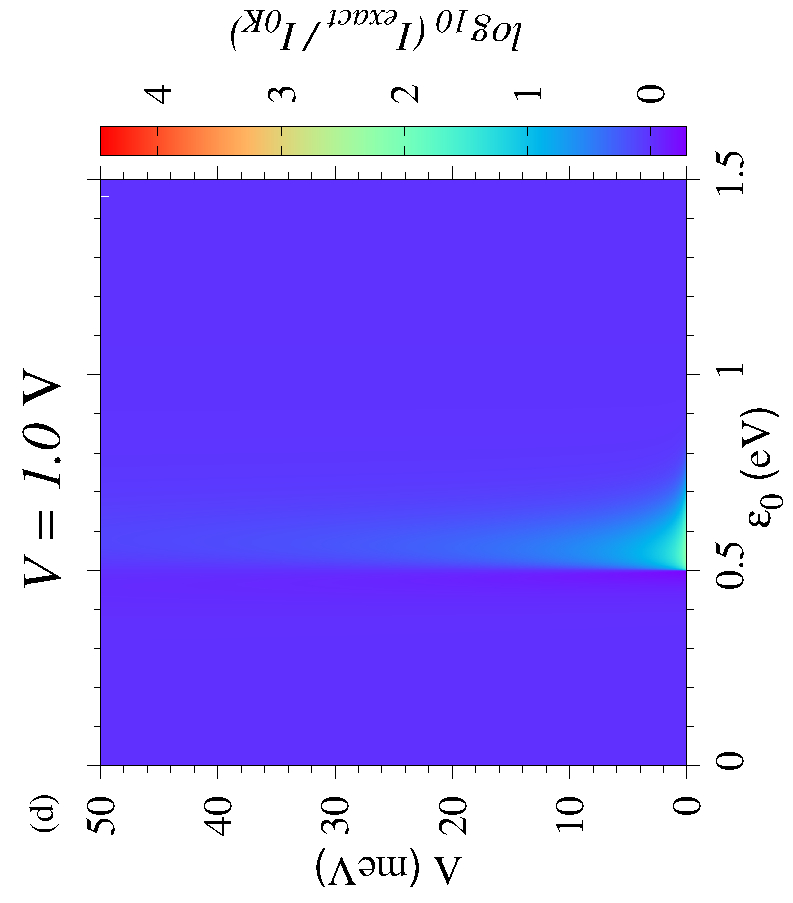} 
      \includegraphics[width=0.22\textwidth,height=0.3\textwidth,angle=-90]{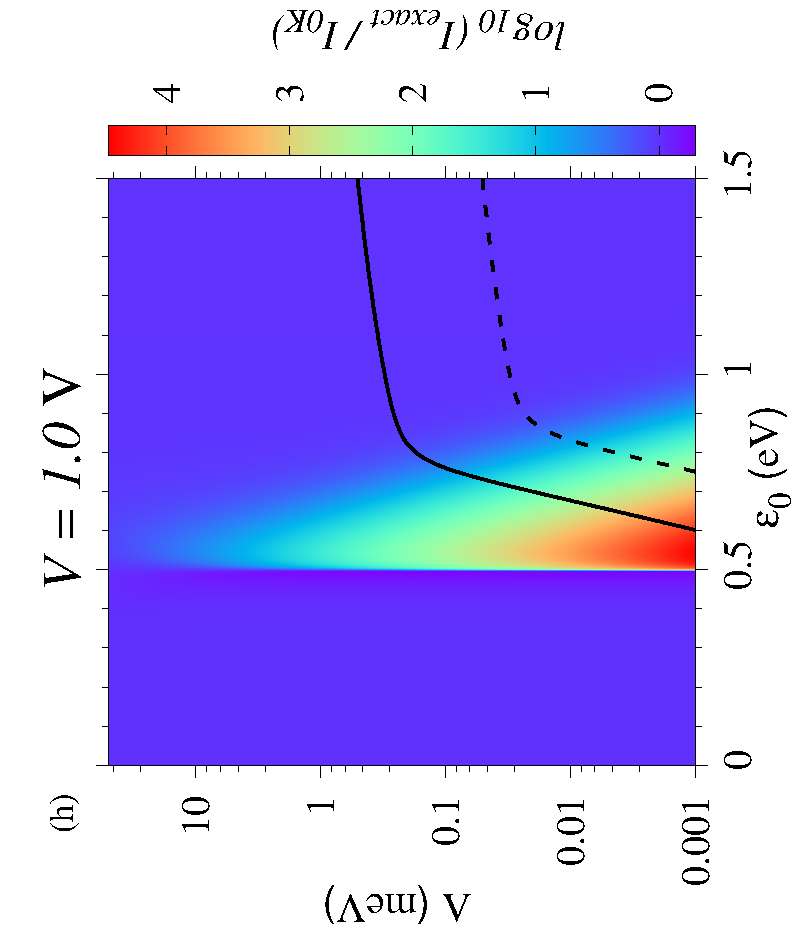} 
      \includegraphics[width=0.22\textwidth,height=0.3\textwidth,angle=-90]{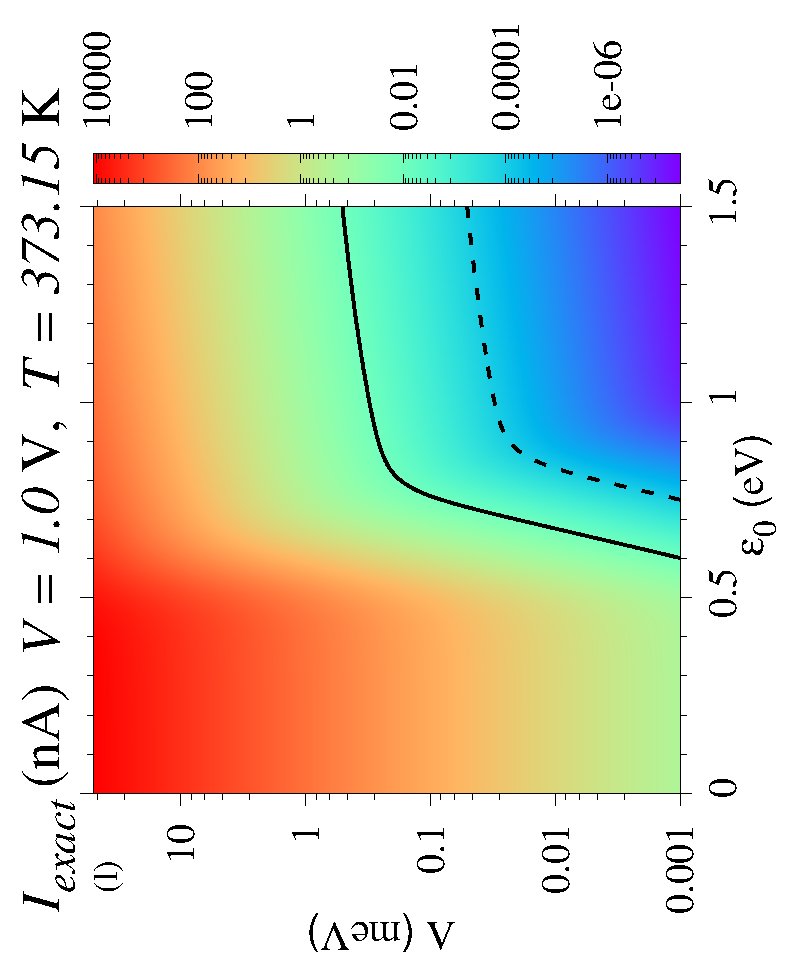} 
    }
    \caption{(a---h) Ratio of the relevant transport properties (conductance $G_{exact}$ and current $I_{exact}$) computed
      at $T = 373.15$\,K and values of bias $V$ indicated relative to their counterparts at $T=0$\,K ($G_{0K}$ and $I_{0K}$, respectively).
      The model parameter $\Lambda$ is depicted both in linear and (to better visualize the order of magnitude)
      logarithmic scale (panels a to d and e to h, respectively).
      By and large, the main effect of the bias $V$ is a horizontal energy shift $e V/2$
      of the parameter ranges corresponding to substantial thermal enhancement, which becomes more pronounced as the
      resonance ($\varepsilon_{0} = e V/2$) is approached from below ($\varepsilon_{0} \agt e V/2$).
      Diagrams for the zero bias conductance in nS (i) and (j to l) current in nA at $T = 373.15$\,K.
      The parameter regions wherein $G_{exact} > 1$\,pS (panel i) and $I_{exact} > 10$\,pA (panels j to l), which can be considered
      reasonable lower bound values for experimental accessibility, lie below the solid and dashed lines. See the main text for details.}
  \label{fig:overview}
\end{figure*}

The numerical results collected in \figname\ref{fig:overview} offer the (positive) answer to this question.
The first and second columns of \figname\ref{fig:overview} (panels a to d and e to h, respectively)
depict the thermal excess quantified by the ratio of the exact conductance $G_{exact}$ (\gl~(\ref{eq-Gex})) and the exact
current $I_{exact}$ (\gl~(\ref{eq-Iex})) computed at $T = 373.15$\,K ($k_B T = 32.2$\,meV)
and several biases ($V=0.1; 0.5; 1.0$\,V) 
and their counterparts $G_{0K}$ and $I_{0K}$ at the same biases but at $T = 0$ (\gls~(\ref{eq-G-0K}) and (\ref{eq-I-0K}), respectively)
for the model parameter ranges
$1\,\mbox{meV} \leq \varepsilon_{0} \leq 1.5$\,eV and $0.001\,\mbox{meV} \leq \Lambda \leq 50\,\mbox{meV}$.
(We chose $T = 373.15$\,K for illustration, as a reasonable value for the highest temperature of any practical interest.)
Inspection of these panels indicates a current thermal enhancement up to three to four orders
of magnitudes for a sufficiently small $\Lambda$ and
MO level outside the Fermi window but no very far away from resonance
(say, $e \vert V\vert/2 < \left\vert \varepsilon_{0} \right\vert \alt e \vert V\vert/2 + 10 k_B T$).

\begin{figure*}[htb]
  \centerline{
    \includegraphics[width=0.22\textwidth,height=0.3\textwidth,angle=-90]{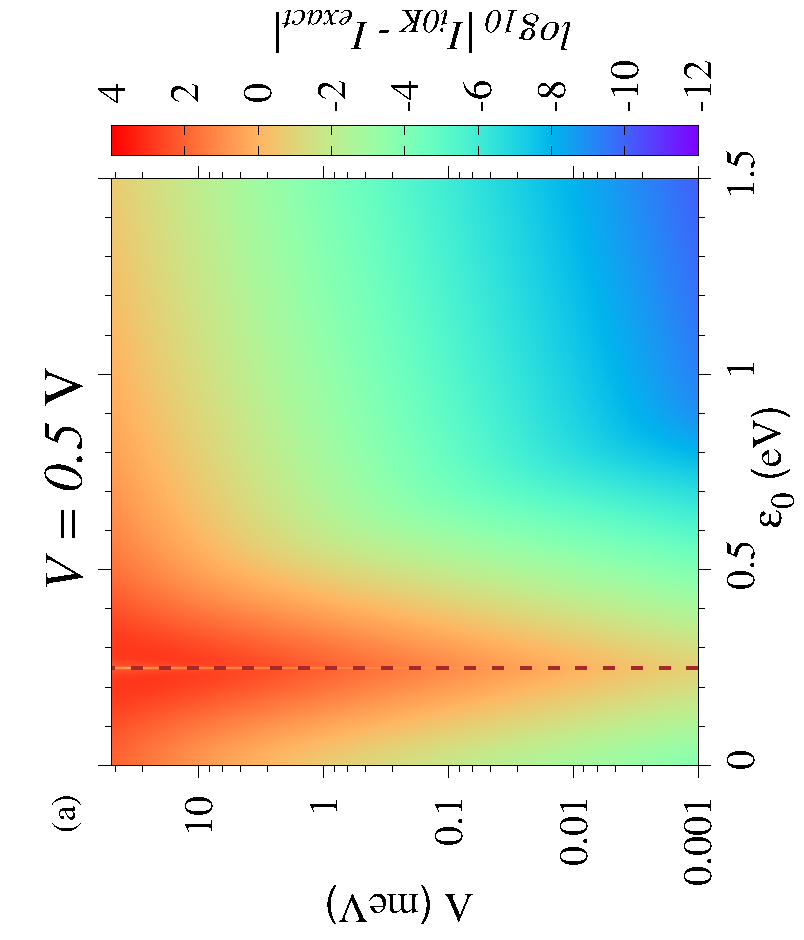} 
    \includegraphics[width=0.22\textwidth,height=0.3\textwidth,angle=-90]{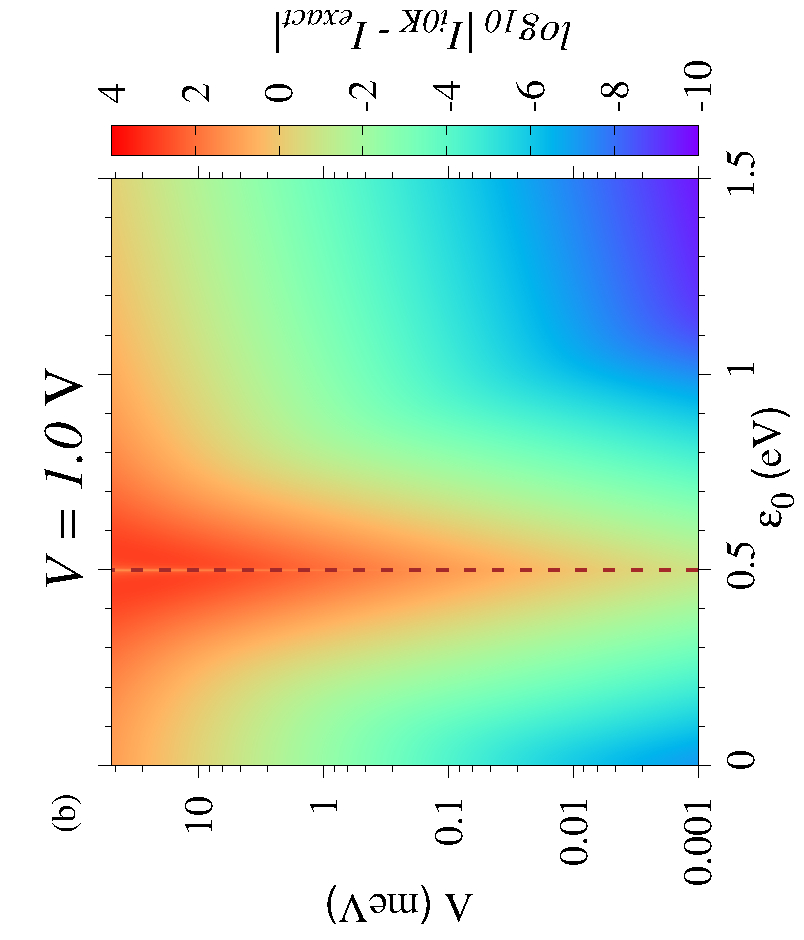} 
  }
  \caption{Unlike the relative deviations depicted in \figname\ref{fig:overview},
    the absolute deviations of $I_{0K}$ from $I_{exact}$ 
    are nearly symmetric around resonance $\left\vert \varepsilon_{0}\right\vert = e\vert V\vert /2 $.
    This is illustrated here for two bias values: (a) $V = 0.5$\,V and (b) $V = 1.0$\,V.}
  \label{fig:abs-I}
\end{figure*}

To avoid misunderstandings related to the appearance of \figname\ref{fig:overview},
we emphasize that, unlike the current thermal excess $I_{exact}/I_{0K}$, which is highly asymmetric around
resonance--- very large below resonance ($I_{exact}/I_{0K} \gg 1$ for $e \vert V\vert/2 \alt \left\vert \varepsilon_{0} \right\vert$)  
but not too small above resonance ($1 > I_{exact}/I_{0K} > 1/2$ for $e \vert V\vert/2 \agt \left\vert \varepsilon_{0} \right\vert$)---
the absolute difference $\left\vert I_{exact} - I_{0K}\right\vert$ is practically symmetric around resonance
($e \vert V\vert/2 = \left\vert \varepsilon_{0} \right\vert$); compare \figname\ref{fig:overview} with \figname\ref{fig:abs-I}. 
\begin{figure}[htb]
  \centerline{\includegraphics[width=0.3\textwidth,height=0.22\textwidth,angle=0]{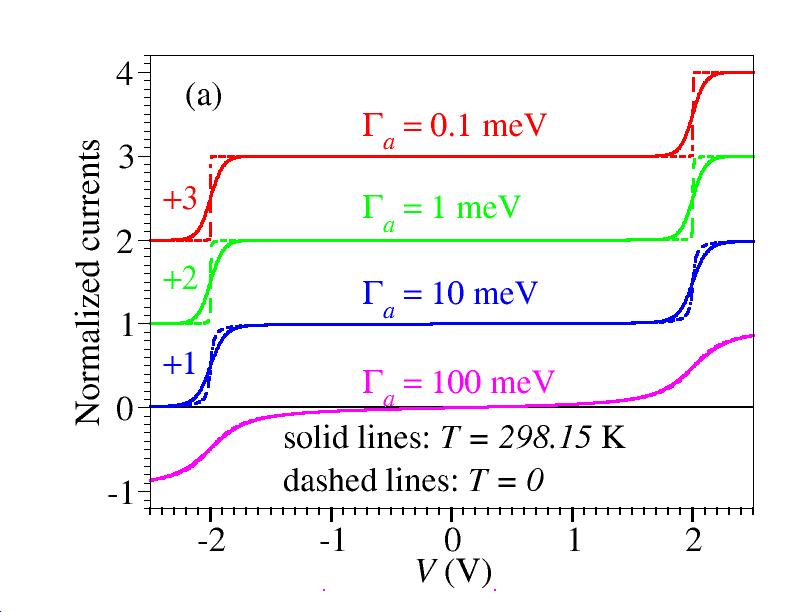}} 
  \centerline{\includegraphics[width=0.3\textwidth,height=0.22\textwidth,angle=0]{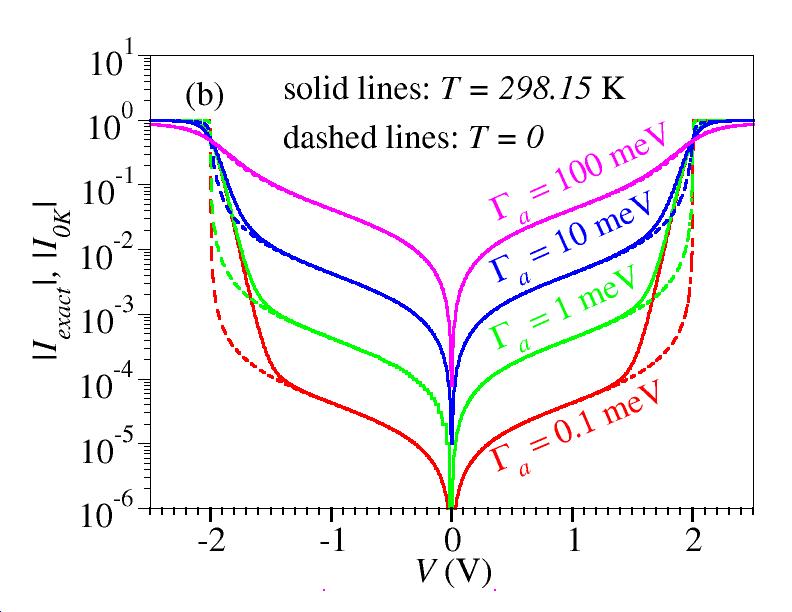}} 
  \caption{Currents at $T = 298.15$\,K (solid lines) and $ T = 0$ (dashed lines) normalized at the plateau value
    $\Lambda /\left(\pi \Gamma^2\right)$ 
    for various values of $\Lambda$ in (a) linear
    and (b) logarithmic y scale. For clarity, in panel a they are shifted vertically as indicated in the legend.
%
  }
  \label{fig:iv-T-0K}
\end{figure}
Along with the strong thermal enhancement, convincing oneself that the conductance/current are reasonable large
to be measurable in the pertaining model parameter ranges is an equally important aspect that deserves consideration. 
Therefore, \figname\ref{fig:overview} also
indicates the parameter ranges wherein the values $G$ and $I$ are larger than $1$\,pS and $0.01$\,pS,
and $10$\,pA and $0.1$\,pA, respectively. These values correspond to the regions above the continuous and dashed lines
depicted in panels e to l. (We do not show these lines in panels a to d because the linear y-scale
would make visibility problematic.) Above, the larger values $1$\,pS and $10$\,pA were chosen
as realistic lowest values still measurable for single molecule junctions. The values one hundred times lower 
($0.01$\,pS and $0.1$\,pA) can be taken as the counterpart for CP-AFM junctions,
which consist of $\sim 100$ molecules per junction.\cite{Baldea:2017e}

To sum up, we conclude that the impact of temperature on the transport via a single step tunneling mechanism
can definitely be as strong as that the impact on the two-step hopping mechanism, which was traditionally associated
to a pronounced temperature dependence.
\section{How accurate are the analytical approximations expressed in terms of elementary functions?}
\label{sec:analytic-approx}
\begin{figure*}[htb]
  \centerline{
    \includegraphics[width=0.22\textwidth,height=0.3\textwidth,angle=-90]{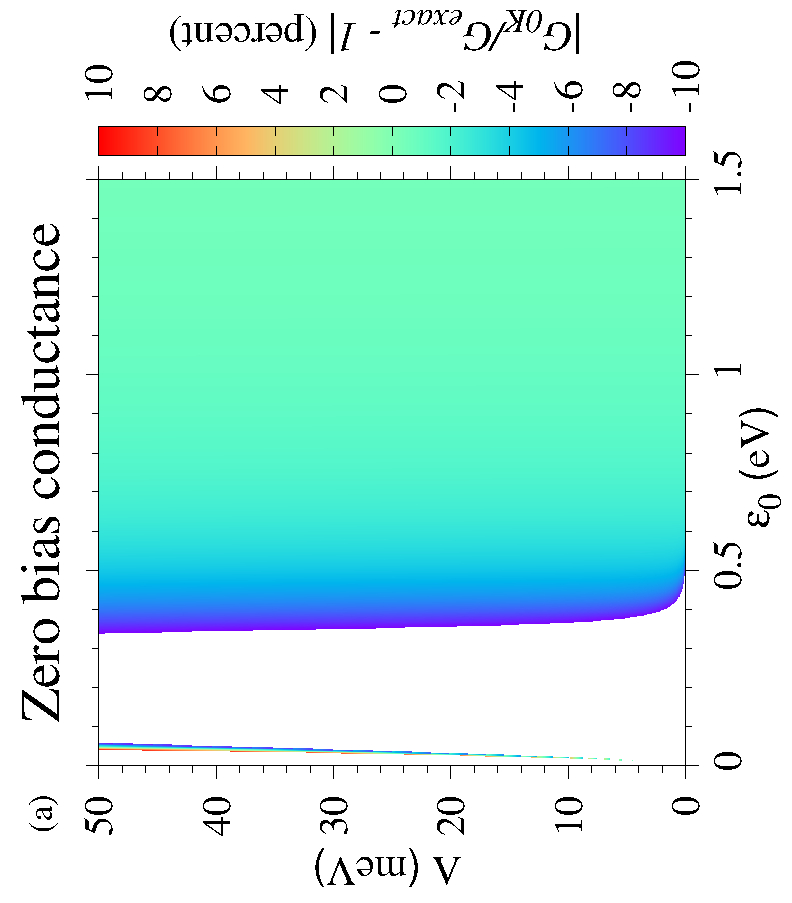} 
    \includegraphics[width=0.22\textwidth,height=0.3\textwidth,angle=-90]{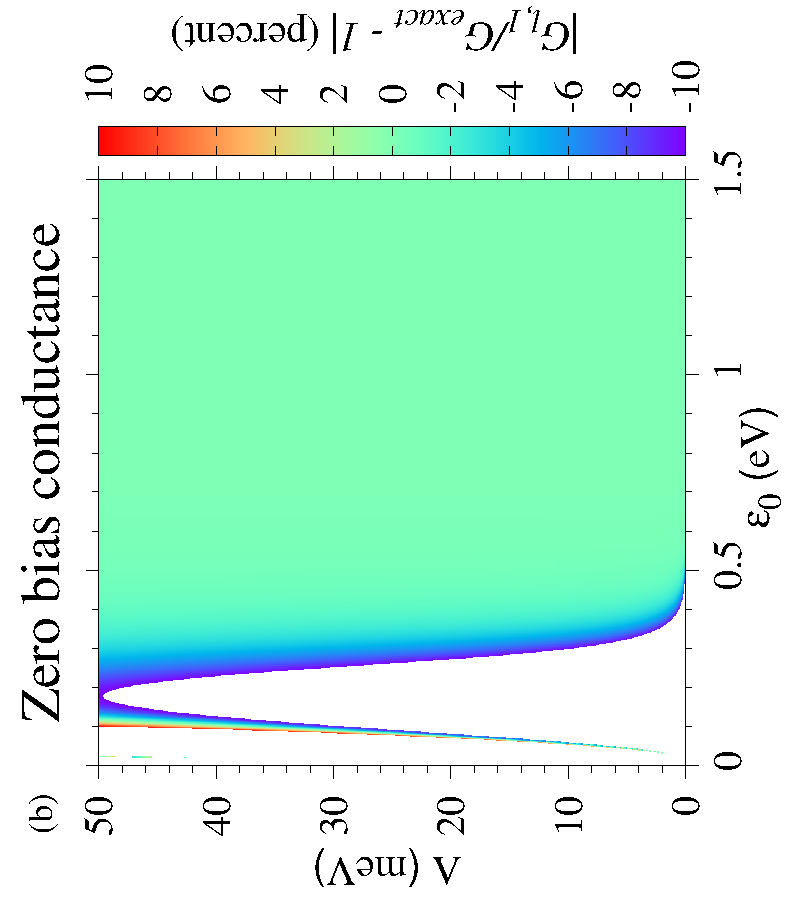}} 
  \centerline{
    \includegraphics[width=0.22\textwidth,height=0.3\textwidth,angle=-90]{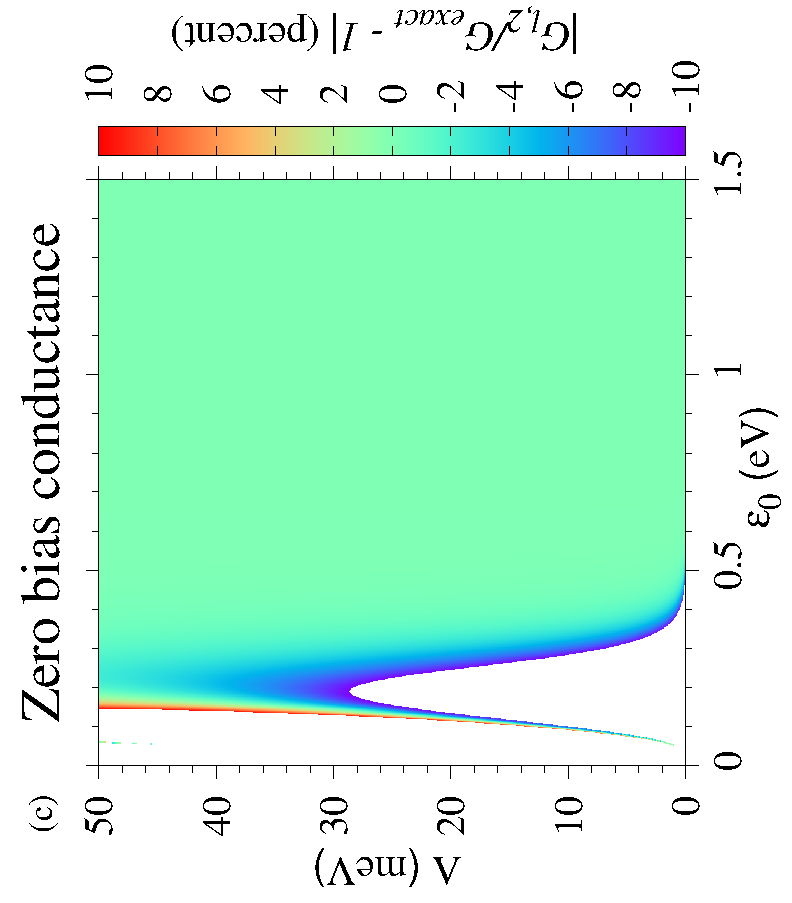} 
    \includegraphics[width=0.22\textwidth,height=0.3\textwidth,angle=-90]{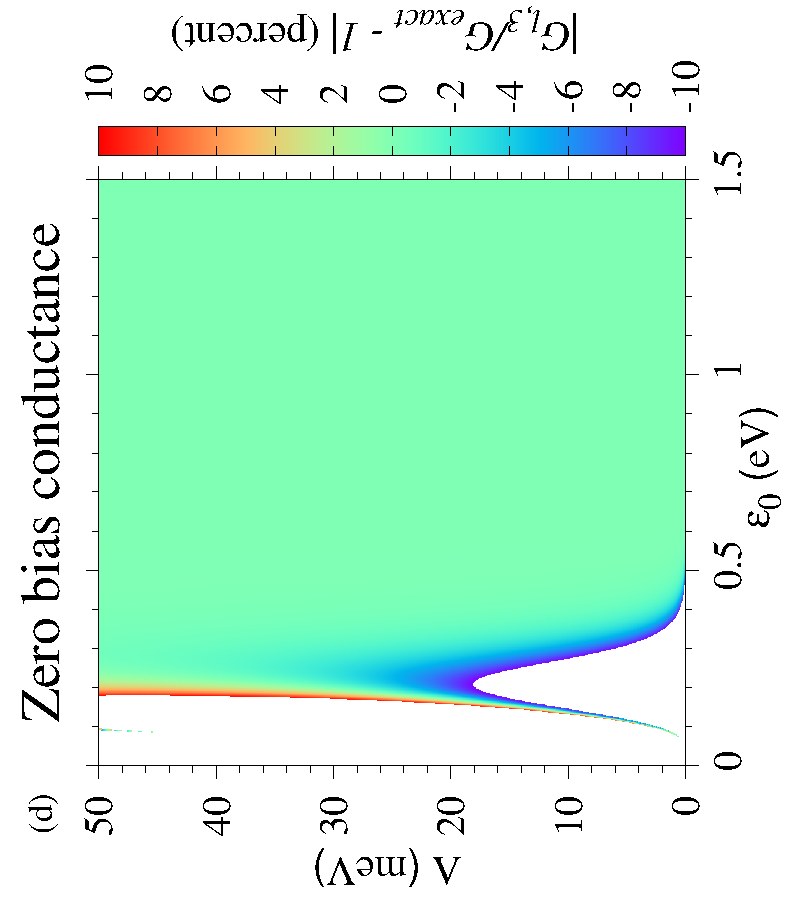}} 
  \centerline{
    \includegraphics[width=0.22\textwidth,height=0.3\textwidth,angle=-90]{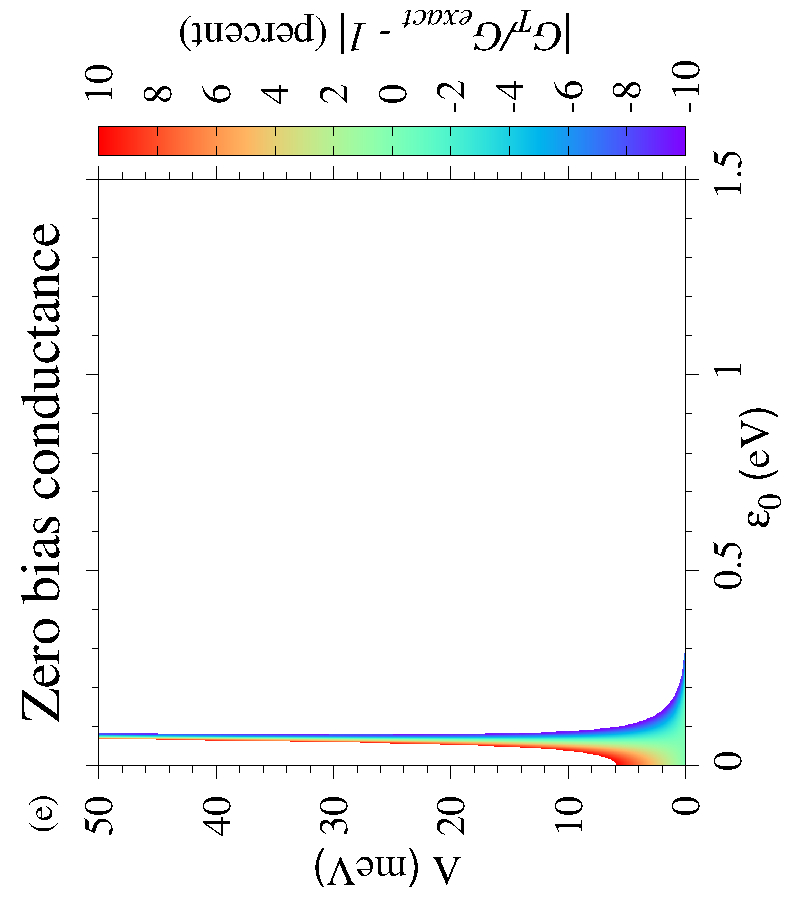} 
    \includegraphics[width=0.22\textwidth,height=0.3\textwidth,angle=-90]{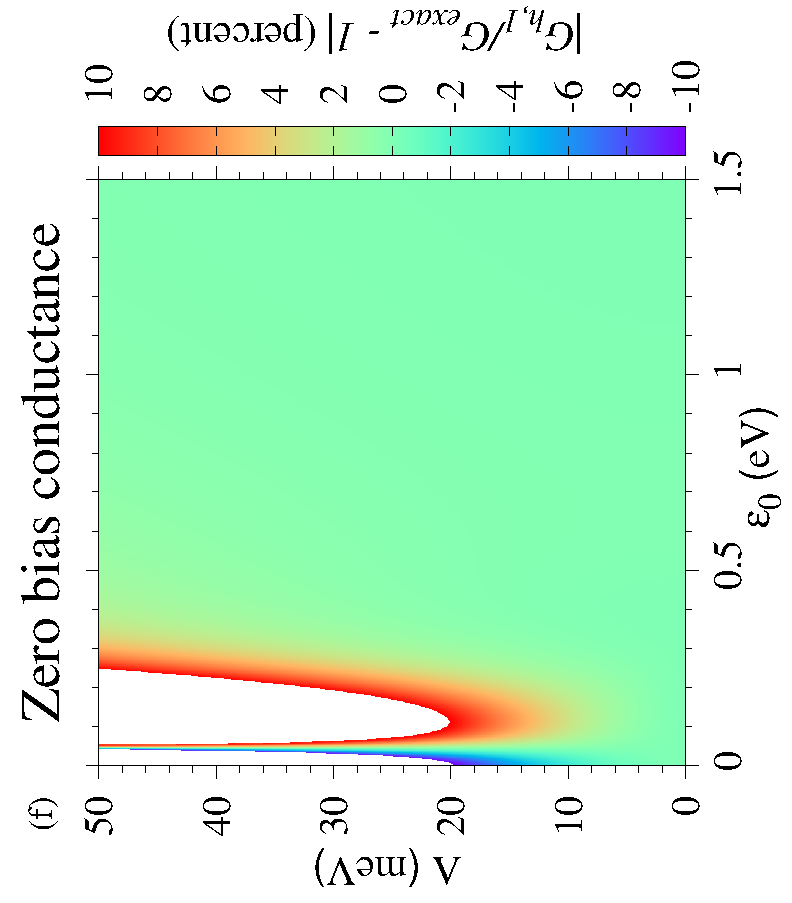} 
    \includegraphics[width=0.22\textwidth,height=0.3\textwidth,angle=-90]{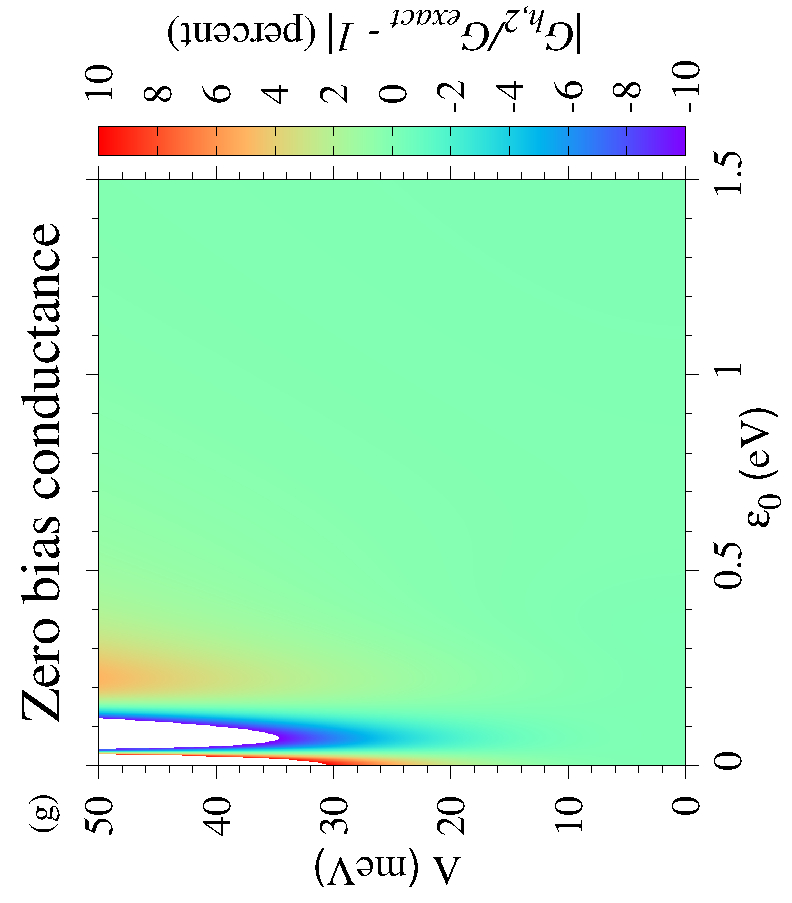}} 
  \caption{Percentage deviations of the zero bias conductance computed by the various methods indicated in each of the panels a to g with respect to
    the value computed exactly at $T = 373.15$\,K.}
  \label{fig:V=0}
\end{figure*}

\begin{figure*}[htb]
  \centerline{
    \includegraphics[width=0.22\textwidth,height=0.3\textwidth,angle=-90]{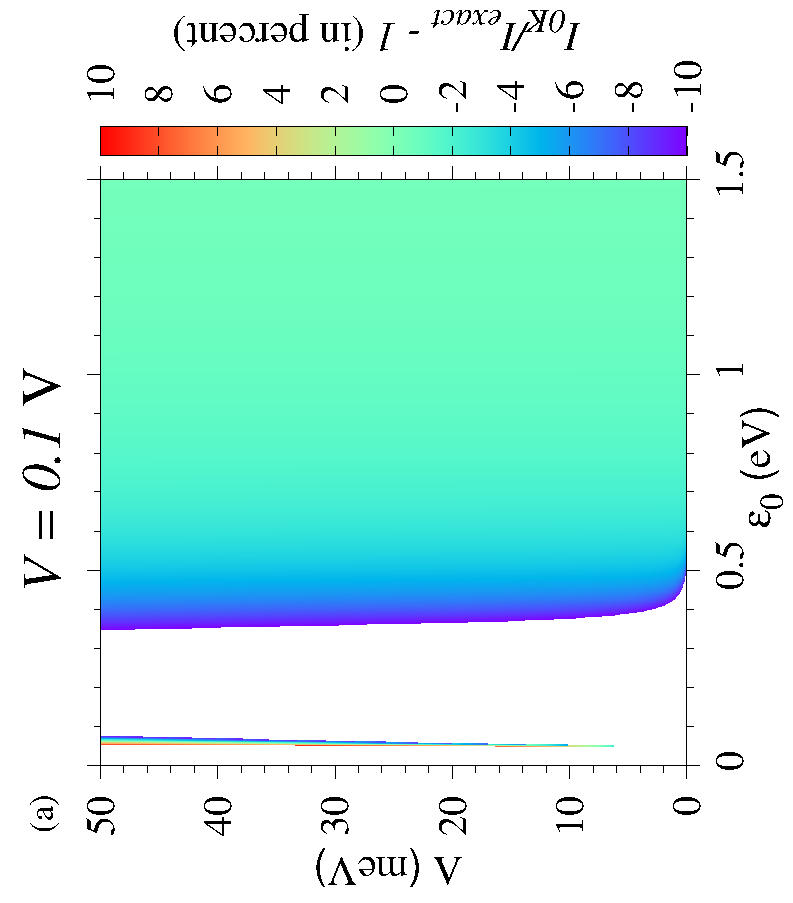} 
    \includegraphics[width=0.22\textwidth,height=0.3\textwidth,angle=-90]{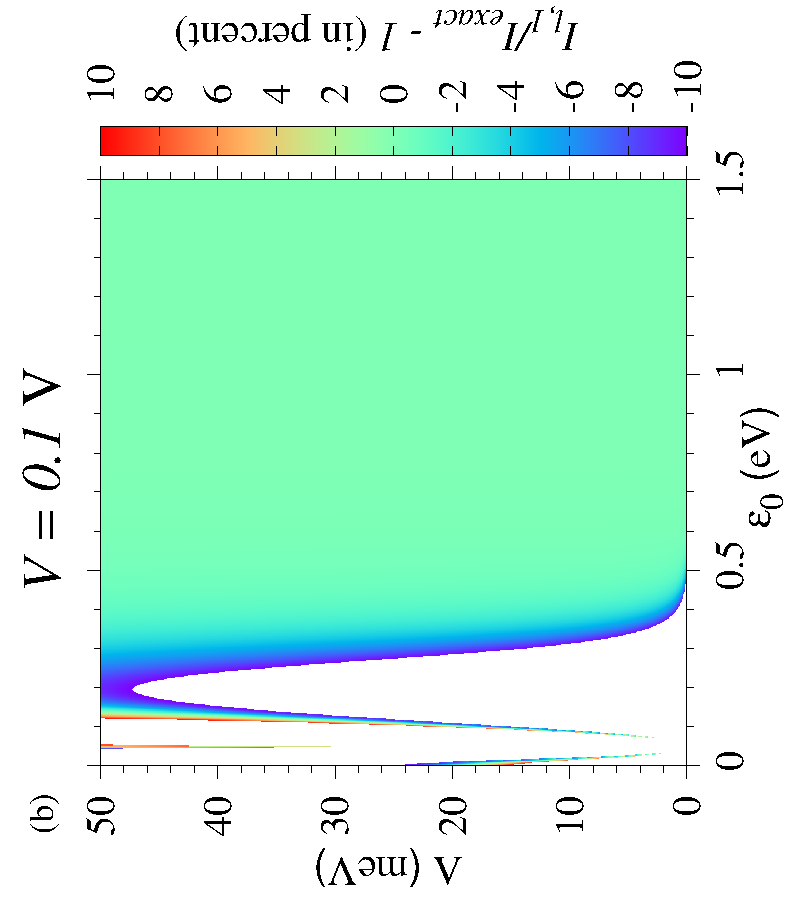} 
    \includegraphics[width=0.22\textwidth,height=0.3\textwidth,angle=-90]{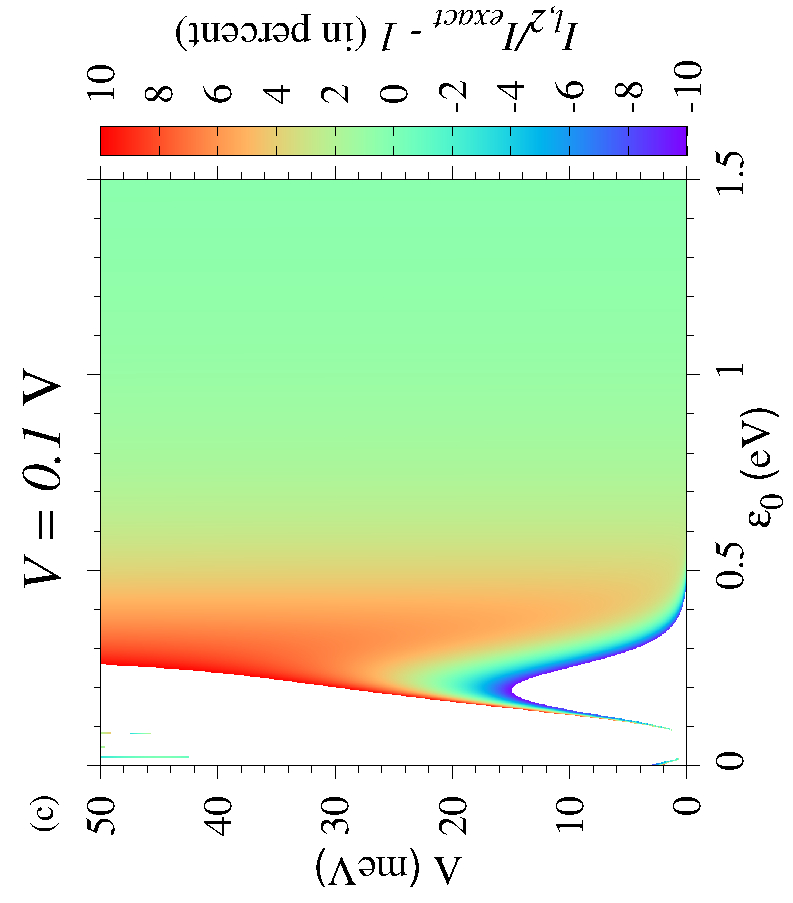} 
}
    \centerline{
    \includegraphics[width=0.22\textwidth,height=0.3\textwidth,angle=-90]{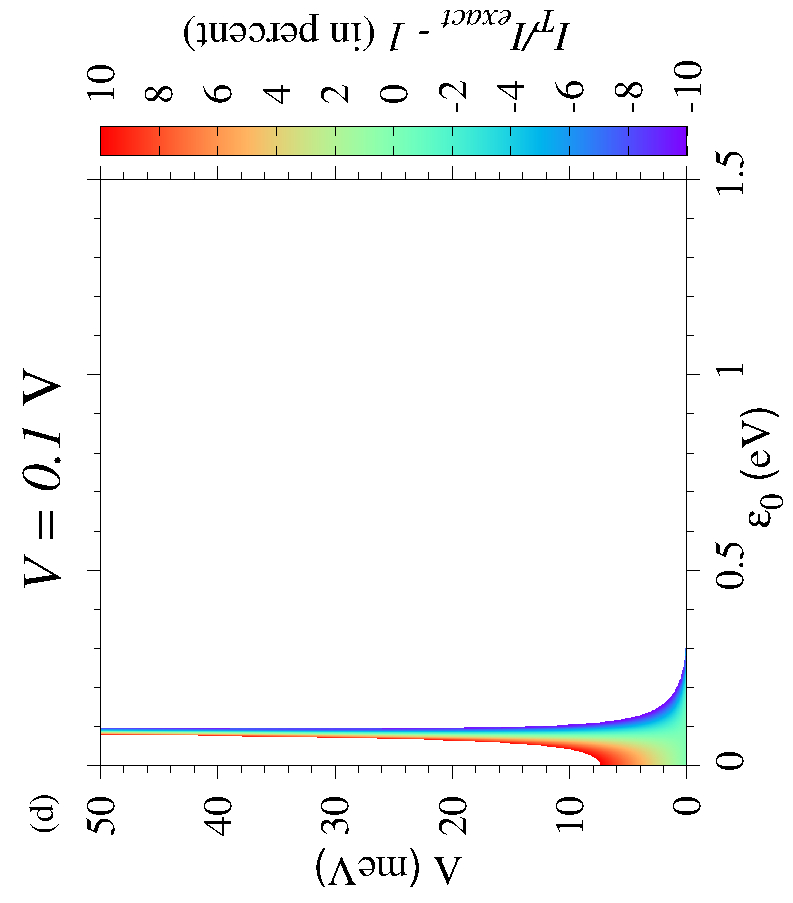} 
    \includegraphics[width=0.22\textwidth,height=0.3\textwidth,angle=-90]{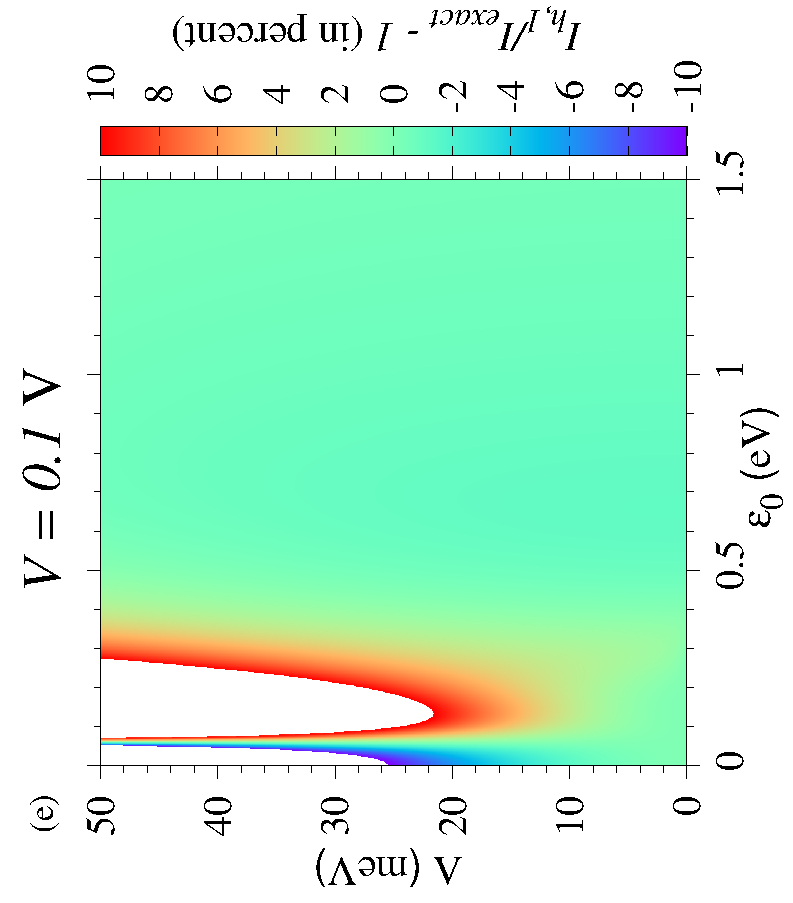} 
    \includegraphics[width=0.22\textwidth,height=0.3\textwidth,angle=-90]{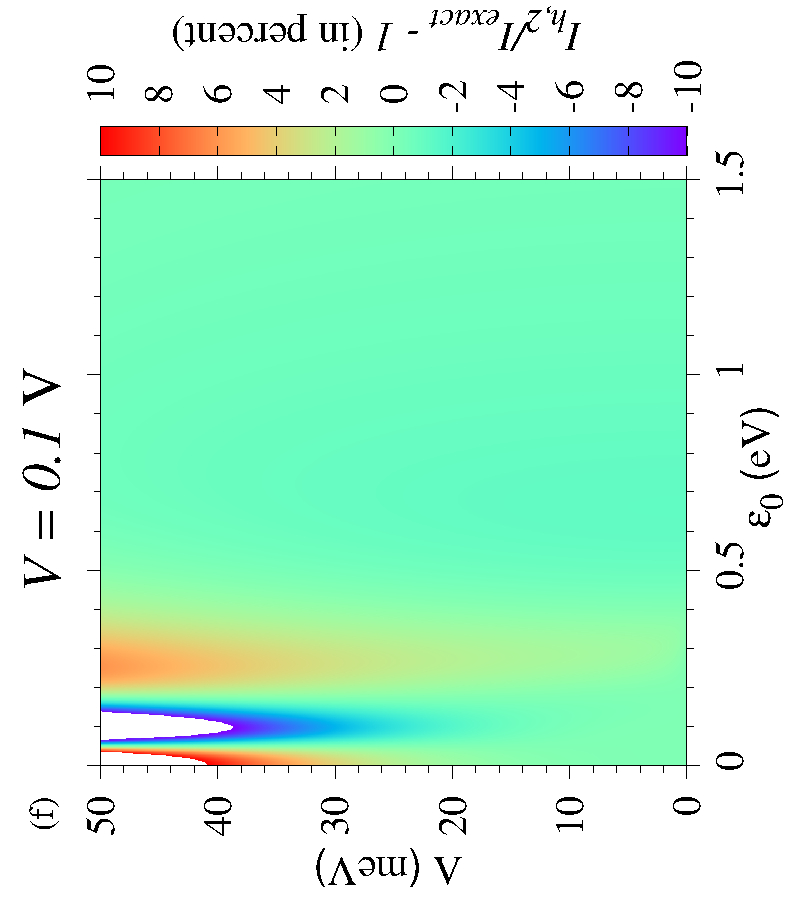}} 
    \caption{Percentage deviations of the current computed at $V = 0.1$\,V and $T = 373.15$\,K by the various methods indicated in each of the panels a to f with respect to
    the value computed exactly.}
  \label{fig:V=0.1}
\end{figure*}

\begin{figure*}[htb]
  \centerline{
    \includegraphics[width=0.22\textwidth,height=0.3\textwidth,angle=-90]{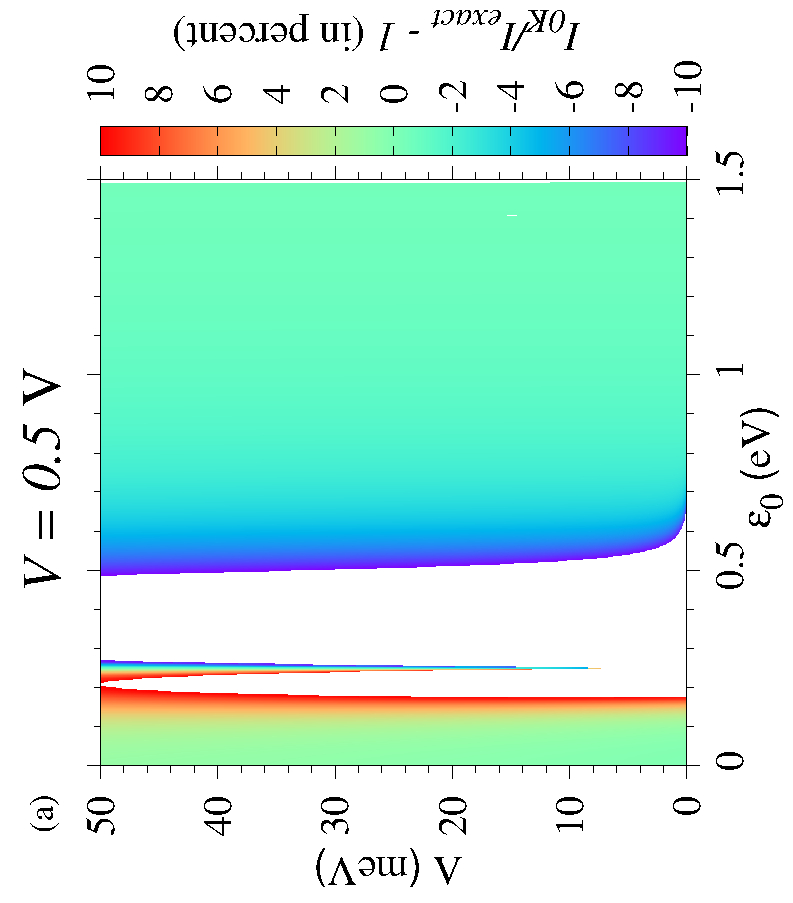} 
    \includegraphics[width=0.22\textwidth,height=0.3\textwidth,angle=-90]{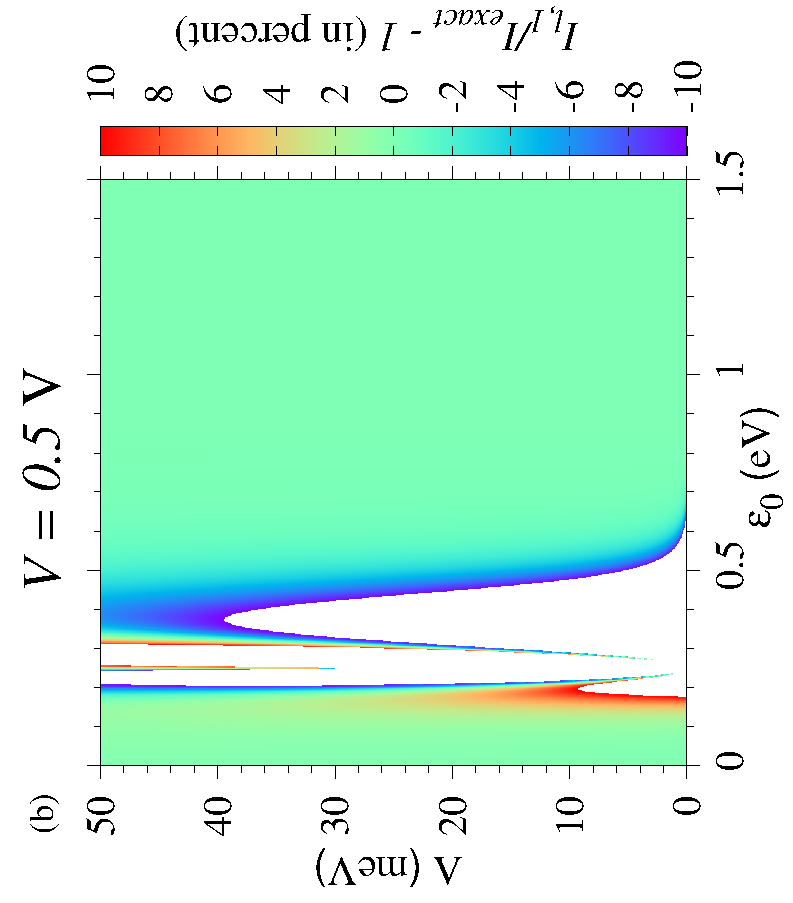} 
    \includegraphics[width=0.22\textwidth,height=0.3\textwidth,angle=-90]{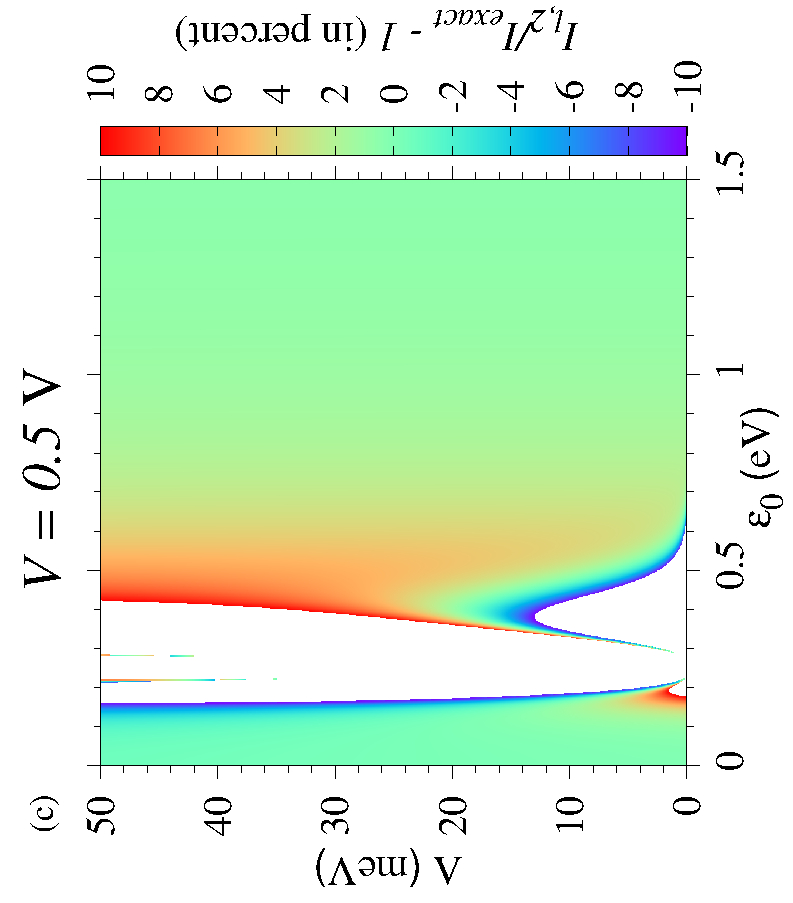} 
}
  \centerline{
    \includegraphics[width=0.22\textwidth,height=0.3\textwidth,angle=-90]{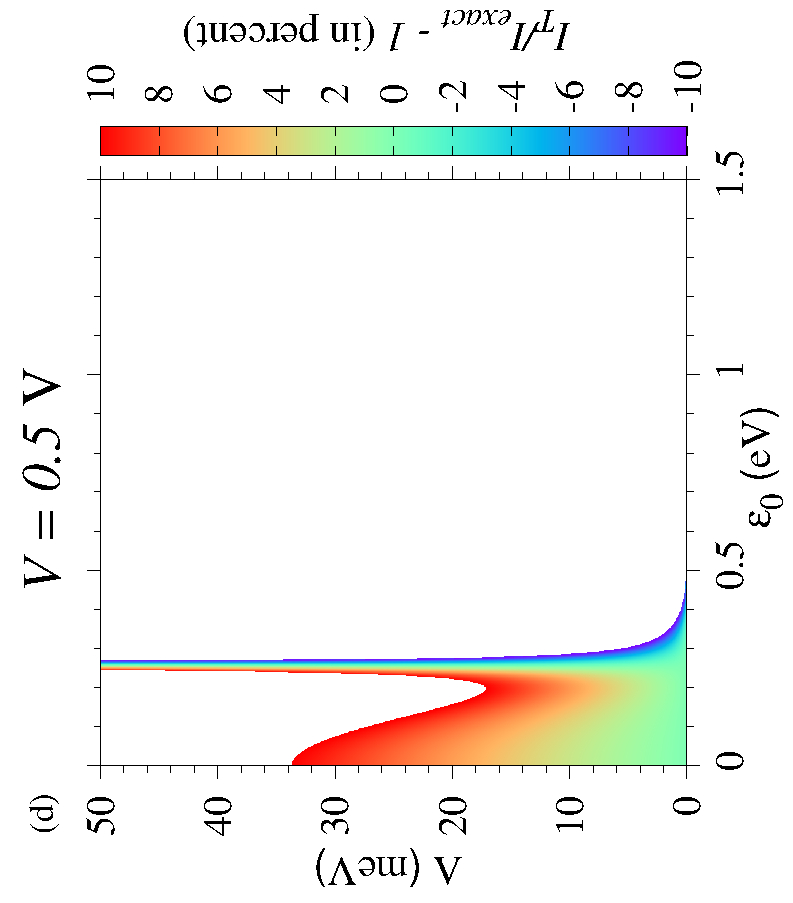} 
    \includegraphics[width=0.22\textwidth,height=0.3\textwidth,angle=-90]{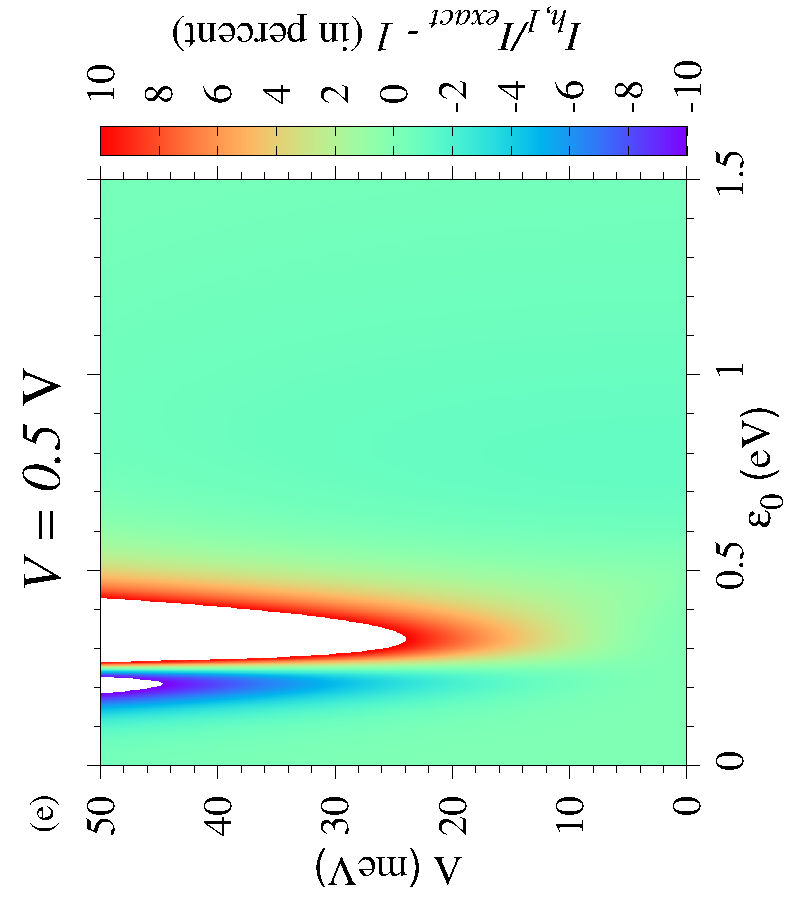} 
    \includegraphics[width=0.22\textwidth,height=0.3\textwidth,angle=-90]{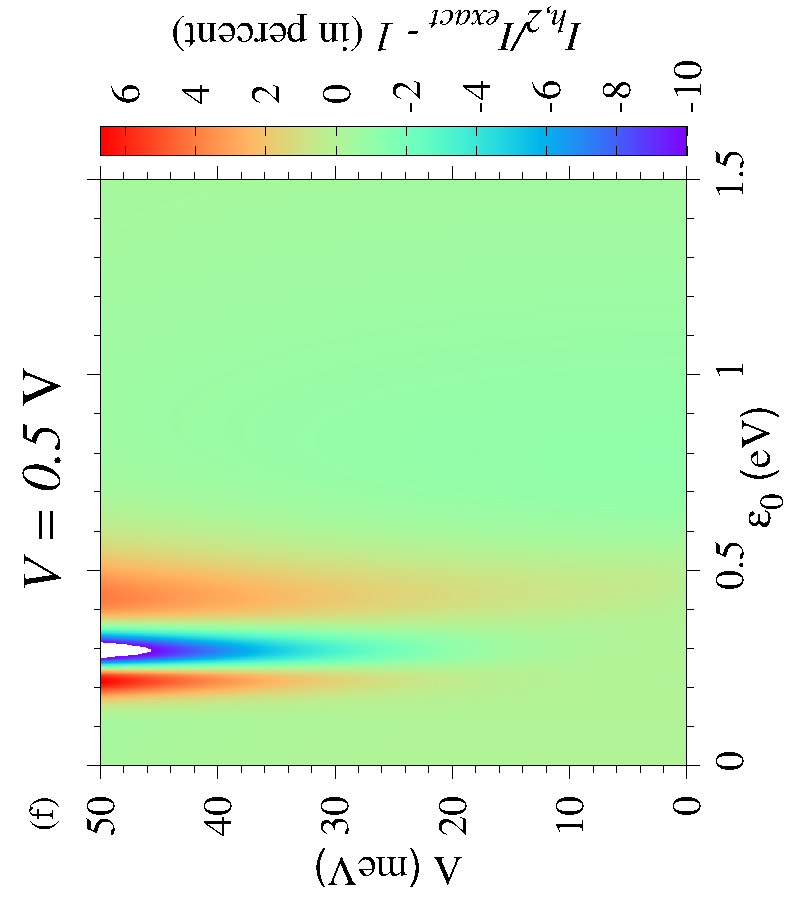} 
  }
  \caption{Percentage deviations of the current computed at $V = 0.5$\,V and $T = 373.15$\,K by the various methods indicated in each of the panels a to f with respect to
    the value computed exactly.}
  \label{fig:V=0.5}
\end{figure*}

\begin{figure*}[htb]
  \centerline{
    \includegraphics[width=0.22\textwidth,height=0.3\textwidth,angle=-90]{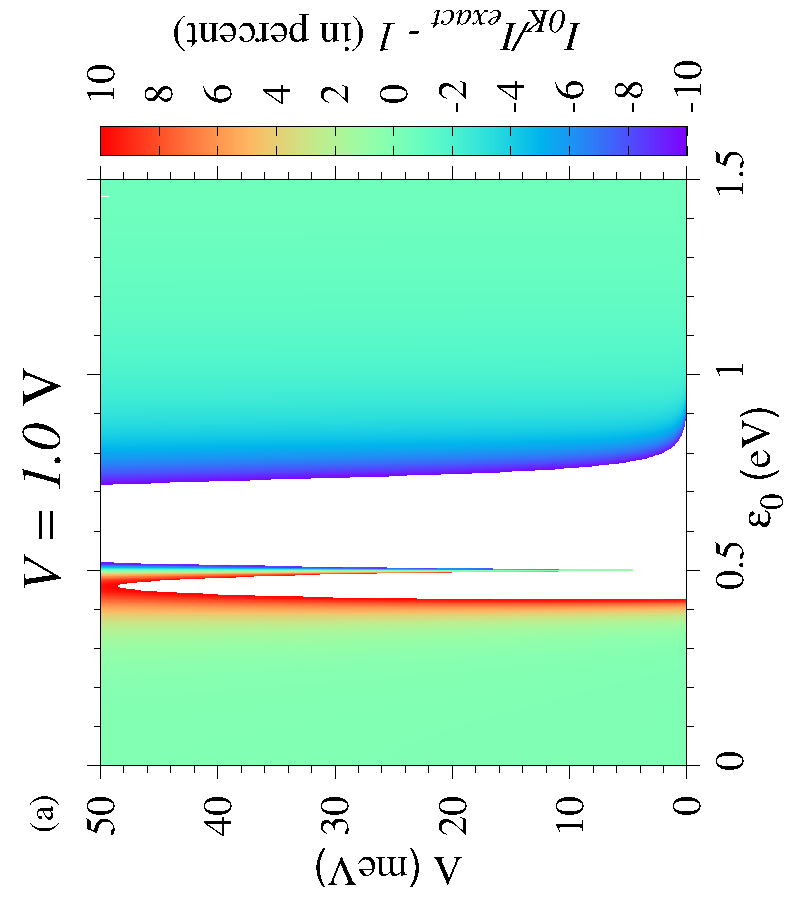} 
    \includegraphics[width=0.22\textwidth,height=0.3\textwidth,angle=-90]{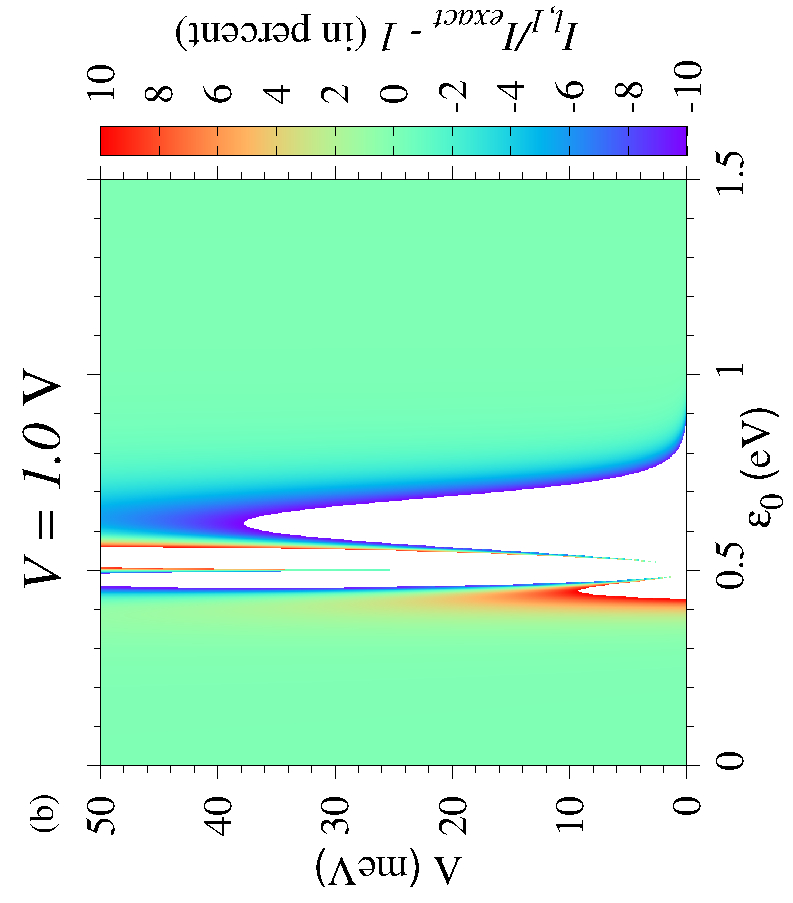} 
    \includegraphics[width=0.22\textwidth,height=0.3\textwidth,angle=-90]{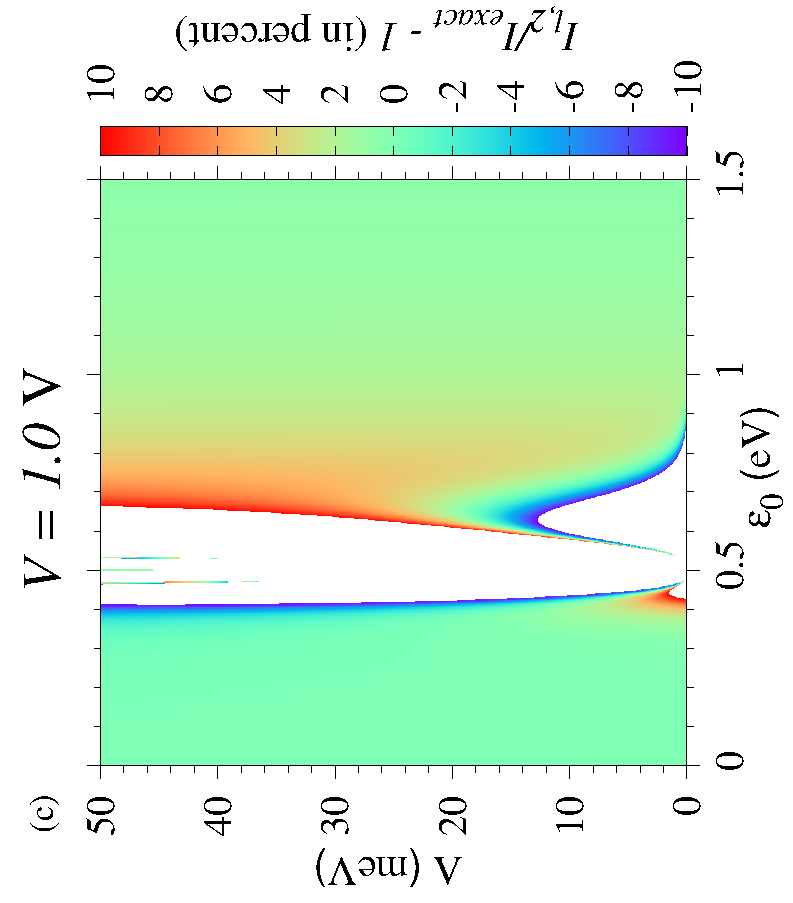} 
}
  \centerline{
    \includegraphics[width=0.22\textwidth,height=0.3\textwidth,angle=-90]{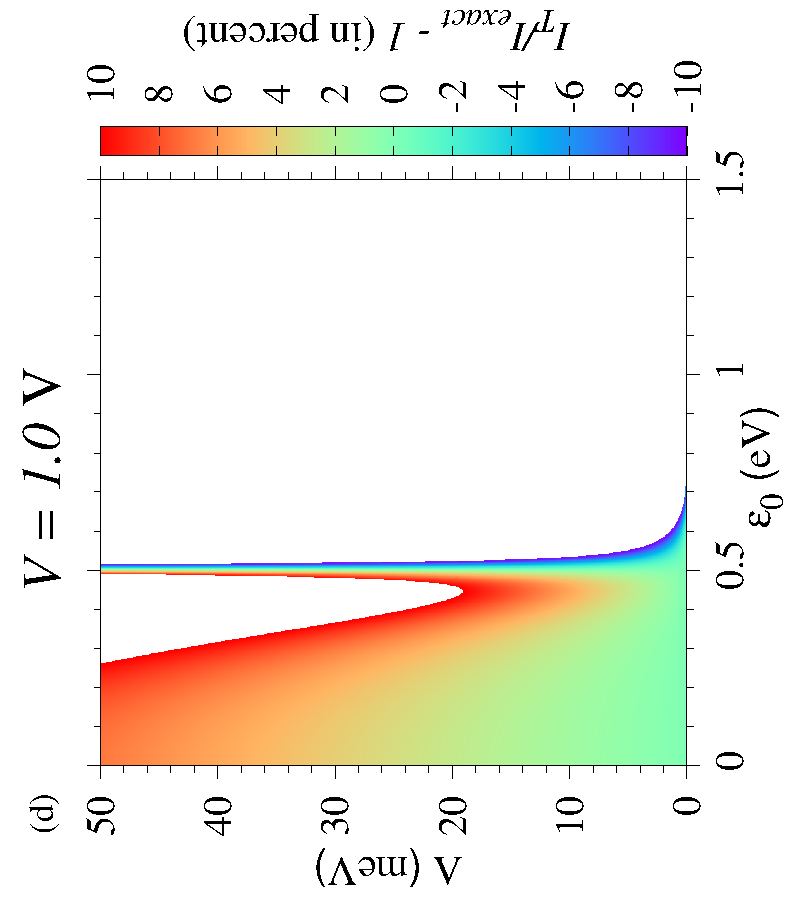} 
    \includegraphics[width=0.22\textwidth,height=0.3\textwidth,angle=-90]{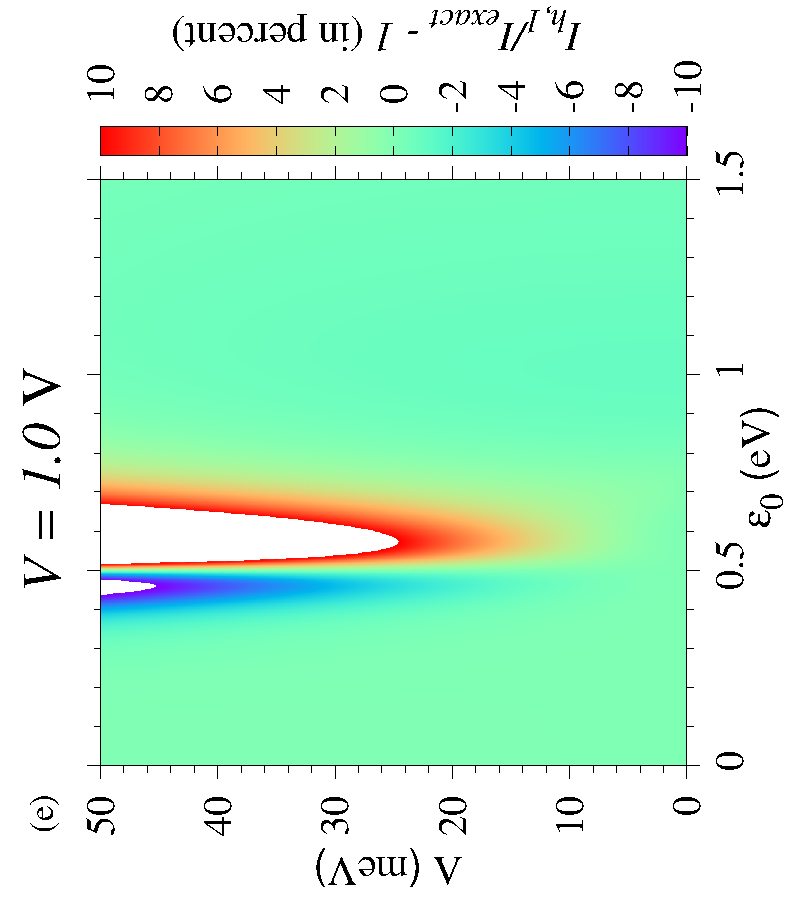} 
    \includegraphics[width=0.22\textwidth,height=0.3\textwidth,angle=-90]{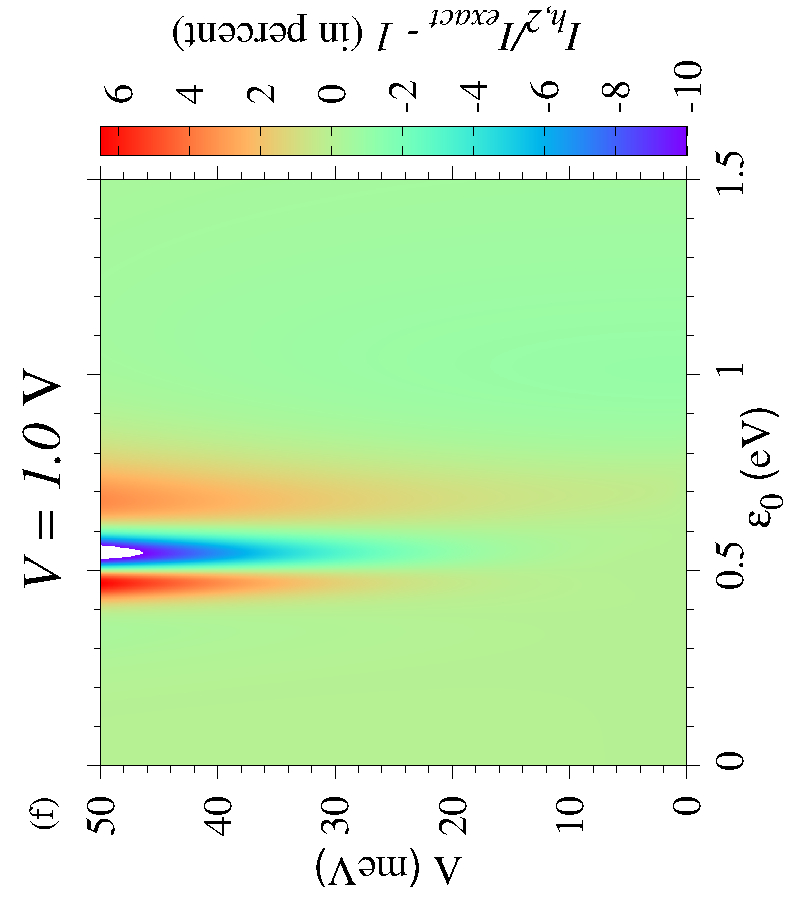}}  
  \caption{Percentage deviations of the current computed at $V = 1.0$\,V and $T = 373.15$\,K by the various methods indicated in each of the panels a to f with respect to
    the value computed exactly.}
  \label{fig:V=1.0}
\end{figure*}

To illustrate the accuracy of the approximate formulas for conductance and current deduced in \secname\ref{sec:g-0K-T},
we present comparison with the exact \gls~(\ref{eq-Gex}) and (\ref{eq-Iex}) at $T = 373.15$\,K at several biases:
$V = 0; 0.1; 0.5; 1.0$\,V (\figsname\ref{fig:V=0} to \ref{fig:V=1.0}, respectively).

Methodologically, performing (Sommerfeld) expansions in powers of $T$ (more precisely, in even powers of $T$
because odd powers do not contribute)
is the most straightforward way to account for thermal effects on the tunneling current.\cite{Sommerfeld:33,AshcroftMermin}
Doing so, \gls~(\ref{eq-I-l1}) to (\ref{eq-I-l3}) and \gls~(\ref{eq-G-l1}) to (\ref{eq-G-l3}) can be recovered
after some tedious calculations.
Inspection reveals that, including the first order correction ($\propto T^2$),
$G_{l1,}$ and $I_{l,1}$ represent a better description with respect to the zero temperature
quantities $G_{0K}$ and $I_{0K}$; compare
\figsname\ref{fig:V=0}b, \ref{fig:V=0.1}b, \ref{fig:V=0.5}b, and \ref{fig:V=1.0}b
with \figsname\ref{fig:V=0}a, \ref{fig:V=0.1}a, \ref{fig:V=0.5}a, and \ref{fig:V=1.0}a, respectively.  
Still, although the second order correction ($\propto T^4$) depicted in
\figsname\ref{fig:V=0}c, \ref{fig:V=0.1}c, \ref{fig:V=0.5}c, and \ref{fig:V=1.0}c
represents a further improvement,
expansion in powers of $ k_B T/\Lambda$ inherently fails for sufficiently small $\Lambda$,
when the transport strongly depends on temperature.

Unfortunately, although appealing simple, the expressions for $G_T$ and $I_{T}$ deduced in the high temperature limit 
(\gls~(\ref{eq-G_T})) and (\ref{eq-I_T}))
appear to have a very restricted validity; see \figsname\ref{fig:V=0}d, \ref{fig:V=0.1}d,
\ref{fig:V=0.5}d, and \ref{fig:V=1.0}d. Below resonance, $I_{T}$ is only reasonably in a very narrow parameter range.
And yet, the formulas deduced by adding first and second high
temperature corrections ($G_{h,1}$, $G_{h,2}$, \gls~(\ref{eq-G-h1}) and (\ref{eq-G-h2})
and $I_{h,1}$, $I_{h,2}$, \gls~(\ref{eq-I-h1}) and (\ref{eq-I-h2})) turn out to be very accurate
even in the parameter ranges where the thermal impact is very pronounced;
see \figsname\ref{fig:V=0}e and f, \figsname\ref{fig:V=0.1}e and f, \figsname\ref{fig:V=0.5}e and f, and \figsname\ref{fig:V=1.0}d and f.

Although the main aim of the present paper is to present general analytical formulas for the
current through molecular junctions mediated by a single (dominant) channel
rather than discussing applications to real junctions, in the next two sections we will analyze $I$-$V$ curves measured for
\ib{junctions fabricated using two extreme platforms:
  a large area junction with top electrode of eutectic gallium indium alloy (EGaIn)\cite{Whitesides:15}
  (\secname\ref{sec:sc14}) and a single-molecule junction \cite{Zotti:10} (\secname\ref{sec:mcbj}).}

\ib{
\section{Analyzing transport data of a specific large area EGaIn-based molecular junction}
\label{sec:sc14}
In the above presentation, emphasis was on 
the temperature dependence of the tunneling current.
However, we also want to draw attention 
on an effect that did not explicitly entered the foregoing discussion:
the possibility that the MO of the active molecule acquires a significant width from
the surrounding medium ($\Gamma_{env} \neq 0$). 
Furthermore, by focusing on a specific large area EGaIn junction below, we also aim at illustrating 
that/how the single level model can help to determine the fraction $f$
of the current carrying molecules $N_{\mbox{\small eff}}$, which represent only tiny amount of the total (nominal) number
of molecules $N_{n}$ in junction ($f \equiv N_{\mbox{\small eff}} / N_{n} \ll 1$).\cite{Selzer:05,Milani:07,Whitesides:13,Frisbie:16d,Cahen:20}

To better underline the importance of the aforementioned, we have
intentionally chosen experimental results reported for a real junction
wherein the impact of temperature is negligible (\emph{cf.}~\figname\ref{fig:sc14}c):
a large area EGaIn-based junction fabricated with 1-tetradecanethiol SAMs
(C14T) chemisorbed on gold electrode \ce{Au-S-(CH2)13-CH_3 / EGaIn}.\cite{Whitesides:15}

Typically, experiments on large area molecular junctions report values for the current density $J = J(V) $
(in A/cm\(^2\)) rather than $I = I(V)$ (in A). For this reason, 
we recast \gls~(\ref{eq-Iex}), (\ref{eq-I-0K}), and (\ref{eq-I-0K-off}) needed below
in terms of current densities
\begin{subequations}
  \label{eq-j}
\begin{eqnarray}
  J_{exact} & \equiv & \frac{N_{\mbox{\small eff}}}{\mathcal{A}_n} I_{exact} = \overline{\Gamma}^2 \frac{G_0}{e \Lambda} 
  \left[
    \mbox{Im}\, \psi\left( \frac{1}{2} + \frac{\Lambda}{2 \pi k_B T} + i \frac{\varepsilon_{V} + e V/2}{2 \pi k_B T}\right) \right . \nonumber \\
  \label{eq-jex}
    & - & \left .
    \mbox{Im}\, \psi\left( \frac{1}{2} + \frac{\Lambda}{2 \pi k_B T} + i \frac{\varepsilon_{V} - e V/2}{2 \pi k_B T}\right)
    \right] 
\end{eqnarray}
\begin{equation}
  J_{0K} = 
  \overline{\Gamma}^2 \frac{G_0}{e \Lambda}
  \left(\arctan \frac{\varepsilon_{V} + e V/2}{\Lambda} -
  \arctan \frac{\varepsilon_{V} - e V/2}{\Lambda} \right)
  \label{eq-j-0K}
\end{equation}
\begin{equation}
  J_{0K,off} = \frac{G_0}{e} \frac{\overline{\Gamma}^2}{\varepsilon_{V}^2 - (e V/2)^2} V
  \label{eq-j-0K-off}
 \end{equation}
\end{subequations}
Here
\begin{equation}
  \label{eq-N-eff}
  \overline{\Gamma}^2 \equiv \frac{N_{\mbox{\small eff}}}{\mathcal{A}_n} \Gamma^2 =
  \underbrace{\frac{N_{\mbox{\small eff}}}{N_n}}_{f} \underbrace{\frac{N_n}{\mathcal{A}_n}}_{\Sigma} \Gamma^2 = f \Sigma \Gamma^2
\end{equation}  
and $\Sigma$ is the SAM coverage.
Recall that $\varepsilon_{V}$ entering \gl~(\ref{eq-j}) is a shorthand for $\varepsilon_0(V)$, \emph{cf.}~\gl~(\ref{eq-gamma}). 

Reliable values of the parameters $\varepsilon_{0}$, $\overline{\Gamma}$, and $\gamma$
can be obtained by fitting the experimental $I$-$V$ data
with any of \gl~(\ref{eq-j}),
The pertaining results are presented in Table~\ref{table:sc14} and \Figname\ref{fig:sc14}.
Noteworthily, as reiterated again and again,\cite{Baldea:2017g,baldea:comment,Baldea:2023a} 
to correctly employ our model proposed in ref.~\citenum{Baldea:2012a},
\gl~(\ref{eq-j-0K-off}) should be applied to bias ranges $\vert V \vert \alt 1.2\,V_t$ (\emph{cf.}~\figname\ref{fig:sc14}d).
Here, the transition voltage $V_t$ is defined as the bias where $V^2/\vert I \vert $ is maximum (\figname\ref{fig:sc14}a).
All three fitting curves successfully reproduce the measurements (\emph{cf.}~\Figname\ref{fig:sc14}b).
The good agreement between these methods demonstrates that
in the large area Au-C14T/EaGaIn considered 
thermal effects are negligible ($j_{0K} \approx j_{exact}$, \emph{cf.}~\figname\ref{fig:sc14}c),
and that the transport in off-resonant situations can be reliably described by
$j_{0K,off}$($\approx j_{exact}$, see green curve in \figname\ref{fig:sc14}b).
\begin{figure*}[htb]
  \centerline{
    \includegraphics[width=0.3\textwidth]{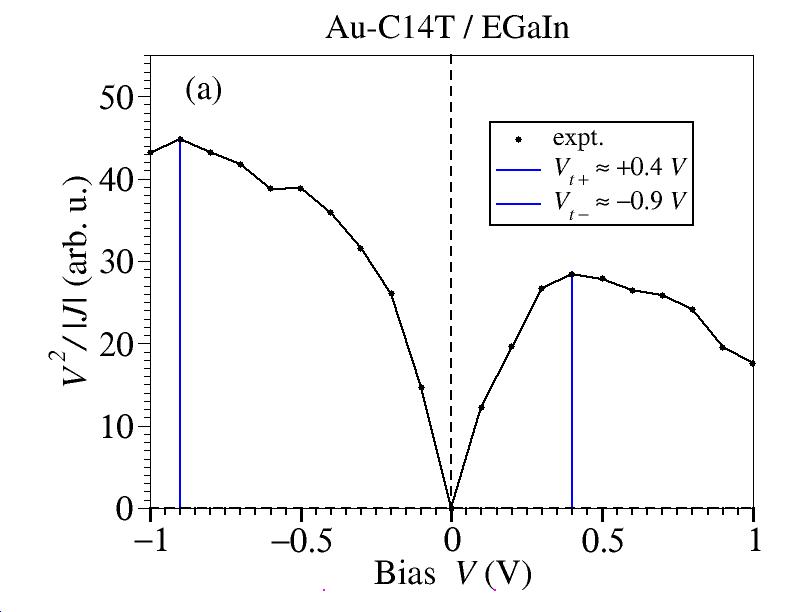} 
    \includegraphics[width=0.3\textwidth]{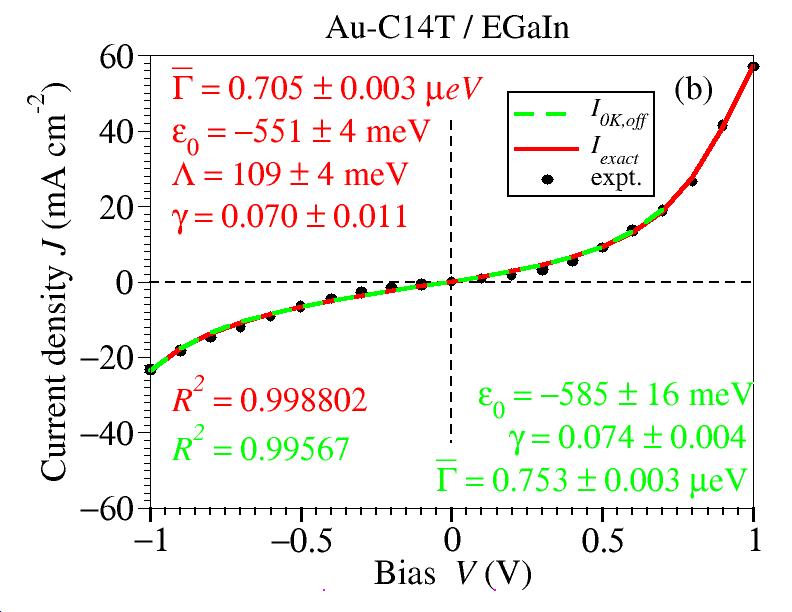} 
    }
  \centerline{
    \includegraphics[width=0.22\textwidth,height=0.3\textwidth,angle=-90]{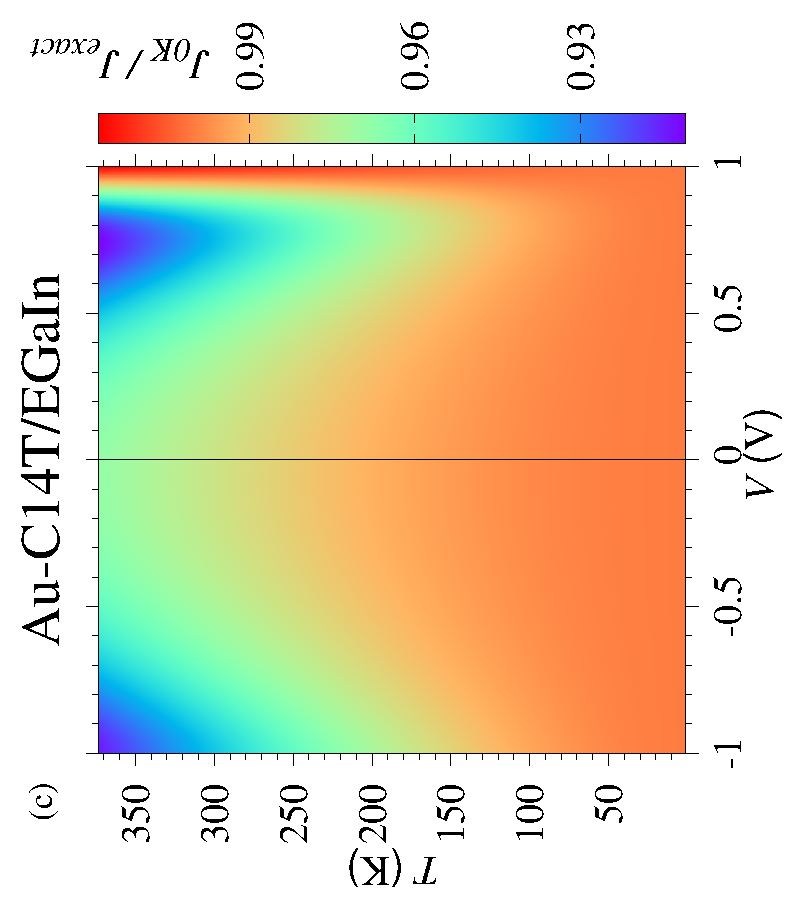}
    \includegraphics[width=0.22\textwidth,height=0.3\textwidth,angle=-90]{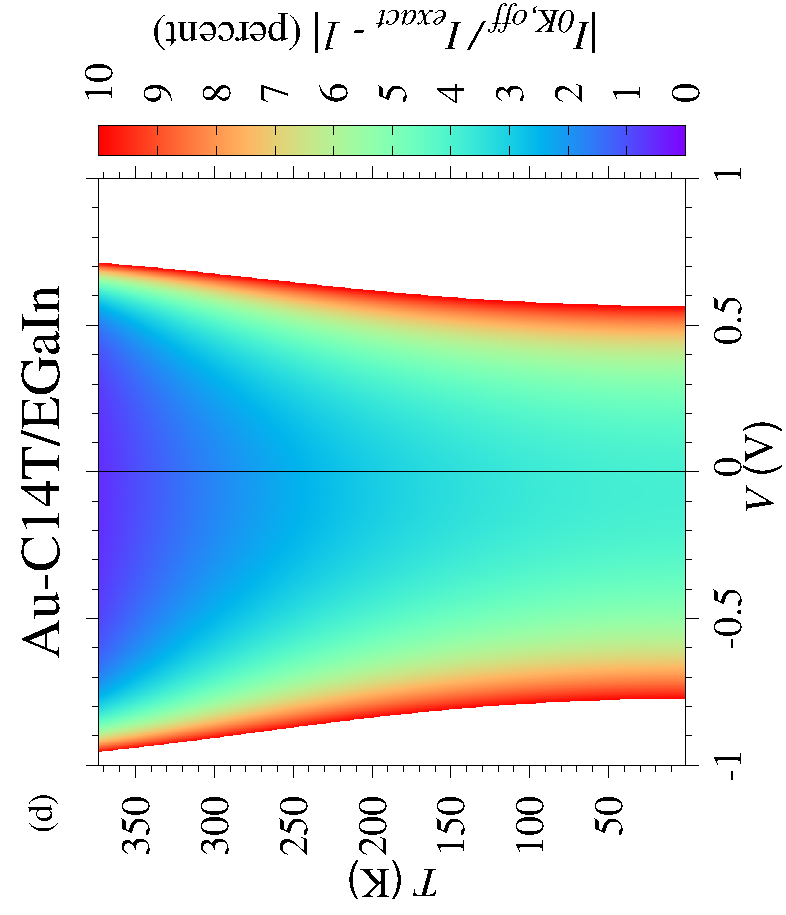}
  }
  \caption{(a) Plots of $V^2/|I|$ \emph{versus} $V$
    enabling the determination of the values of the transition voltages $V_{t\pm}$
    needed to properly define the bias range where \gl~(\ref{eq-j-0K}) applies.
    (b) Correctly utilized, \gl~(\ref{eq-j-0K}) is able to reproduce (green curve)
    the experimental data (black
    points) for the large area Au-C14T/EGaIn junction of ref.~\citenum{Whitesides:15}.
    with an accuracy comparable to that of the exact \gl~(\ref{eq-jex} (red curve).
    Diagrams revealing (c) that thermal effects are negligible in the Au-C14T/EGaIn junction envisaged
    and (d) that, when applied for the bias range for it was devised,\cite{Baldea:2012a}
    \gl~(\ref{eq-j-0K}) is able to describe quantitatively $I$-$V$ data in large area junctions.
    See the main text for details.
  }
  \label{fig:sc14}
\end{figure*}
\begin{table*}
\small
\centering
\caption{
  {Model parameter values obtained by fitting the experimental $J$-$V$ data \cite{Whitesides:15}
    to the formulas for the current indicated in the first column}
}
\footnotesize
\begin{tabular*}{0.98\textwidth}{@{\extracolsep{\fill}}rccccccccc}
  \hline
  Method       &   $\varepsilon_{0}$ (eV)  &  $\Lambda $ (meV) & $\overline{\Gamma}$ (\(\mu\)eV/nm) &  $\gamma$ & $f$ & $\Gamma_{env}$ & $\Gamma_{env} /(2 \Lambda) $ & $R^2$  \\
\hline
$J_{exact}$   &   $ -0.551 \pm 0.004 $   &  $ 109 \pm 0.004 $  & $ 0.705 \pm 0.003 $ & $ 0.07  \pm 0.01 $  & $4.3\times 10^{-4}$ & 217.3 & 99.7\% & 0.998802 \\
$J_{0K}$      &  $ -0.535 \pm 0.004 $   &  $ 115 \pm  0.004 $ & $ 0.695 \pm 0.002 $ & $ 0.07  \pm 0.01 $  & $4.2\times 10^{-4}$ & 229.3 &  99.7\%  & 0.99884 \\
$J_{0K,off}$  &   $ -0.585 \pm 0.004 $   &  ---                & $ 0.753 \pm 0.003 $ & $ 0.074 \pm 0.004 $ & $4.9\times 10^{-4}$ &  ---   & ---  & 0.99567\\
  \hline
\end{tabular*}
\label{table:sc14} 
\end{table*}
\subsection{Working equations for large area EGaIn junctions}
\label{sec:working-eq}
Let us first recast \gls~(\ref{eq-Gamma_g}), (\ref{eq-Gamma_a}), and (\ref{eq-N-eff})
in a form more specific for junctions whose molecules are chemisorbed on a gold substrate (label $c, Au$) 
and physisorbed on EGaIn top electrode (label $p, EGaIn$) or CP-AFM gold tip (label $p, Au$)
\begin{subequations}
  \label{eq-w.eq}
\begin{eqnarray}
  \label{eq-splitting-Gamma}
  \Gamma^2 & \to & \Gamma_{c, Au|p, EGaIn}^2 = \Gamma_{c,Au} \Gamma_{p,EGaIn} = \frac{\overline{\Gamma}^2}{f \Sigma}\\
  \label{eq-relationship-Gammas}
  \Gamma_{p,EGaIn} & = &
  \Gamma_{c,Au} \underbrace{\frac{\Gamma_{p,Au}}{\Gamma_{c,Au}}}_{r_1 = 1/37}
  \underbrace{\frac{\Gamma_{p,EGaIn}}{\Gamma_{p,Au}}}_{r_2 = 1/7} \mbox{ (from ref.~\citenum{Baldea:2022j})}\\
\label{eq-CP-AFM}
  \Gamma_{c, Au|p, Au} & \equiv & \sqrt{\Gamma_{c, Au} \Gamma_{p, Au} } = 118\, \mu\mbox{eV (} via \mbox{ ref.~\citenum{Baldea:2019h})} \\
  \Lambda & = & \frac{1}{2}\left(\Gamma_{c, Au} + \Gamma_{p,EGaIn} + \Gamma_{env}\right)
\label{eq-Lambda-EGaIn}
\end{eqnarray} 
\end{subequations}
Above, $\Gamma_{c, Au|p, EGaIn}$ and $ \Gamma_{c, Au|p, Au} $
are effective couplings (\emph{cf.}~\secname\ref{sec:model}) of
junctions chemisorbed on gold substrate and physisorbed
on EGaIn and gold top electrodes, respectively.
$\Gamma$'s with subscripts without vertical bar stand for individual HOMO-electrode couplings
(recall that alkanethiols CnT exhibit p-pype HOMO-mediated conduction\cite{Baldea:2019h});
\emph{e.g}, $\Gamma_{p,X}$ means p(hysical) coupling to electrode X($=Au, EGaIn$).

Let us now explain how we arrived at the numerical values entering \gls~(\ref{eq-relationship-Gammas})
and (\ref{eq-CP-AFM}).

The coupling ratios $r_{1,2}$ of \gl~(\ref{eq-relationship-Gammas}) were estimated recently.\cite{Baldea:2022j}
The value of $r_1 \approx 1/37$ was deduced by comparing the couplings of alkane thiols and alkane dithiols.
The value of $r_2 \approx 1/7$ relies on the dependence of the coupling on the electrode work function.

To get $\Gamma_{c,Au|p, Au} = 118\,\mu$eV in \gl~(\ref{eq-CP-AFM}), we extrapolated 
the values of the effective coupling $\Gamma_{c,Au|p, Au}^{(n)} \to \Gamma_n \propto \exp(-\beta n /2)$
for CP-AFM alkanethiols CnT (\ce{H-S-(CH2)_{n-1}-CH_3}) junctions (Au-CnT/Au).
That is, we used the values of $\Gamma_n$ (for $n=7,8,9,10,12$)
from Table~1 of ref.~\citenum{Baldea:2019h}.
The exponential scaling of $\Gamma_n \ versus \ n$ follows from the exponential falloff $G_n \propto \exp(-\beta n)$
of the low bias conductance $ G_n \propto \Gamma_n^2 /(\varepsilon_{0}^{n})^2$,
given the fact that the HOMO offset $\varepsilon_{o}^{(n)}$ of the CnT series is independent of size ($n$).\cite{Baldea:2019h}
\subsection{Estimating the fraction of current carrying molecules and the extrinsic HOMO relaxation}
\label{sec:f-GammaEnv}
We will now explain how to use the values of $\overline{\Gamma}$ and $\Lambda$ obtained from $I$-$V$ data fitting
and the above working equations to estimate the fraction of the current carrying molecules $f$
and the extrinsic contribution to the (HO)MO broadening $\Gamma_{env}$.
\subsubsection{Fraction of current carrying molecules $f$}
\label{sec:f}
From the definition of $r_1$ of \gl~(\ref{eq-relationship-Gammas}), we use \gl~(\ref{eq-CP-AFM})
to compute $\Gamma_{c,Au} = \Gamma_{c, Au|p, Au} /\sqrt{r_1} = 718\,\mu\mbox{eV}$. Using again the definition of $r_1$,
we deduce $\Gamma_{p, Au} = \Gamma_{c, Au|p, Au} \sqrt{r_1} = 3.2\,\mu\mbox{eV}$.
From the definition of  $r_2$ of \gl~(\ref{eq-relationship-Gammas}) we get
$\Gamma_{p, EGaIn} = r_2 \Gamma_{p, Au} = 0.456\,\mu\mbox{eV}$.
From \gl~(\ref{eq-splitting-Gamma}) we can now compute
$\Gamma_{c, Au| p, EGaIn} = \sqrt{\Gamma_{c,Au} \Gamma_{p, EGaIn}} = 18.1\,\mu\mbox{eV}$.
With the value $\Sigma = 3.5\,\mbox{molec\,nm}^{-2}$ measured for the molecular coverage of alkanethiol SAMs,\cite{Frisbie:16e}
and the fitting parameter value $\overline{\Gamma} = 0.705\,\mu\mbox{eV}$ (Table~\ref{table:sc14}),
we can finally estimate \emph{via} \gl~(\ref{eq-splitting-Gamma}) the fraction of the current carrying molecules
\[ f_{\mbox{\small Au-C14T/EGaIn}} \approx 4.3\times 10^{-4}\]
This value is on the same order of magnitudes of other values of $f$
reported for large area EGaIn-based junctions using methos completely different
from the present one.\cite{Whitesides:14,Frisbie:16a,Baldea:2022j} Definitely, out of the total/geometric/nominal
area at the EGaIn contact $\mathcal{A}_{n}$, the area of the physical/electric/effective contact
$\mathcal{A}_{\mbox{\small eff}} = f \mathcal{A}_{n}$ is but a tiny fraction ($f\ll 1$).
\subsubsection{Extrinsic contribution $\Gamma_{env}$ to the (HO)MO width}
\label{sec:GammaEnv}
With the values for $\Lambda$ (Table~\ref{table:sc14}), $\Gamma_{c, Au}$, and $\Gamma_{p,EGaIn}$ in hand,
we use \gl~(\ref{eq-Lambda-EGaIn}) and obtain $\Gamma_{env} \approx 217.3\,\mbox{meV}$.
This is a very important result.
It demonstrates that the extrinsic contribution represents
$(\Gamma_{env}/2)/\Lambda \approx 99.7$\% of the total HOMO width.
Compared to the contribution of the surrounding, the contribution of the electrodes the the HOMO broadening is completely negligible.

To obtain the aforementioned values of $f$ and $\Gamma_{env}$
we used the values of $\overline{\Gamma}$ and $\lambda$
obtained by fitting the experimental $I$-$V$ data \cite{Whitesides:15}
to the exact \gl~(\ref{eq-jex}). 
The values collected in Table~(\ref{table:sc14}) reveal that these values
do not significantly differ from
the estimates for $f$ and $\Gamma_{env}$ based on \gl~(\ref{eq-j-0K}).

To end, it is worth emphasizing that
the off-resonant single level model (\gl~(\ref{eq-j-0K-off}\cite{Baldea:2012a}) 
is also useful for large area EGaIn junctions. \Gl~(\ref{eq-j-0K-off}) does not only reproduce
$I$-$V$ curves in nonlinear regime (green line in \figname\ref{fig:sc14}b) but also allows to reasonably estimate
the fraction of current carrying molecules. The rather minor
difference between the value $f = 4.9 \times 10^{-4}$
based on \Gl~(\ref{eq-j-0K-off}) and the ``exact'' value $f = 4.3 \times 10^{-4}$
deduced \emph{via} \gl~(\ref{eq-jex})
is due to the fact, albeit smaller than $\vert \varepsilon_{0}\vert$, $\Lambda$ is not smaller
that $\vert \varepsilon_{0}\vert / 10$, which is the condition for \gl~(\ref{eq-j-0K-off})
to be very accurate.\cite{baldea:comment,Baldea:2023a} 
}
\section{Predicting a Bias-Tuned Sommerfeld-Arrhenius Transition in a Real Molecular Tunnel Junction}
\label{sec:mcbj}
\begin{figure*}[htb]
  \centerline{
    \includegraphics[width=0.3\textwidth,height=0.22\textwidth,angle=0]{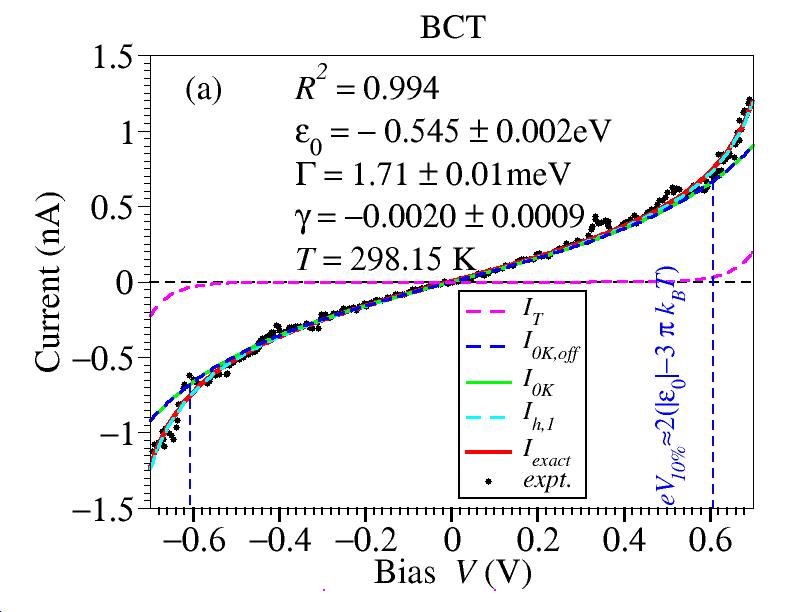} 
    \includegraphics[width=0.3\textwidth,height=0.22\textwidth,angle=0]{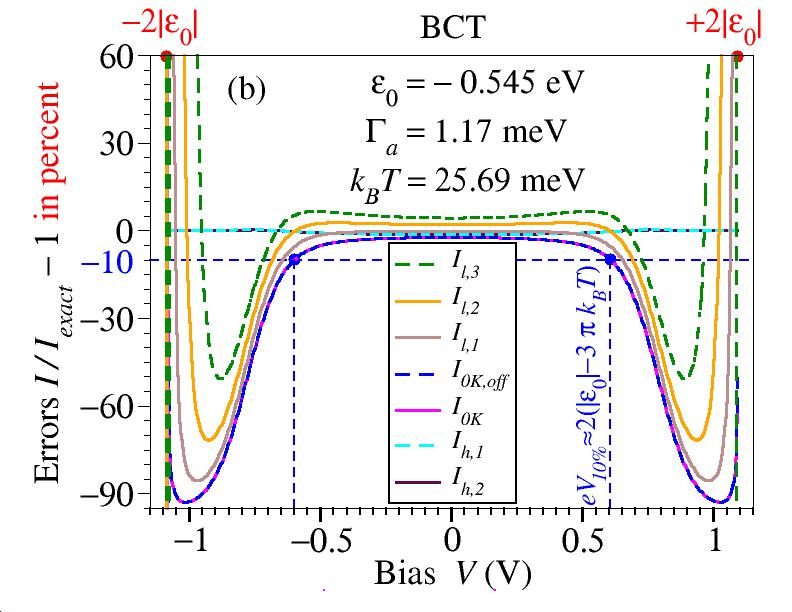} 
  }
  \centerline{
    \includegraphics[width=0.3\textwidth,height=0.22\textwidth,angle=0]{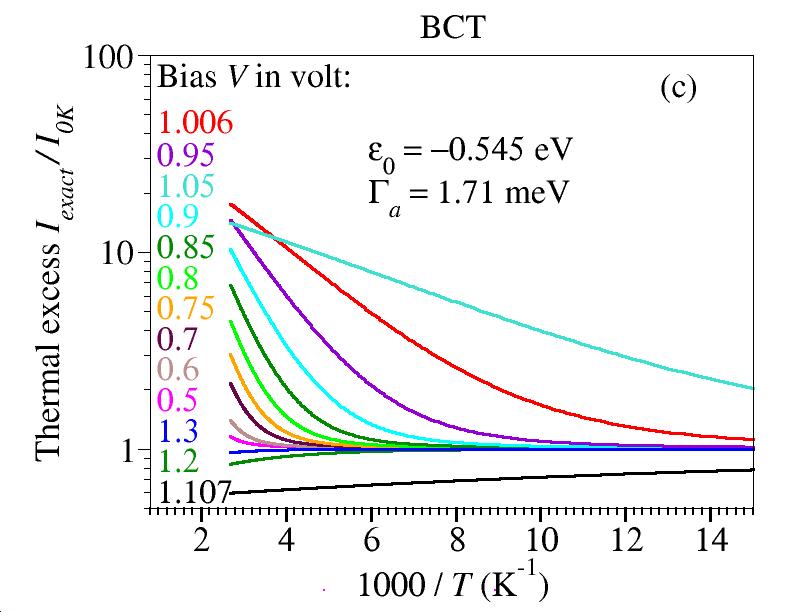}\hspace*{39ex} 
  }
  $ $\\[-32.5ex]
  \centerline{\hspace*{40ex}\includegraphics[width=0.22\textwidth,height=0.34\textwidth,angle=-90]{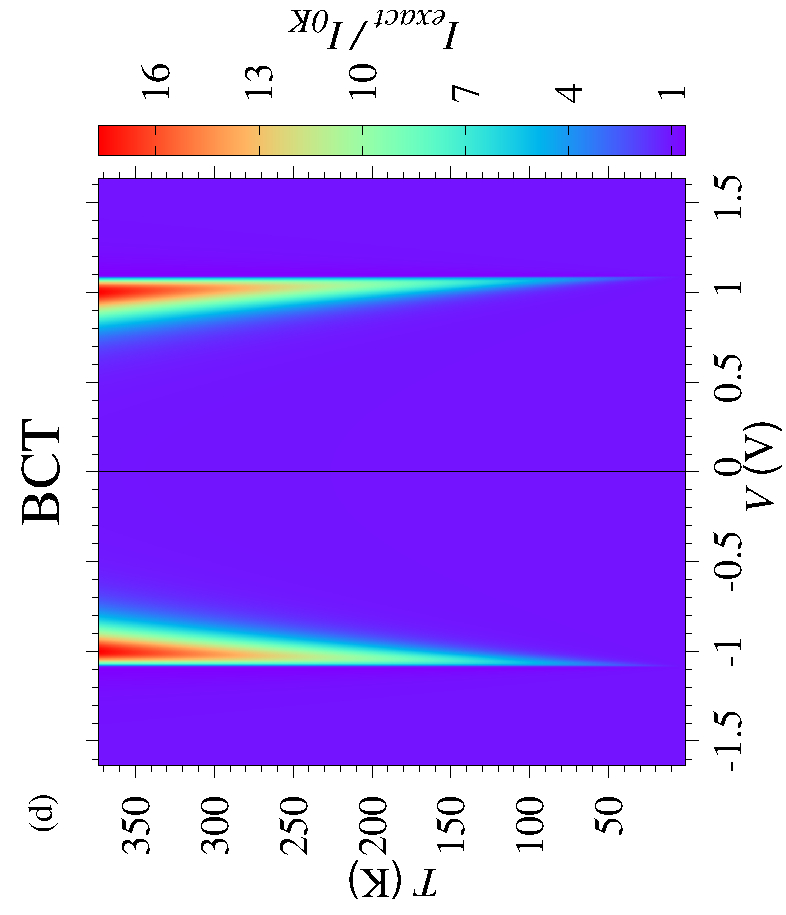} 
  }
  \centerline{
    \includegraphics[width=0.3\textwidth,height=0.22\textwidth,angle=0]{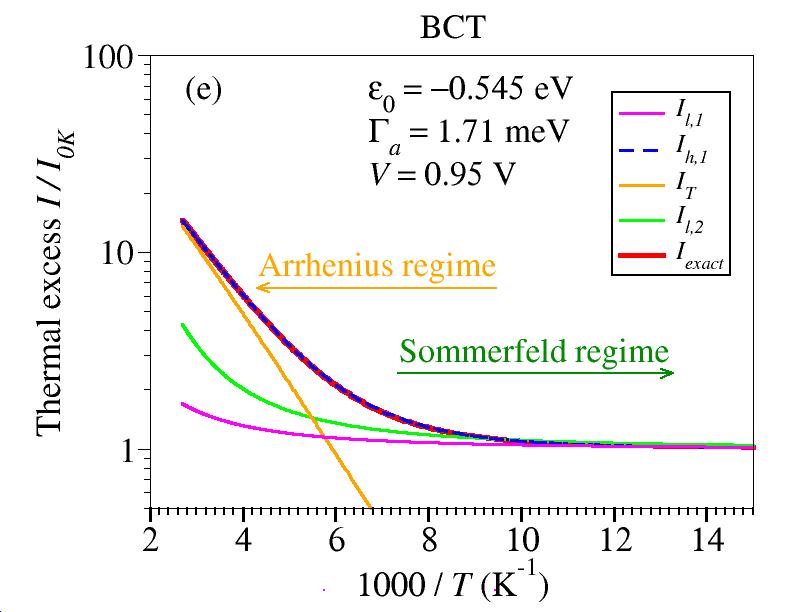} 
    \hspace*{39ex}}
  $ $\\[-32.5ex]
  \centerline{\hspace*{40ex}\includegraphics[width=0.22\textwidth,height=0.34\textwidth,angle=-90]{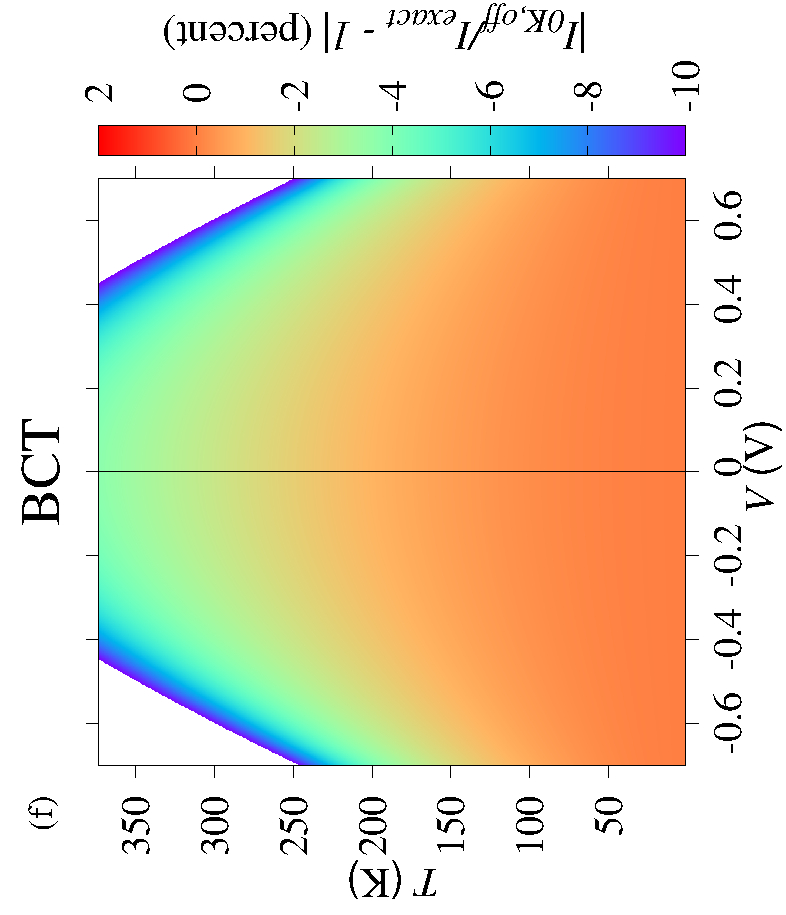} 
  }
  \caption{(a) The experimental $I$-$V$ curve for a 4,4'-biscyanotolane (BCT)
    mechanically controlled break junction (MC-BJ)
    obtained by digitizing \figname{1} of Ref.~\citenum{Zotti:10} is excellently fitted
    by the theoretical current $I_{exact}$ given by \gl~(\ref{eq-Iex}) at $T = 298.15$\,K.
    Curves computed using various approximate formulas for current
    with the optimized model parameters (see legend) are also presented.
    (b) Percentage deviations from the exact current of the current computed using the methods indicated in the legend
    shown in a bias range broader than the the bias range explored in experiment (depicted by vertical dashed lines).
    The thermal excess current $I_{exact}/I_{0K}$ depicted (c) as a function of temperature ($T$) at several bias values ($V$)
    and (d) in the ($V$, $T$)-plane. Notice the large values of $I_{exact}/I_{0K}$ in situations slightly below
    resonance.
    (e) At biases slightly below resonance
    ($V = 0.95$\,V, $e V \alt 2 \left\vert\varepsilon_{0}\right\vert$) $I_{exact}$ switches from a weakly $T$-dependent 
    (Sommerfeld) regime to a strongly $T$-dependent (Arrhenius-like) regime wherein it can be approximate by the
    expressions deduced for the low and
    high $T$ limits, respectively.
    (f) Regions in the ($V$, $T$)-plane where \gl~(\ref{eq-I-0K-off}) is accurate within the experimental accuracy.
    }
  \label{fig:bct}
\end{figure*}

Let us now switch to a mechanically controlled break junction (MC-BJ) of 4,4'-biscyanotolane (BCT)
with gold electrodes.\cite{Zotti:10}
Results for this junction are collected in \figname\ref{fig:bct}.
Fitting the digitized $I$-$V$ data measured at room temperature (\figname{1} of Ref.~\citenum{Zotti:10})
to \gl~\ref{eq-Iex}) at $T = 298.15$\,K yielded
an excellent fitting curves ($R^2 = 0.994$, cf.~\figname\ref{fig:bct}a).

Within the bias range sampled in experiment ($ \vert V\vert < 0.7$\,V),\cite{Zotti:10}
thermal effects are non-negligible (thermal excess higher than 10\%)
only at the ends of the experimental range:
$\vert V \vert > V_{10\%} = 0.607\,\mbox{V}$
($e V_{10\%} \approx 2 \left(\left\vert\varepsilon_{0}\right\vert - 3 \pi k_B T\right) \approx
2 \left(\left\vert\varepsilon_{0}\right\vert - 10 k_B T\right)$);
see \figname\ref{fig:bct}b. Except for biases very close to the highest experimental bias ($V = 0.7$\,V),
even the zero temperature expressions ($I_{0K}$ and $I_{0K,off}$) are accurate; see \figsname\ref{fig:bct}e and f.
For such situations, the low temperature expansion ($I_{l,1}$, $I_{l,2}$, and $I_{l,3}$) provides
an improved description over the zero temperature description ($I_{0K}$ and $I_{0K,off}$, see \figname\ref{fig:bct}b).
More significant is however the high accuracy of the
high-temperature approximations:
at the small value $\Lambda = 1.71 \pm 0.01\mbox{\,meV} \ll k_B T = 25.7 \mbox{\,meV}$
obtained by data fitting,
even the lowest order expansion ($I_{h,1}$) excellently reproduces the exact current $I_{exact}$;
higher order corrections embodied in $I_{h,2}$ are superfluous.

Beyond the bias range explored in Ref.~\citenum{Zotti:10}, the thermal current enhancement $I_{exact}/I_{0K}$ becomes
more and more pronounced as the resonance value is approached from below ($e V \alt 2 \left\vert \varepsilon_{0}\right\vert$)
and the temperature is increased. Results up to the highest temperature of
potential practical interest ($T = 373.15$\,K) are presented in \figsname\ref{fig:bct}c, d, and e.
They show that for biases sufficiently close to resonance, values up to
$I_{exact}/I_{0K} \approx 20$ can be reached. In such situations,
a gradual Arrhenius-Sommerfeld transition \cite{Baldea:2022c} can become observable (\figname\ref{fig:bct}d).

In Table~\ref{table:fit-bct}, we collect the model parameters $\varepsilon_{0}$($<0$, HOMO conduction),
$\Lambda $($= \Gamma$, single-molecule junction
fabricated with a symmetric molecular species) estimated from data fitting to the various formulas for
the current.
\begin{table}
\small
\centering
\caption{
  {Model parameter values obtained by fitting the experimental BCT data \cite{Zotti:10} to the various formulas for the current, as indicated in the first column}
}
\footnotesize
\begin{tabular*}{0.48\textwidth}{@{\extracolsep{\fill}}rccccc}
  \hline
  Method       &   $\varepsilon_{0}$ (eV)  &  $\Gamma = \Lambda$        &  $\gamma$ &  $R^2$  \\
\hline
  $I_{exact}$   &   $ -0.545 \pm 0.002 $   &  $ 1.71 \pm 0.01 $   & $ -0.0021  \pm 0.0009 $ & 0.993551 \\
  $I_{h,1}$     &   $ -0.547 \pm 0.002 $   & $ 1.73 \pm  0.01 $  & $ -0.0021 \pm 0.0009 $ & 0.993513 \\
  $I_{h,2}$     &   $ -0.547 \pm 0.002 $   & $ 1.72 \pm  0.01 $  & $ -0.0021 \pm 0.0009 $ & 0.993513 \\
  $I_{0K}$      &   $ -0.472 \pm 0.004 $   & $ 1.46 \pm  0.02 $  & $ -0.003  \pm 0.001 $ & 0.992632 \\
  $I_{0K,off}$  &   $ -0.472  \pm 0.004 $   & $ 1.46 \pm 0.02 $   & $-0.003 \pm 0.001 $ & 0.992632 \\
  $I_{l,1}$     &   $ -0.492 \pm  0.004 $  & $ 1.50 \pm  0.02 $  & $ -0.003 \pm 0.001 $ & 0.992729 \\
  $I_{l,2} $    &   $ -0.511 \pm  0.003 $  & $ 1.55 \pm  0.02 $  & $ -0.003 \pm 0.001 $ & 0.992809 \\
  $I_{l,3} $    &   $ -0.526 \pm  0.003 $  & $ 1.58 \pm  0.02 $  & $ -0.003 \pm 0.001 $ & 0.992860 \\
  \hline
\end{tabular*}
\label{table:fit-bct} 
\end{table}
As visible there, applying the zero temperature formulas ($I_{0K}$ and $I_{0K, off}$) to situations
not enough far away from resonance and $\Lambda \ll k_B T$ 
yielded an error of 22\% in the value estimated for the HOMO offset and 15\% for $\Gamma$.
\ib{
\section{Remarks on the applicability of presently deduced formulas}
\label{sec:remarks}
Going through the plethora of equations presented above, 
an experimentalist less interested in how they were mathematically deduced
may ask
what is the appropriate formula needed and how to proceed in transport data processing
for a specific junction, fabricated with a specific molecule, specific electrodes, and a specific platform.
This section and the next two sections aim at addressing such legitimate questions.

Basically, within the model parameter ranges indicated, the formulas of
\secsname\ref{sec:model}, \ref{sec:j-0K-T}, and \ref{sec:analytic-approx} are general:

\subsection*{With regard to electrodes:} They apply to all electrodes
having flat conduction bands around the Fermi energy $E_F$
\emph{i.e.}, having a density of states practically independent of energy $\rho(\varepsilon) \approx \rho(E_F)$.
This translates into energy independent MO-electrode couplings
($\Gamma_{s,t} (\varepsilon) \approx \Gamma_{s,t} (E_F) \propto  \rho(E_F) \tau_{s,t}^2$, see below).
While this is not the case of semiconductors or graphene, it is definitely
the case of all typical metals (like Ag, Au, Pt).\cite{Papaconstantopoulos:15} 

\subsection*{With regard to molecules:}
The formulas apply to all molecules wherein the charge transport is dominated by a single level (MO),
which is in most cases either the HOMO or the LUMO. The applicability of the
single level description is often justified intuitively by the sufficiently large
energy separation from the adjacent MOs.
In fact, MOs' energetic separation may be a sufficient condition but is not a necessary condition. 
If, in some cases, a few highest occupied (\emph{e.g.}, HOMO and HOMO-1)
or a few lowest unoccupied (\emph{e.g.}, LUMO and LUMO+1) have energies close to each other,
they typically have different symmetries and/or different spatial (de)localization.
The result is that, out of each group (of occupied or unoccupied MOs),
only one MO has significantly overlap with the electronic states in metals.
Due to different symmetry (and/or (de)localization) it can happen, \emph{e.g.},
that the (more distant from $E_F$) HOMO-1 rather than the (closer to $E_F$) HOMO has a
larger hybridization with the single-electron states in the substrate (or tip/top) electrode.
A larger hybridization means a transfer integral
$\tau_s$ quantifying the charge transfer between
HOMO-1 and substrate larger than that of the charge transfer between HOMO and substrate.
This translates into MO-substrate (or tip) coupling 
$\Gamma_{s,t} \propto  \rho(E_F) \tau_{s,t}^2$ dominated by the HOMO-1 and not by the HOMO.

\subsection*{With regard to the fabrication platform:} 
With appropriate model parameter values (to be adequately estimated from data fitting),
these formulas can be applied
for molecular junctions fabricated using any platform, let it be MC-BJ, STM-BJ, CP-AFM or
(as we have just seen in \secname\ref{sec:sc14}) EGaIn.

It may be helpful to recall \cite{Baldea:2022j} here that,
like the model proposed in ref.~\citenum{Baldea:2012a},
the present single level model does \emph{not} assume
negligible intermolecular interactions. An electron that tunnels
across a molecule can interact with the adjacent molecules. Provided that 
electron exchange (transfer) between adjacent molecules is absent, 
the single level model ``allows'' intermolecular interactions to give rise to an extra MO shift
(thence possible different values of $\varepsilon_0$ for
single-molecule, CP-AFM, and EGaIn junctions, a fact which will reflect itself in the best fitting estimates)
and an extra level broadening expressed by $\Gamma_{env}$.\cite{Baldea:2022j}
This is the rationale of ``simply'' multiplying the present expressions for current
through a single molecule by the number of molecules per junction, as done earlier in case of CP-AFM junctions \cite{Baldea:2015d,Baldea:2019d,Baldea:2019h}
or in arriving at \gl~(\ref{eq-j}) above.
\section{Protocol for studying a specific junction}
\label{sec:protocol}
In view of the results presented above, assuming that room temperature $I$-$V$ measurements for a specific molecular
junction are available (which is the common case), we recommend the following steps for data processing:

(i) Start by plotting the measured data in coordinates ($V^2/|I|, V$)
and determine the transition voltages $V_{t\pm}$ from the peak location
(\emph{cf.}~\figname\ref{fig:sc14}a). Use our \gl~(6) and (7) of ref.~\citenum{Baldea:2012a}
(reproduced below for the reader's convenience) to estimate the magnitude of the
MO offset $\vert \varepsilon_{0}\vert$ and $\gamma$
\begin{eqnarray*}
\left\vert \varepsilon_{0} \right\vert
& = & 2 \frac{V_{t+}\left\vert V_{t-}\right\vert}{\sqrt{V_{t+}^2 + 10 V_{t+}\left\vert V_{t-}\right\vert /3 + V_{t-}^2}} \\
\gamma & = & \frac{\mbox{sign}\,\varepsilon_{0}}{2} \frac{V_{t+} + \left\vert V_{t-}\right\vert}{\sqrt{V_{t+}^2 + 10 V_{t+}\left\vert V_{t-}\right\vert /3 + V_{t-}^2}}
\end{eqnarray*}

The MO offset $\varepsilon_{0}$ is negative for p-type (normally, HOMO) conduction and
positive for n-type (normally, LUMO) conduction.
Recall that, whatever the formulas for the current utilized, the sign of $ \varepsilon_{0}$
cannot be merely determined from $I$-$V$ curves measured for a given junction (= given molecule, given electrodes). 
Thermopower experiments to determine the sign of the Seebeck coefficient $ S $ or
investigating junctions fabricated with a given molecule and electrodes having different work functions $\Phi$.
\cite{Baldea:2015d,Baldea:2019h,Baldea:2019d} 
may settle this issue; positive (negative) $S$ means negative (positive) $\varepsilon_0$,\cite{Datta:03b}
higher (lower) current for larger $\Phi$ means negative \cite{Baldea:2015d,Baldea:2019h,Baldea:2019d} (positive \cite{Baldea:2018a})
$\varepsilon_0$; 

Determine next the value of $\Gamma = \left\vert \varepsilon_{0}\right\vert \sqrt{G/G_0}$ using
the low bias conductance $G$ obtained from the slope of the measured $I$-$V$ curve at low bias. With the values of 
$ \varepsilon_{0}$, $\Gamma$ and $\gamma$ thus obtained compare the measured curve with the simulated curve computed using our \gl~(\ref{eq-I-0K-off}) in the bias range $-1.2\,\left\vert V_{t-}\right \vert \alt V \alt 1.2\,V_{t+}$.
Alternatively, always in the bias range indicated (for which \gl~(\ref{eq-I-0K-off}) applies),
one can fit the measured data to $I_{0K-off}$ of \gl~(\ref{eq-I-0K-off}) wherein $ \varepsilon_{0}$, $\Gamma$ and $\gamma$
are adjustable parameters. 

(ii) If attempt (i) fails to produce a good theoretical curve, fit data to $I_{0K}$ of \gl~(\ref{eq-I-0K})
in the entire bias accessed experimentally with
$ \varepsilon_{0}$, $\Gamma$, $\gamma$, and $\Lambda$ as adjustable parameters.
If the corresponding point ($ \varepsilon_{0}^{0K}$, $\Lambda^{0K}$) falls in the colored regions of 
\figsname\ref{fig:V=0}a, \ref{fig:V=0.1}a, \ref{fig:V=0.5}a, or \ref{fig:V=1.0}a,
this is evidence that the best fit values $ \varepsilon_{0}^{0K}$, $\Gamma^{0K}$, $\gamma^{0K}$ and $\Lambda^{0K}$
are reliable and that, at the corresponding bias, \gl~(\ref{eq-I-0K}) is reliable. In other words, at the envisaged bias,
thermal corrections to the current are negligible. Notice that (like \gl~(\ref{eq-I-0K-off}) \gl~(\ref{eq-I-0K}))
can hold up to a certain bias (\emph{e.g.}, $V = 0.5$\,V) but can become inadequate at higher biases
(\emph{e.g.}, $V = 1$\,V) (see the example of \secname\ref{sec:mcbj}).

(iii) The failure of attempt (ii) indicates that (possibly only beyond certain biases) the temperature dependence
of the current is significant. Generally speaking, whenever possible (\emph{e.g.}, using MATHEMATICA)
fitting data to the exact \gl~(\ref{eq-Iex}) is then most advisable.
Otherwise the various analytical formulas for the current $I_{l,1}$, $I_{l,2}$, $I_{l,2}$ of
\gl~(\ref{eq-sommerfeld-exp}) or $I_{h,1}$, $I_{h,2}$ of \gl~(\ref{eq-T}) can be easier applied.
Because they merely contain
elementary functions the approximate formulas for $I_{l,1}$, $I_{l,2}$, $I_{l,2}$, $I_{h,1}$, and $I_{h,2}$
are implementable in any fitting software package.

Loosely speaking, the deeper the point ($ \varepsilon_{0}^{0K}$, $\Lambda^{0K}$) into the white (non-colored) regions mentioned under
(ii) above, the stronger is the impact of $T$ on the current at the corresponding bias.

(iii$_a$) For cases wherein the point ($ \varepsilon_{0}^{0K}$, $\Lambda^{0K}$)
lies near the edge between colored and non-colored regions of
\figsname\ref{fig:V=0}a, \ref{fig:V=0.1}a, \ref{fig:V=0.5}a, or \ref{fig:V=1.0}a, thermal effects are rather weak,
and using the formula of $I_{l,1}$ of \gl~(\ref{eq-I-l1}) for data fitting is advisable. If the best fitting point
($ \varepsilon_{0}^{l,1}$, $\Lambda^{l,1}$) obtained fitting the experimental $I$-$V$ curve to \gl~(\ref{eq-I-l1})
is found in the colored regions of \figsname\ref{fig:V=0}b, \ref{fig:V=0.1}b, \ref{fig:V=0.5}b, or \ref{fig:V=1.0}b
$I_{l,1}$ provides an accurate description and the fitting parameters $ \varepsilon_{0}^{l,1}$, $\Gamma^{l,1}$, $\gamma^{l,1}$, and $\Lambda^{l,1}$
are good estimates of the corresponding quantities.
Otherwise, performing and inspecting the results of data fitting based on
$I_{l,2}$ of \gl~(\ref{eq-I-l2}) (and then $I_{l,3}$ of \gl~(\ref{eq-I-l3}) if the case) follows in the next step.

(iii$_b$) If the point ($ \varepsilon_{0}^{0K}$, $\Lambda^{0K}$) lies deep inside the white (non-colored) regions of
\figsname\ref{fig:V=0}a, \ref{fig:V=0.1}a, \ref{fig:V=0.5}a, or \ref{fig:V=1.0}a, thermal effects are strong, and the formula deduced
via high temperature expansions ($I_{h,1}$ of \gl~(\ref{eq-I-h1}) and $I_{h,2}$ of \gl~(\ref{eq-I-h2}) should be employed for
reliably estimate the parameters $ \varepsilon_{0}$, $\Gamma$, $\gamma$, and $\Lambda$ for the junction under investigation.

In cases where (iii$_b$) applies, performing $I$-$V$ measurements at variable temperature is highly desirable,
and the issues related to the Sommerfeld-Arrhenius transition addressed in the next section deserve consideration.

\section{Tunneling-throughout Sommerfeld-Arrhenius scenario versus tunneling-hopping transition: a too challenging a dilemma?}
\label{sec:dilemma}
For a Sommerfeld-Arrhenius transition to occur in a specific molecular junction,
the parameters $\varepsilon_0$ and $\Lambda$ at given bias $V$
should fall in regions in the plane ($\varepsilon_0$, $\Lambda$)
where the current thermal enhancement $I_{exact}/I_{0K}$ is sufficiently large
(say, at least $I_{exact}/I_{0K} \agt 10$), like the red curve in \figname\ref{fig:bct}e).
The color coding of \figsname\ref{fig:V=0}a, \ref{fig:V=0.1}a, \ref{fig:V=0.5}a, or \ref{fig:V=1.0}a
precisely depicts which combination of values ($\varepsilon_0, \Lambda, V$)
is ``eligible'' for a Sommerfeld-Arrhenius transition (\emph{i.e.}, $I_{exact}/I_{0K}$ is reasonably large).

Experimental curves for current versus inverse temperature having a shape
resembling the red curve in \figname\ref{fig:bct}e have been measured in the past for several
molecular junctions.
Examples include junctions fabricated with:
conjugated oligo-tetrathiafulvalenepyromelliticdiimideimine (OPTIn; OPTI2 in \figname{6a} in ref.~\citenum{Choi:10b}),
bis-thienylbenzene (BTB; \figsname{3c and 4} in ref.~\citenum{McCreery:13b}), 
cytochrome C and hemin-doped human serum albuminum (HSA-hemin)
(Cyt C and HSA-hemin; \figsname{6a and b} in ref.~\citenum{Amdursky:13b}),  
and closed and open diarylethene isomers (\figsname{2c and d} in ref.~\citenum{Guo:17}).
Out of the cases cited, only ref.~\citenum{McCreery:13b} reported attempts (that failed)
of quantitatively fitting the measured curve of $I$ \emph{versus} $1/T$ to
a theoretical model that could come into question, namely the variable hopping range model \cite{Mott:61,Hill:71,Hill:76}
and the classical Poole-Frenkel model for transport between coulombic traps.\cite{Poole:16,Frenkel:38}

Based on the previous knowledge in the field, all the aforementioned data were interpreted as evidence of a
tunneling-hopping transition. However, this interpretation needs to be reconsidered,
and all the more so since the present theory of the Sommerfeld-Arrhenius scenario provides a unique formula
that can be applied for the entire temperature and bias ranges accessed experimentally.

A dilemma may arise in cases where the measured current $I$ follows an Arrhenius pattern at high $T$
and eventually switches to T-independent values at low $T$: is the dependence
\(I \propto \exp\left( - \frac{E_a}{k_B T}\right)\) at high $T$ due to thermally activated hopping
or evidence for the strongly $T$-dependent tunneling current discussed in this paper?

To differentiate between these two possibilities, 
in previous studies we suggested investigations of junctions having
electrodes with significantly different work function \cite{Baldea:2017d,Baldea:2018a}
or subject to mechanical stretching.\cite{Baldea:2022j}
The present results indicate two alternatives easier to implement experimentally in order
to differentiate between strongly $T$-dependent tunneling and hopping:

(a)
To unravel the physics underlying an observed switching from strong $T$-dependent to $T$-independent (or weak $T$-dependent) currents,
$I$-$V$ curves measured at low temperatures can be fitted to \gl~(\ref{eq-0K}) assuming $T$-independent $I$
(or \gl~(\ref{eq-sommerfeld-exp}) assuming weak $T$-dependent $I$). The value \(\vert\varepsilon_0\vert\) of the MO offset
thus obtained can be compared:

(a$_1$) with the value of the ``activation energy'' $E_a$ obtained by fitting the high temperature data to the Arrhenius law
$I \propto \exp\left(-\frac{E_a}{k_B T}\right)$. An $E_a$ significantly different from \(\vert\varepsilon_0\vert\) is evidence for
hopping-tunneling transition, \(\vert\varepsilon_0\vert\ \approx E_a\) pleads for a tunneling-throughout Sommerfeld-Arrhenius scenario.

(a$_2$) with the value of the MO offset \(\vert\varepsilon_{1/2}\vert\) obtained \emph{via} ultraviolet photoelectron spectroscopy (UPS).
Obviously, this makes sense in case of p-type conduction, where evidence exists that the ``second'' (tip) electrode negligibly
affect the level alignment (\(\varepsilon_{1/2} \approx \varepsilon_{0}\)), which is basically the same for full and ``half''
(\emph{i.e.} only SAM adsorbed on substrate without tip/top electrode) junctions.\cite{Baldea:2015d,Baldea:2019d,Baldea:2019h}
For n-type conduction determining the (LU)MO offset \(\varepsilon_{1/2}\)
\emph{via} inverse photoelectron spectroscopy (IPS) could come into question.

(b)
To start with a specific example, let us consider again the BCT junction of \secname\ref{sec:mcbj} and compare
the ratio of the currents at room temperature and at liquid nitrogen
$\left . I_{RT}/I_{LN}\right\vert_{V} \equiv I(V, 298.15\,\mbox{K})/I(V, 77\,\mbox{K})$
computed using the exact \gl~(\ref{eq-Iex}) for several biases $V$:
$\left . I_{RT}/I_{LN}\right\vert_{V = 1\,\mbox{V}} = 11.38 $;
$\left . I_{RT}/I_{LN}\right\vert_{V = 0.9\,\mbox{V}} = 5.55 $; 
$\left . I_{RT}/I_{LN}\right\vert_{V = 0.8\,\mbox{V}} = 2.20 $; 
$\left . I_{RT}/I_{LN}\right\vert_{V = 0.7\,\mbox{V}} = 1.31 $;
$\left . I_{RT}/I_{LN}\right\vert_{V = 0.6\,\mbox{V}} = 1.10 $; 
$\left .  I_{RT}/I_{LN}\right\vert_{V = 0.5\,\mbox{V}} = 1.05 $;
$\left . I_{RT}/I_{LN}\right\vert_{V = 0.4\,\mbox{V}} = 1.03 $;
$\left . I_{RT}/I_{LN}\right\vert_{V = 0.4\,\mbox{V}} = 1.02 $.
These values show an almost exponential decrease in the thermal enhancement $ I_{RT}/I_{LN} $
in the narrow bias range $0.7\,\mbox{V} \alt V \alt 1\,\mbox{V}$ slightly below resonance ($e V_{resonance} = 2 \vert \varepsilon_{0} \vert \simeq 1.1\,\mbox{V}$).
By contrast, sufficiently away from resonance ($ V \alt 0.6\,\mbox{V}$)
the thermal enhancement is practically negligible ($ \left . I_{RT}/I_{LN} \right\vert_{V\alt  0.6\,\mbox{V}} \approx 1$).
This example illustrates that the Sommerfeld-Arrhenius transition is bias sensitive. In a given junction,
a Sommerfeld-Arrhenius transition is observable at some biases while at other biases it cannot be observed.
Such a bias sensitivity could not be expected if hopping dominated at higher temperature. The thermal activation energy
is mainly determined by molecular properties (\emph{e.g.}, reorganization energy) which are little sensitive to external bias.

To sum up, by and large, 
data measured over a broad range of biases exhibiting a similar pattern (gradual transition between two distinct regimes)
plead for a tunneling-hopping transition. Otherwise
(i.e., if switching between a weak temperature dependence at low $T$ to a strong $T$ dependence at high $T$
mainly occurs in a narrow range of $V$, see \figname\ref{fig:bct}c) an overall tunneling picture applies;
the temperature dependence is merely the result of thermally broadened Fermi distributions.
}
\section{Conclusion}
\label{sec:conclusion}
We hope that the various analytical formulas for the tunneling current reported here will assist
experimentalists to adequately process transport data measured at nonvanishing (room) temperature
and to more properly assign the transport mechanism at work in a specific
molecular junction. 
Letting alone the aspect of elegance, the present analytical formulas have an important practical advantage  
over the often utilized brute force method of numerically integrating \gl~(\ref{eq-I}) using a uniform energy grid.
To resolve both the thermal smearing of the Fermi distributions and the variation of transmission with energy, a
sufficiently fine uniform energy grid requires an energy step size 
$\delta \varepsilon \approx \min\left(k_B T, \Lambda\right)/10$. In an energy range
$\Delta \varepsilon \sim 10 $\,eV this easily translates into a grid with $N \sim 10^4 - 10^6$ points,
which may make data fitting time consuming in some cases.

To be sure, adaptive numerical quadrature can alternatively be employed,
and it is to be preferred to the brute force approach.
Still, in spot checks to reproduce the present exact results obtained via \gl~(\ref{eq-Iex}),
we had to conclude that applying adaptive numerical quadrature to \gl~(\ref{eq-I})
with default settings (precision goal, working precision, etc)
using current distributions of MATHEMATICA (13.2.1) and MATLAB (R2021a)
also poses nontrivial numerical problems for small $\Lambda$ and $k_B T$
($k_B T, \Lambda \alt 1$\,meV).

Based on the results reported in this paper, for experimentalists unable to use
\gl~(\ref{eq-Iex}) to data fitting---
e.g., because neither MATLAB nor ORIGIN can handle digamma function of complex argument---
we recommend utilization of \gl~(\ref{eq-I-h1}) and \gl~(\ref{eq-I-l2}),
which provide a accurate description that complimentary covers most situations of practical interest.
Albeit somewhat lengthy, they merely comprise elementary functions and can be used with any
routine fitting software.

A cursory glance at an experimental curve looking like that
depicted in \figname\ref{fig:bct}d
can easily be taken as ``evidence'' of 
a tunneling-hopping crossover or--- in view of the $T$-dependent slope (``activation energy'')---
of a variable variable-range hopping. In reality, we have seen that,
it is the tunneling that is at work there both at low and high $T$.
Emphasizing this point, the results reported above aim at aiding the molecular electronics community
in not rushing conclude a transition from tunneling to hopping merely
based on measured transport properties that switch from nearly temperature independent (Sommerfeld regime)
to strongly temperature dependent (Arrhenius-like regime) upon rising the temperature.

The fact that the presently discussed Sommerfeld-Arrhenius can be tuned by bias can be an important
theoretical finding in correctly assessing the physics underlying a curve like that of \figname\ref{fig:bct}d.

\ib{
  Not to forget, we have also shown that the single level model considered in this paper
  is useful not only for single-molecule \cite{Zotti:10,Baldea:2012g,LukaGuth:16} and CP-AFM
  \cite{Tan:10,Baldea:2015d,Baldea:2017e,Baldea:2019h}
  junctions for which it was previously validated. It can also
  decipher charge transport in large area ($\mathcal{A}_{n} \approx 500\,\mu$m$^2$) junctions with EGaIn top electrode.
  For example, we have shown that, when adequately applied, even our off-resonant single level model \cite{Baldea:2012a} 
  can provide a reasonable estimate of the fraction of the current carrying molecules
  ($f = N_{\mbox{\small eff}} /N_{n} = \mathcal{A}_{\mbox{\small eff}} /\mathcal{A}_{n}\approx 4 \times 10^{-4}$)
  in a large area EGaIn junction fabricated with 1-tetradecanethiol.
}
\section*{Acknowledgments}
Financial support from the German Research Foundation
(DFG Grant No. BA 1799/3-2) in the initial stage of this work and computational support by the
state of Baden-W\"urttemberg through bwHPC and the German Research Foundation through
Grant No.~INST 40/575-1 FUGG (bwUniCluster 2, bwForCluster/HELIX, and JUSTUS 2 cluster) are gratefully acknowledged.
\balance

\renewcommand\refname{References}
\providecommand*{\mcitethebibliography}{\thebibliography}
\csname @ifundefined\endcsname{endmcitethebibliography}
{\let\endmcitethebibliography\endthebibliography}{}

\end{document}